\documentclass{emulateapj}
\usepackage{amsmath}
\usepackage{amssymb}
\usepackage{ulem}
\usepackage{longtable}
\usepackage{lineno, blindtext}
\usepackage{graphicx}
\usepackage{epstopdf}
\usepackage{natbib}
\usepackage[colorlinks, citecolor=blue]{hyperref}
\usepackage{hyperref}
\usepackage[all]{hypcap}
\usepackage{url}
\usepackage{etoolbox}
\usepackage{multirow}
\usepackage{xpatch}
\usepackage{lineno}

\shorttitle{}

\begin{document}

\title{Evidence of photosphere emission origin for gamma-ray burst
prompt emission}
\author{Yan-Zhi Meng\altaffilmark{1,2} }

\begin{abstract}
The physical origins of gamma-ray burst (GRB) prompt emission (photosphere or synchrotron) are still
subject to debate, after more than five decades. Here, we find that many of the observed characteristics of 15 long GRBs, which have the highest prompt emission efficiency $\epsilon _{\gamma}$ ($%
\epsilon_{\gamma }\gtrsim 80\%$), strongly support photosphere (thermal)
emission origin, in the following ways: (1) the relation between $E_{\text{p}}$ and $E_{\text{iso}}$
is almost $E_{\text{p}}\propto (E_{\text{iso}})^{1/4}$ and the dispersion
is quite small; (2) the simple power-law shape of the X-ray afterglow light
curves and the presence of significant reverse shock signals in the optical afterglow
light curves; (3) the best fits using the cutoff power-law model for the
time-integrated spectra; and (4) the consistent efficiency from observations
(with $E_{\text{iso}}/E_{k}$) and the predictions from the photosphere emission
model (with $\eta /\Gamma $). We then further investigate the
characteristics of the long GRBs for two distinguished samples ($\epsilon
_{\gamma }\gtrsim 50\%$ and $\epsilon _{\gamma }\lesssim 50\%$). It is found
that the different distributions for $E_{\text{p}}$ and $E_{\text{iso}}$,
and the similar observed efficiency (from the X-ray afterglow) and
theoretically predicted efficiency (from the prompt emission or the optical
afterglow), follow the predictions of the photosphere emission model well. Also,
based on the same efficiency, we derive an excellent correlation of $\Gamma
\propto E_{\text{iso}}^{1/8}E_{\text{p}}^{1/2}/(T_{90})^{1/4}$ to estimate $%
\Gamma $. Finally, we show that different distributions for $E_{\text{p}}$ and $E_{%
\text{iso}}$, and the consistent efficiency, exist for short GRBs.
We also give a natural explanation of the extended emission ($\epsilon
_{\gamma }\lesssim 50\%$) and the main pulse ($\epsilon _{\gamma }\gtrsim
50\%$).
\end{abstract}


\affil{\altaffilmark{1}School of Astronomy and Space Science, Nanjing University, Nanjing
210023, China; yzmeng@nju.edu.cn} 
\affil{\altaffilmark{2}Key Laboratory of Modern Astronomy and Astrophysics (Nanjing
University), Nanjing 210023, Ministry of Education, China}

\section{INTRODUCTION}

More than 50 yr after its discovery, the radiation mechanism of gamma-ray burst (GRB)
prompt emission (photosphere emission or synchrotron emission) still remains unidentified %
\citep[e.g.,][]{Me2002,ZhangYan2011,Uhm2014,Geng18,Geng2019,Lin2018,ZhaBB2018,ZhangBB18b,ZhangBB2021,Li2019b,Li2021,Burgess2020,Yang2020,ZhaB2020,Iyya2021,Vyas2021,ZhangZ2021,Vere2022}%
. The photospheric emission is the basic prediction of the classical
fireball model \citep{Good1986,Pac1986} for a GRB, because the optical depth $%
\tau $ at the jet base is much larger than unity \citep[e.g.][]{Pi1999}. As
the fireball expands, the optical depth drops down. The internally
trapped thermal photons finally escape at the photosphere ($\tau =1$).
Indeed, based on the spectral analysis, a quasi-thermal component has been
found in several \textit{Swift} GRBs \citep{Ry2004,Ry2005,Ry2009} and 
\textit{Fermi} GRBs \citep{Gui2011,Gui2013,Axel2012,Ghir2013,Lar2015,Tang2021,Deng2022,Wang2022,Zhao2022},
especially in GRB 090902B \citep{Abdo2009,Ry2010,ZhaB2011}. But whether the
typically observed Band function (a smoothly joint broken power law; %
\citealt{Band1993}) or a cutoff power law (CPL) can be explained by the photosphere
emission, namely the photospheric emission model, remains unknown %
\citep[e.g.,][]{Abra1991,Thom1994,Me2000,Ree2005,Pe2011,Fan2012,Lazz2013,Ru2013,Be2015,Gao2015,Pe2015,Ry2017,Acun2018,Hou2018,Meng2018,Meng2019,Meng2021,Li2019a,Li2019c,Li2019d,Acun2020,Dere2020,Vere2020,Wang2020,Pars2022,Song2022}%
. If this scenario is true, the quasi-thermal spectrum should be broadened.
Theoretically, two different broadening mechanisms have been proposed (see
Appendix A): subphotospheric dissipation (namely, the dissipative photosphere
model; \citealt{Ree2005,Gian2007,Vur2016,Belo2016,Bhat2018,Bhat2020}) or geometric broadening
(namely, the probability photosphere model; %
\citealt{Pe2008,Pe2011,Lund2013,Deng2014,Meng2018,Meng2019,Meng2021}).

Previously, some of the implications of the statistical properties of the
spectral analysis results for a large GRB sample have appeared to support the
photosphere emission model. First, lots of bursts have a low-energy spectral
index $\alpha $ that is harder than the death line (or the maximum value, $\alpha $= 
$-$2/3) of the basic synchrotron model, especially for short GRBs and
the peak-flux spectrum \citep{Kan2006,ZhaBB2011,Bur2017}. Second, the
spectral width is found to be quite narrow for a significant fraction of
GRBs \citep{AxBo2015,Yu2015}. Third, for half or more of GRBs, the CPL is the best-fit empirical model \citep{Gold2012,Grub2014,Yu2016},
indicating that the photosphere emission model can naturally interpret their
high-energy spectra. Here, we find more convincing evidence for the
photosphere emission origin (especially the probability photosphere model) of GRB prompt emission, by obtaining the prompt efficiency $\epsilon _{\gamma }$.

The paper is organized as follows. In Section \ref{sec:data}, we state the data accumulation and the scaling relations predicted by the photosphere model. In Section \ref{sec:extreme}, we describe the evidence from long GRBs with extremely high prompt efficiency $%
\epsilon _{\gamma }$ ($\epsilon _{\gamma }\gtrsim 80\%$). Then, in
Section \ref{sec:long}, the evidence from long GRBs with $\epsilon
_{\gamma }\gtrsim 50\%$ and $\epsilon _{\gamma }\lesssim 50\%$ is shown. In
Section \ref{sec:short}, we illustrate the evidence from short GRBs. A
brief summary is provided in Section \ref{sec:summa}.

\section{DATA ACCUMULATION AND THE SCALING RELATIONS EXPECTED BY THE
FIREBALL MODEL}

\label{sec:data}

\subsection{Data Accumulation}

\begin{table*}[tbp]
\caption{The observed quantities for the bursts with extremely high
efficiency $\ $($\protect\epsilon _{\protect\gamma }\gtrsim 80\%$).}
\label{Tab_1}
\begin{center}
%\begin{tiny}%
\par
\renewcommand\arraystretch{0.8} 
\resizebox{\linewidth}{!}{
\begin{tabular}{cccccccccc}
\hline
GRB & z & $L_{\text{iso}}$ & $E_{\text{iso}}$ & $L_{\text{X,11}}$ & Efficiency & $E_{p,z}$ & $\alpha $ & $\beta $ & best-fitted \\ 
&  & (10$^{52}$ erg s$^{-1}$) & (10$^{52}$ erg) & (10$^{45}$ erg s$^{-1}$) &  & (keV) &  &  & 
model \\ \hline
990705$^{a}$ & 0.8424 & 1.61 $\pm $ 0.15$^{c}$ & 21.8 $\pm $ 0.8 & 1.14 & 0.83 (0.99$^{f}$) &  551 $\pm $ 17 &  -0.72 $\pm $ 0.03& -2.68  $\pm $ 0.15$^{e}$&  \\ 
000210$^{a}$ & 0.8463 & 9.76 $\pm $ 0.5$^{c}$ & 19.3 $\pm $ 0.5 & 1.67 & 0.74 (0.97$^{f}$) & 687 $_{-37}^{+39}$ &  &  &  \\ 
060927$^{b}$ & 5.47 & 10.8 $\pm $ 0.8 & 7.56 $\pm $ 0.46 & 0.28 & 0.96 & 459 $\pm $ 90 & -0.81 $\pm $ 0.36 &  &  \\ 
061007$^{b}$ & 1.261 & 10.9 $\pm $ 0.9 & 101 $\pm $ 1.4 & 1.11 & 0.97 & 965 $\pm $ 27 & -0.75 $\pm $ 0.02 & -2.79 $\pm $ 0.09 &  \\ 
080319B$^{b}$ & 0.9382 & 10.2 $\pm $ 0.9 & 142 $\pm $ 3 & 3.90 & 0.91 & 1307 $\pm $ 43 & -0.86 $\pm $ 0.01 & -3.59 $\pm $ 0.45 &  \\ 
080607$^{b}$ & 3.0363 & 225.9 $\pm $ 45.3 & 186 $\pm $ 10 & 3.46 & 0.97 & 1691 $\pm $ 169 & -1.08 $\pm $ 0.06 &  &  \\ 
081203A$^{b}$ & 2.05 & 2.82 $\pm $ 0.19 & 35.0 $\pm $ 12.8 & 1.52 & 0.90 & 1541 $\pm $ 756 & -1.29 $\pm $ 0.14 &  &  \\ 
110205A$^{b}$ & 2.22 & 2.51 $\pm $ 0.34 & 55.9 $\pm $ 5.3 & 2.12 & 0.92 & 715 $\pm $ 238 & -1.52 $\pm $ 0.14 &  &  \\ 
110818A$^{c}$ & 3.36 & 6.76 $\pm $ 0.76 & 21.7 $\pm $ 1.02 & 0.99 & 0.93 & 799.45 $\pm $ 371.90 & -1.19 $\pm $ 0.08 &  & CPL$^{d}$ \\ 
120729A$^{a}$ & 0.80 & & 2.3 $\pm $ 1.5 & 0.034 & 0.94 & 559 $\pm $ 57 &  &  &  \\ 
130606A$^{a}$ & 5.913 & & 28.3 $\pm $ 5.2 & 1.26 & 0.95 & 2032 $_{-346}^{+622}$ &  -1.14 $\pm $ 0.15 &  & CPL$^{e}$ \\ 
131108A$^{a}$ & 2.40 & 26.22 $\pm $ 0.6$^{c}$ & 54 $\pm $ 2.4 & 1.98 & 0.93 & 1217 $_{-88}^{+105}$ & -1.16 $\pm $ 0.07 &  & CPL$^{e}$ \\ 
150821A$^{a}$ & 0.755 & 0.77 $\pm $ 0.03$^{c}$ & 15.5 $\pm $ 1.2 & 1.05 & 0.78 & 765 $_{-126}^{+188}$ & -1.52 $\pm $ 0.05 &  & CPL$^{e}$ \\ 
160410A$^{a,s}$ & 1.717 & & 9.3 $\pm $ 1.8 & 0.44 & 0.89 & 3853$_{-973}^{+1429}$ & -0.71 $\pm $ 0.20 &  & CPL$^{e}$ \\ 
161014A$^{c}$ & 2.823 & 5.21 $\pm $ 0.52 & 9.49 $\pm $ 0.50 & 0.55 & 0.90 & 646.18 $\pm $ 55.13 & -0.76 $\pm $ 0.08 &  & CPL$^{d}$ \\ 
170214A$^{a}$ & 2.53 & 30.32 $\pm $ 0.85$^{c}$ & 318.43 $\pm $ 0.21 & 1.94 & 0.99 & 1810 $\pm $ 34 & -0.98 $\pm $ 0.01 & -2.51 $\pm $ 0.10$^{d}$ &  \\ 
130606A$^{a}$ & 5.913 & & 28.3 $\pm $ 5.2 & 1.26 & 0.95 & 2032 $_{-346}^{+622}$ &  -1.14 $\pm $ 0.15 &  & CPL$^{e}$ \\ 
\hline
\end{tabular}
}
\end{center}
\par
$^{a}$ $E_{\text{iso}}$ and $E_{p,z}$ are taken from \citet{Minaev2019}. $%
^{b}$ $E_{\text{iso}}$, $E_{p,z}$, $L_{\text{iso}}$, $\alpha $ and $\beta $
are taken from \citet{Nava2012}. $^{c}$ $E_{\text{iso}}$,$E_{p,z}$ and $L_{%
\text{iso}}$ (or $L_{\text{iso}}$ alone) are taken from \citet{Xue2019}. $%
^{d}$ The spectral analysis results are from the Fermi GBM Burst Catalog %
\citep{Kien2020}. $^{e}$ The spectral analysis results are from the Konus/Wind Burst
Catalog \citep{Tsve2017}. $^{f}$ The extremely high efficiency claimed in %
\citet{Lloy2004}. $^{s}$ The short GRB.
\end{table*}

Generally, the radiation efficiency of the prompt emission\ $\epsilon_{\gamma }$
is defined as $E_{\gamma }/(E_{\gamma}+E_{k})$. Here, $E_{\gamma }$ is the
radiated energy in the prompt phase and $E_{k}$ is the remaining kinetic
energy in the afterglow phase.

To obtain $\epsilon _{\gamma }$, the isotropic energy $E_{\text{iso}}$
(namely $E_{\gamma }$) and the $L_{\text{X,11h}}$ (the late-time X-ray
afterglow luminosity at 11 hr) data should be accumulated for the GRB
sample with the redshift $z$\footnote{%
The redshift data are publicly available at %
\url{http://www.mpe.mpg.de/jcg/grbgen.html}.}. Because that, $L_{\text{X,11h}%
}$ is roughly proportional to the $E_{\text{k}}$ (see Appendix B.1).

(1) For 46 long bursts before GRB 110213A, we use the $L_{\text{X,11h}}$
data given in \citet{Avan2012}. Also, $E_{\text{iso}}$,
the isotropic luminosity $L_{\text{iso}}$, the peak spectral energy in the
rest frame $E_{p,z}$ (or $E_{\text{p}}$), the low-energy spectral index $%
\alpha$, and the high-energy spectral index $\beta$ are taken from %
\citet{Nava2012}.

(2) For 117 long bursts after GRB 110213A (see Tables 1–3, with $\epsilon
_{\gamma }\gtrsim 80\%$, $\epsilon _{\gamma }\lesssim 50\%$, and $\epsilon
_{\gamma }\gtrsim 50\%$, respectively), we calculate the $L_{\text{X,11h}}$
following the method (see Appendix B.1) in \citet{Avan2012}. $E_{\text{iso}}$, $L_{\text{iso}}$, $E_{p,z}$,  $T_{90,i}$ (the intrinsic duration in the
rest frame, or $T_{90}$), and $z$ are mainly taken from %
\citet{Minaev2019} and \citet{Xue2019}.

Our spectral fitting results for the $\epsilon _{\gamma}\gtrsim 80\%$ sample
are given in Table 4. And for several bursts of the $\epsilon_{\gamma
}\lesssim 50\%$ sample, the perfectly consistent observed efficiency (with $E_{\text{%
iso}}/E_{k}$) and the theoretically predicted efficiency of the photosphere
model (from the prompt emission) are given in Table 5.

In Table 6, for the long-GRB sample with the detection of the peak time of the
early optical afterglow (namely $T_{p}$ or $T_{p,op}$, to obtain the Lorentz
factor of the outflow $\Gamma $ and then $(R_{\text{ph}}/R_{s})^{-2/3}$) in %
\citet{Ghirlan2018}, the consistent predicted efficiency (from the prompt
emission and the optical afterglow) of the photosphere model is provided.
Also, $E_{\text{iso}}$, $L_{\text{iso}}$, and $E_{p,z}$ are taken from %
\citet{Ghirlan2018}. In Table 7, its subsample (9 bursts) with the maximum $%
\Gamma $ (for fixed $L_{\text{iso}}$; according to the photosphere model, $%
\epsilon _{\gamma }=50\%$) is provided.

In Table 8, the short-GRB sample (with $L_{\text{X,11h}}$ derived in this
work) is given, along with that possessing extended emission. $E_{\text{iso}}$
and $E_{p,z}$ are taken from \citet{Minaev2019}.

$E_{\text{iso}}$ is generally estimated by $E_{\text{iso}}=4\pi
D_{L}^{2}S_{\gamma }/(1+z)$, where $S_{\gamma }$ is the time-integral
fluence in the $1-10^{4}$ keV energy range in the rest frame (in units of erg cm$%
^{-2}$) and $D_{L}^{{}}$ is the luminosity distance. $L_{\text{iso}}$ is
estimated as $L_{\text{iso}}=4\pi D_{L}^{2}F_{p}$, where $F_{p}$ is the peak
flux (in units of erg cm$^{-2}$ s$^{-1}$). $T_{90,i}$ is calculated as $%
T_{90,i}=T_{90,\text{ob}}/(1+z)$, where $T_{90,\text{ob}}$ is determined by
the time range between the epochs when the accumulated net photon counts
reach the $5\%$ level and the $95\%$ level. And $E_{\text{p}}=(1+z)\cdot E_{%
\text{p},\text{ob}}$, where $E_{\text{p},\text{ob}}$ is determined by the
peak energy in the $\nu F_{\nu }$ spectrum.

Typically, the afterglow peak time $T_{p}$ is estimated from the optical afterglow peak time $T_{p,op}$, since the early X-ray afterglow peak can be produced by “internal" mechanisms (such as the prolonged central engine activity) or bright flares. Also, the bursts with an early multipeaked optical light curve or an optical peak preceded by a decreasing light curve should be excluded (see \citealt{Ghirlan2018}).

\subsection{Scaling Relations Predicted by the Photosphere Model}

For the photosphere emission model, $\epsilon _{\gamma }\gtrsim 50\%$ and $%
\epsilon _{\gamma }\lesssim 50\%$ should correspond to the unsaturated
acceleration case ($\Gamma \lesssim \eta $; $R_{\text{ph}}<R_{s}$; $E_{\text{iso}%
}/E_{\text{k}}=\eta Mc^{2}/\Gamma Mc^{2}=\eta /\Gamma $ $\gtrsim 1$)
and the saturated acceleration case ($\Gamma =\eta $; $R_{\text{ph}}>R_{s}$; $E_{%
\text{iso}}/E_{\text{k}}=[\eta Mc^{2}(R_{\text{ph}}/R_{s})^{-2/3}]/\Gamma
Mc^{2}=(R_{\text{ph}}/R_{s})^{-2/3}$ $\lesssim 1$), respectively.

\subsubsection{$E_{\text{iso}}/E_{\text{k}}=\eta/\Gamma$ for the
unsaturated acceleration case, corresponding to $\epsilon_{\gamma}\gtrsim 50\%$}

\begin{figure*}[th]
\label{Fig_1} \centering\includegraphics[angle=0,height=2.4in]{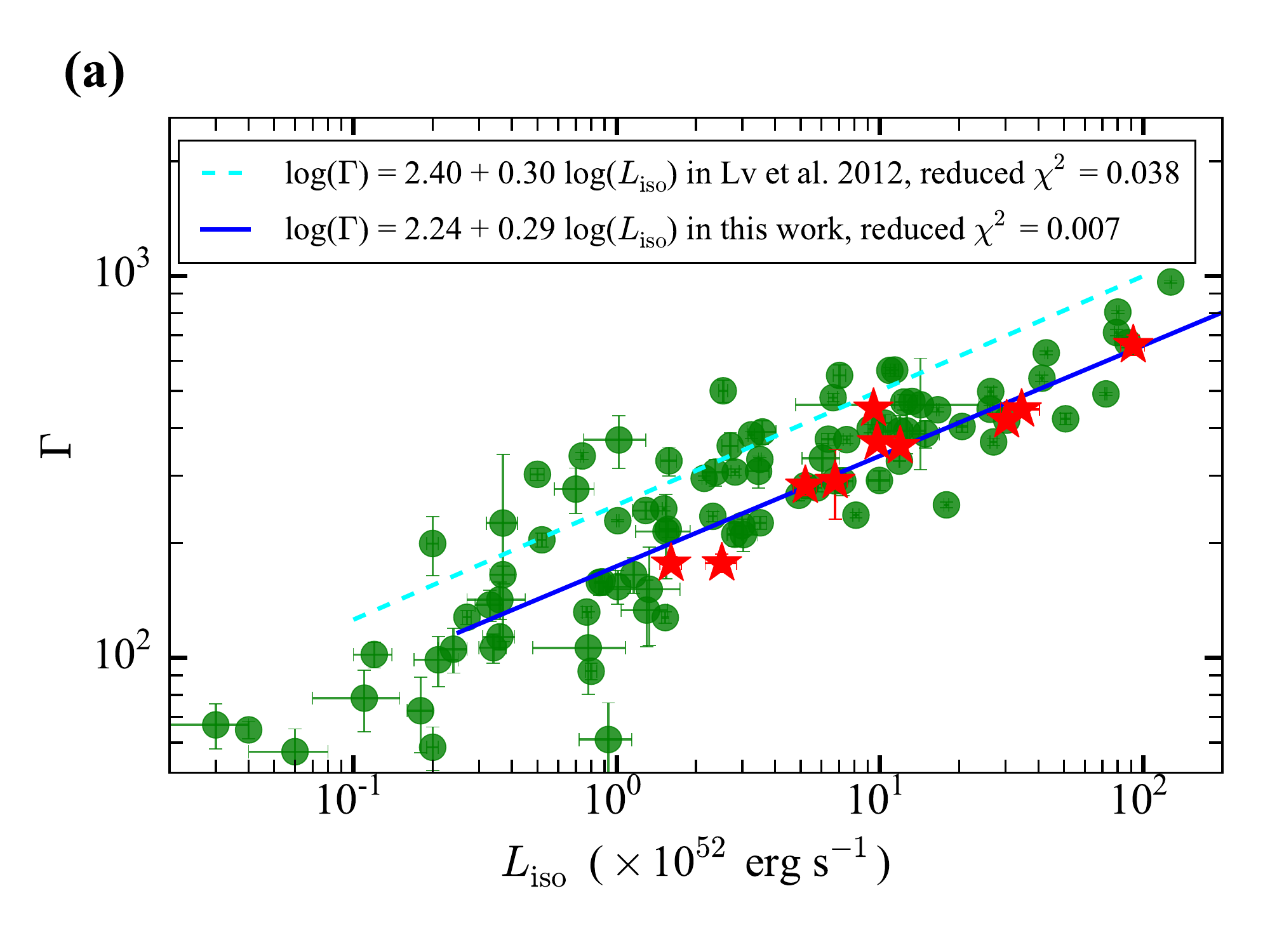} %
\centering\includegraphics[angle=0,height=2.4in]{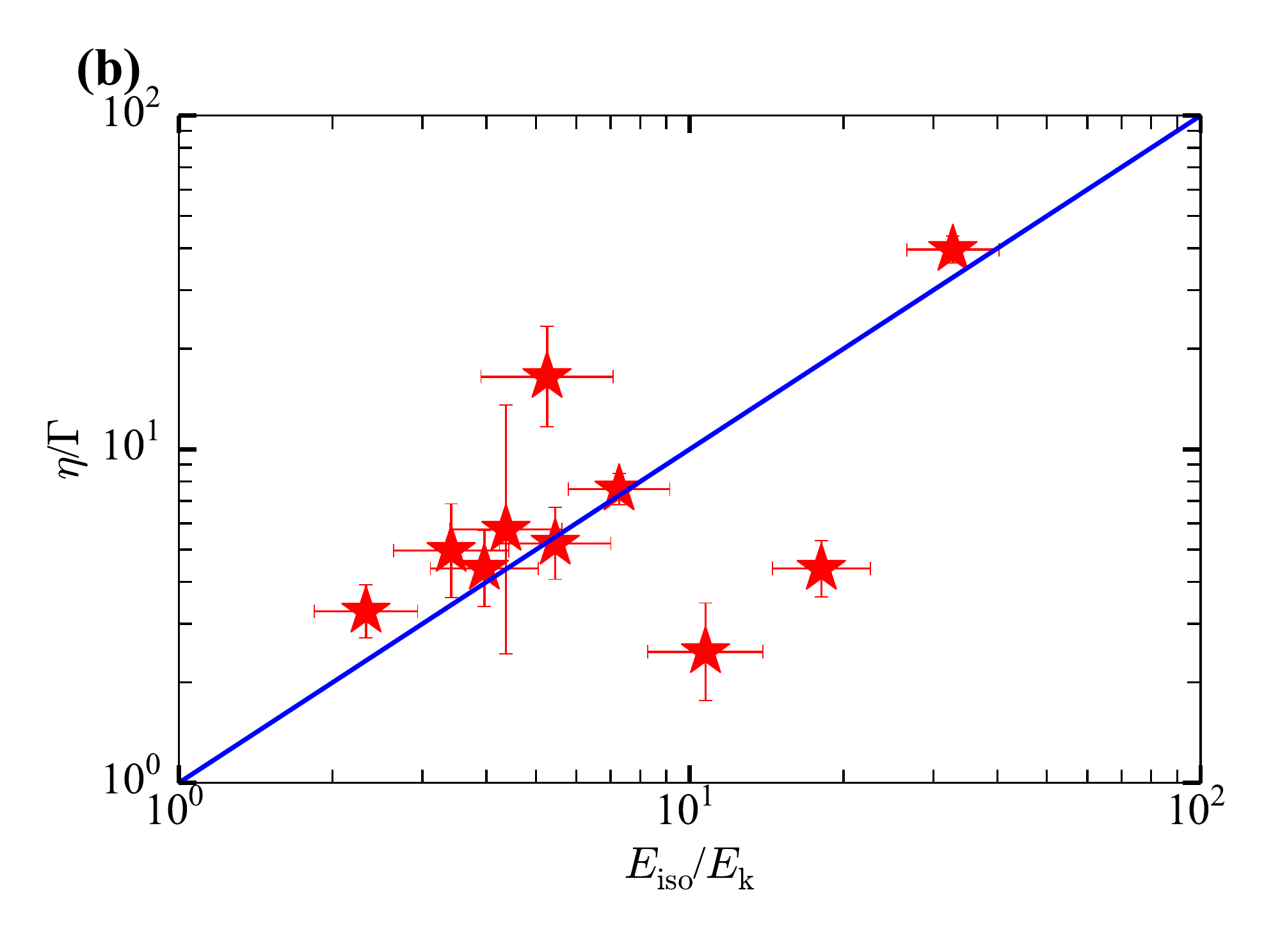} \ \ 
\caption{The $L_{\text{iso}}-\Gamma $ correlation and the $E_{\text{iso}}/E_{%
\text{k}}$ $-$ $\protect\eta /\Gamma $ correlation for the selected
extremely high-efficiency GRBs. (a) The correlation of $L_{\text{iso}}$ and $\Gamma $.
Obviously, the tight correlation of $\Gamma \propto (L_{\text{iso}})^{0.29}$
(the red stars, reduced $\protect\chi^{2}=0.007$) is found, which is well
consistent with the prediction of the neutrino annihilation from the
hyperaccretion disk, $\Gamma \propto (L_{\text{iso}})^{7/27}=(L_{\text{iso}%
})^{0.26}$. Thus, the jet is likely to be thermal-dominated. The cyan dashed line shows the $L_{\text{iso}}-\Gamma $ correlation found in \citet{Lv2012} (reduced $\protect\chi^{2}=0.038$, for the large $\Gamma$ sample of green circles \citep{Xue2019}). (b) The
correlation of $E_{\text{iso}}/E_{\text{k}}$ and $\protect\eta /\Gamma $.
A significant linear correlation is found, and they are almost the same
when we take $E_{\text{k,52}}=5\ast L_{\text{X,45}}$ (reduced $\protect\chi%
^{2}=0.138$). This is well consistent with the predicted $E_{\text{iso}}/E_{%
\text{k}}=\protect\eta Mc^{2}/\Gamma Mc^{2}=\protect\eta /\Gamma $ by the
photosphere emission model in the unsaturated acceleration regime.}
\end{figure*}

\begin{figure*}[th]
\label{Fig_2} \centering\includegraphics[angle=0,height=2.4in]{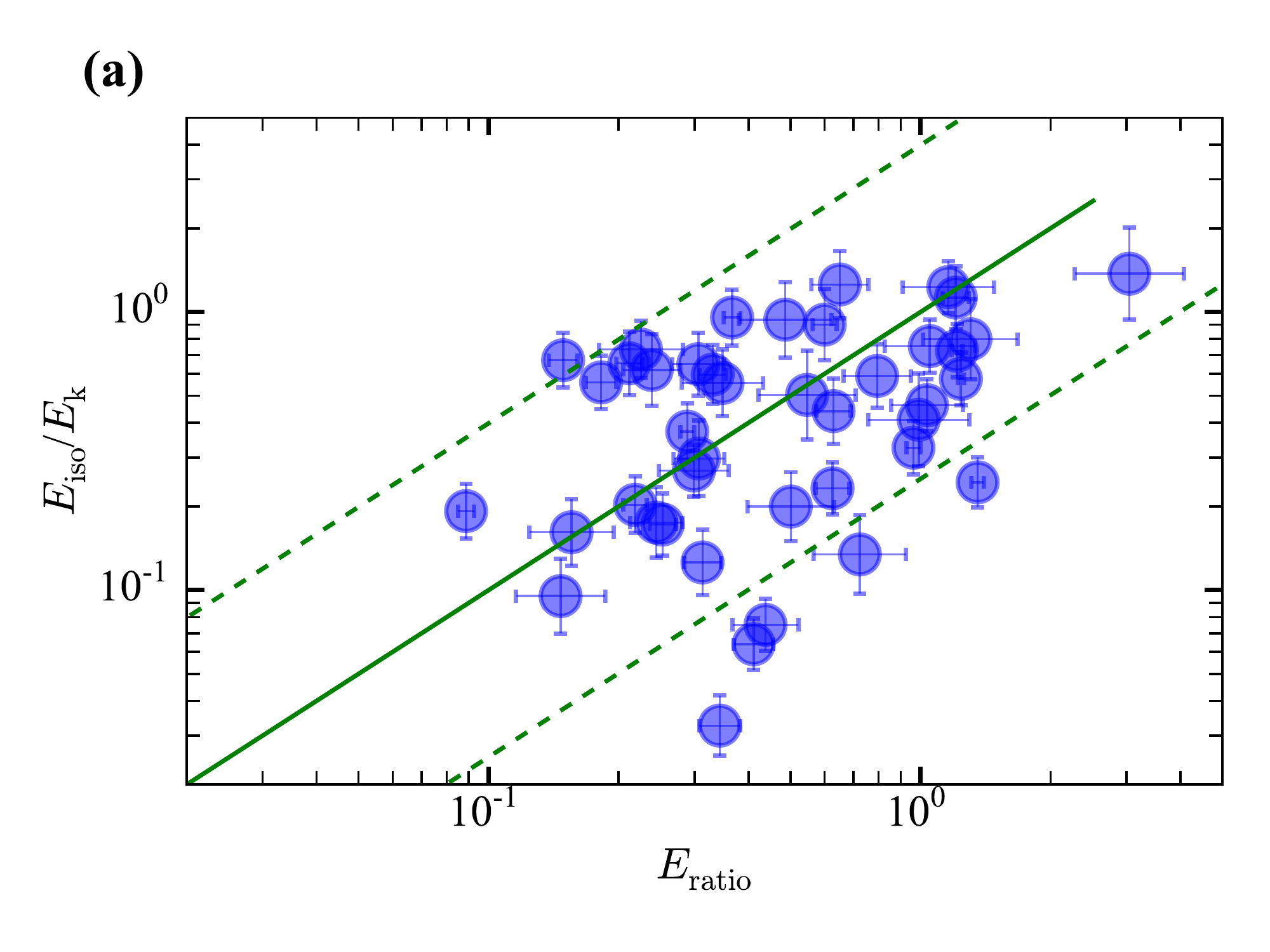} %
\centering\includegraphics[angle=0,height=2.4in]{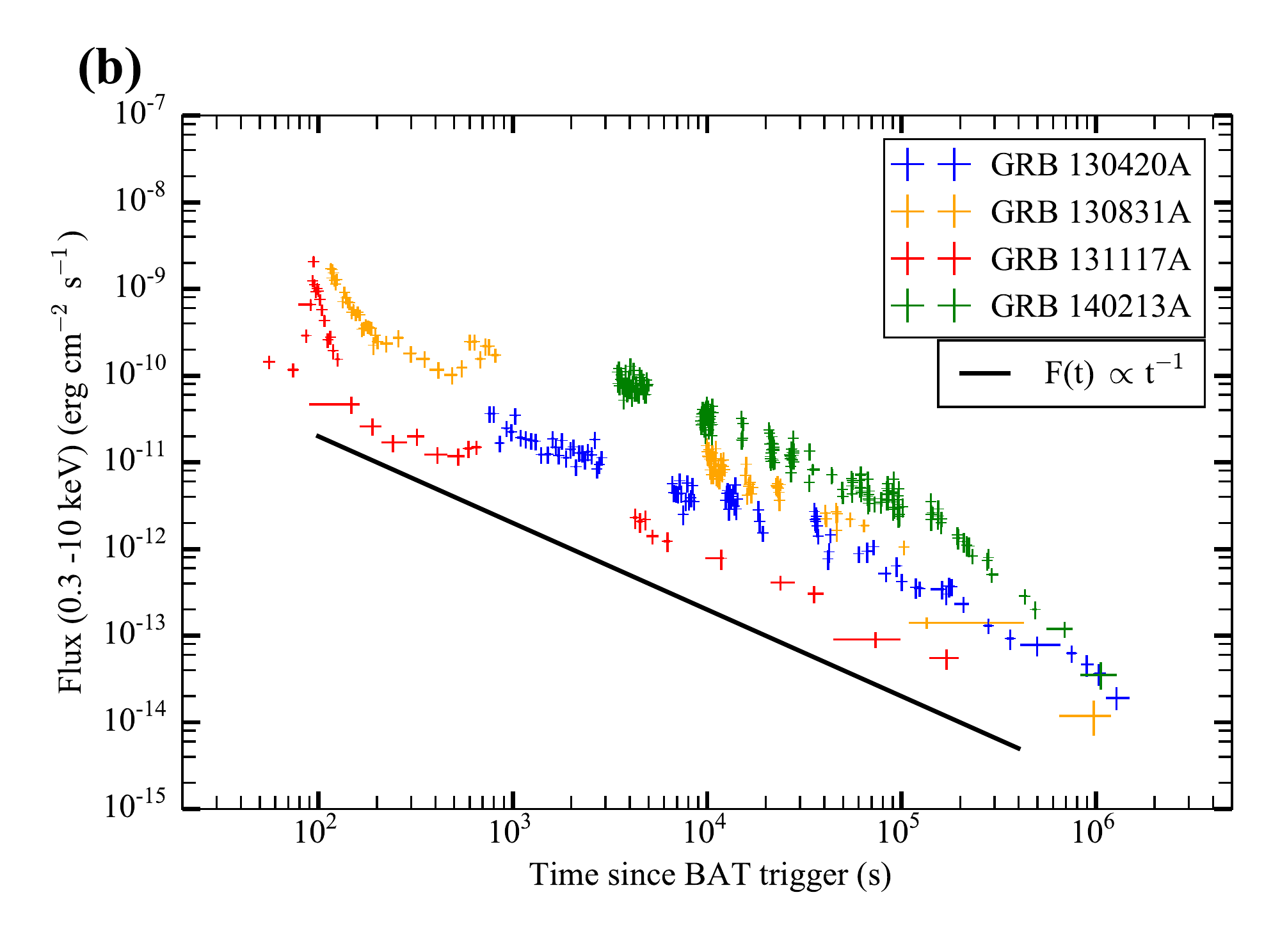} \centering%
\includegraphics[angle=0,height=2.4in]{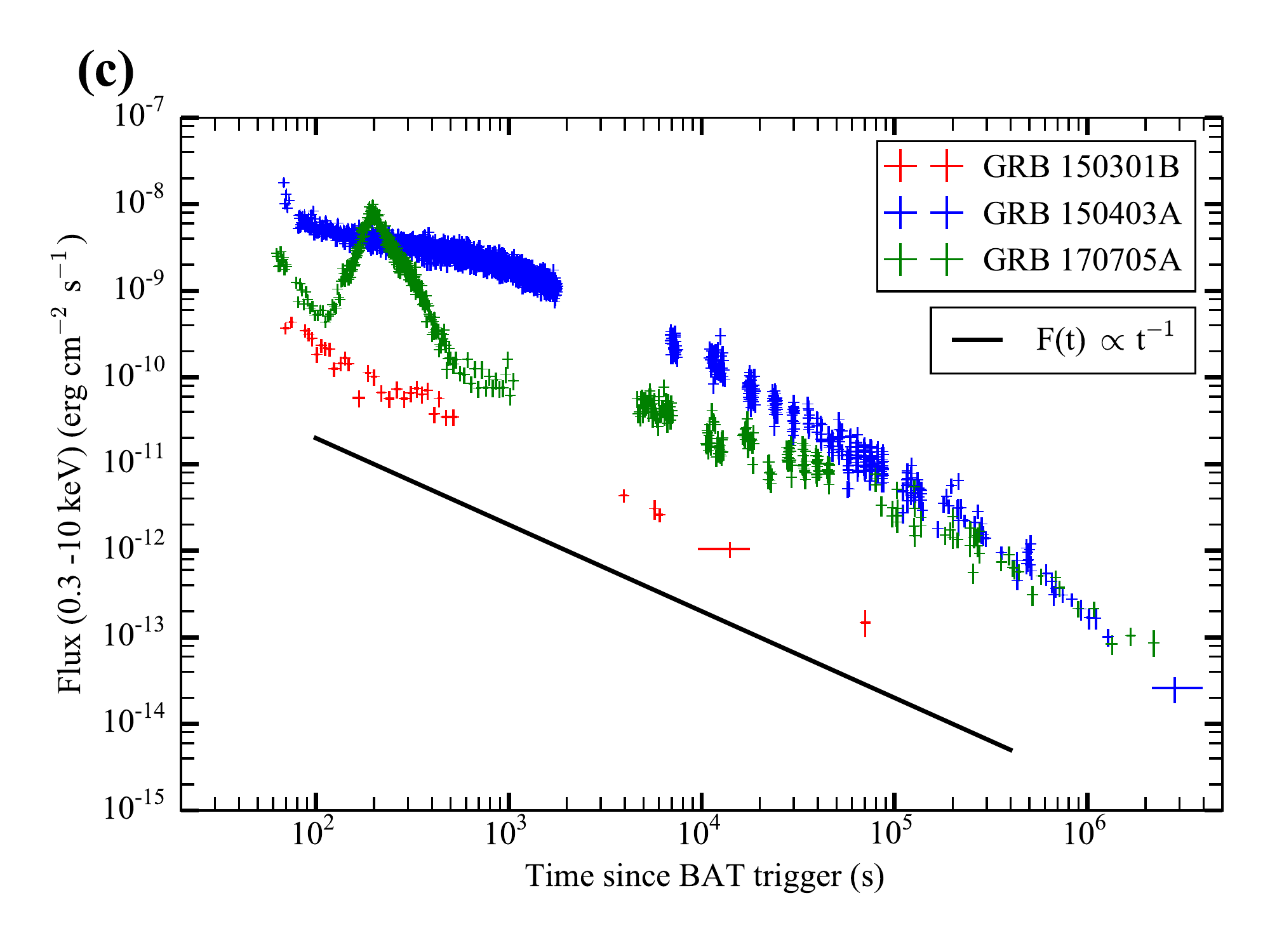} \centering%
\includegraphics[angle=0,height=2.4in]{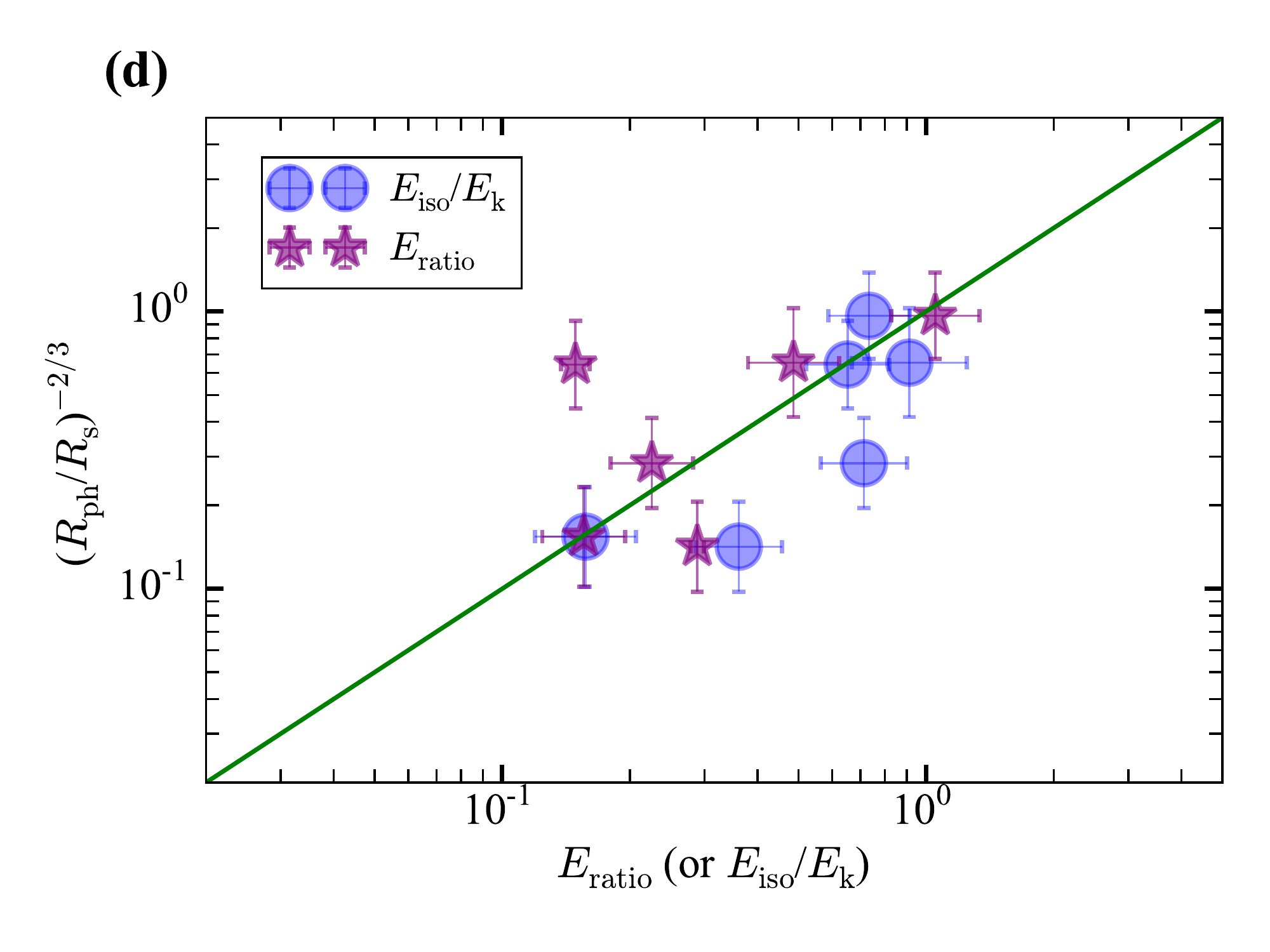} \centering%
\includegraphics[angle=0,height=2.4in]{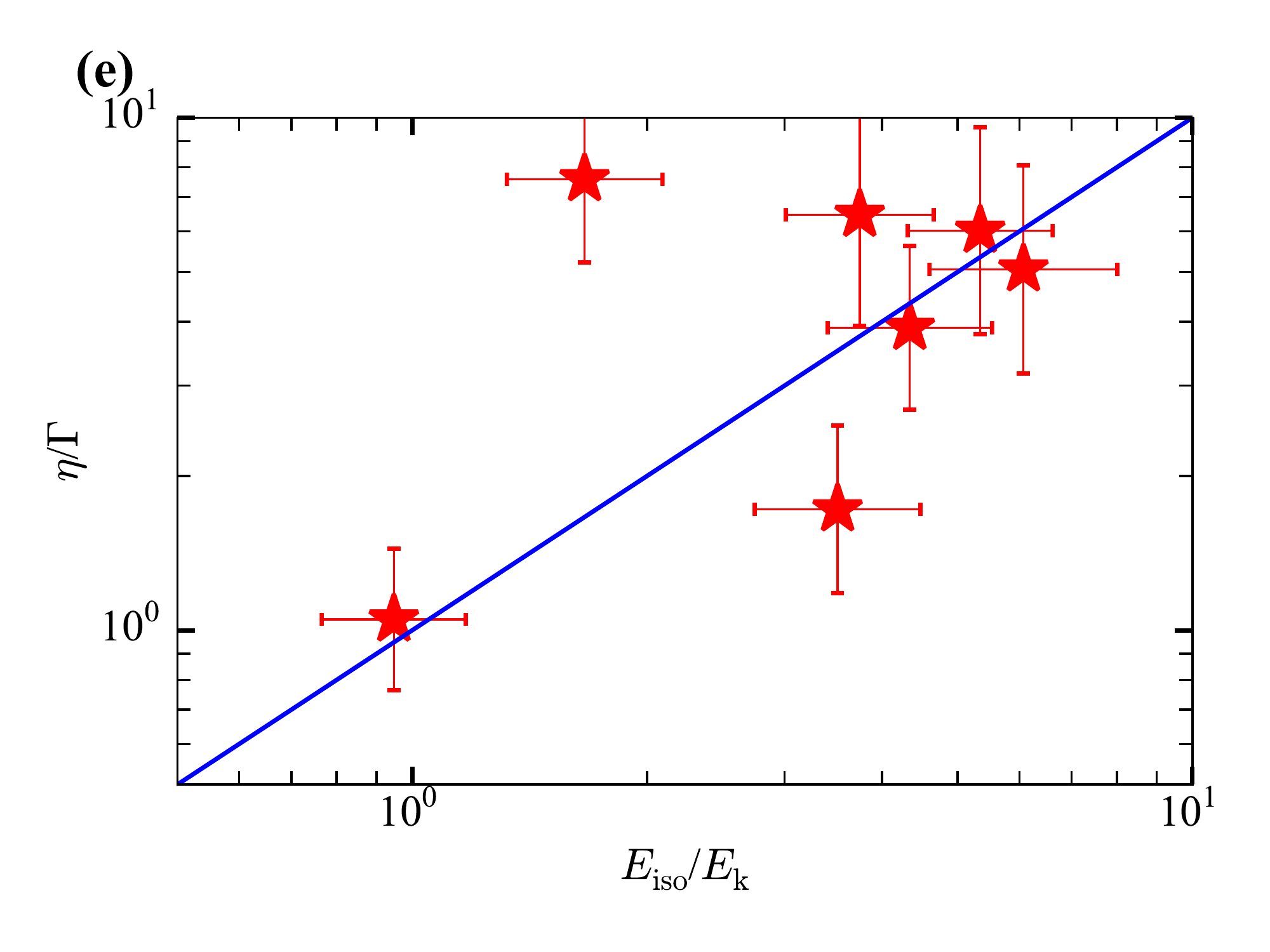} \ \ 
\caption{The similar observed efficiency (with $E_{\text{iso}}/E_{\text{k}}$
from the X-ray afterglow) and theoretically predicted efficiency (from the
prompt emission or the optical afterglow) by the photosphere emission model
for the bursts after GRB 110213A. (a) The distribution of $E_{\text{ratio}}$
($E_{\text{ratio}}=[(E_{\text{p}}/2.7k)^{4}\ast (4\protect\pi %
r_{1}^{2}ac)^{{}}/E_{\text{iso}}]^{1/3}$, from the prompt emission) and $E_{%
\text{iso}}/E_{\text{k}}$ for the $\protect\epsilon _{\protect\gamma %
}\lesssim 50\%$ sample (see Table 2). They are found to be well centered
around the equal-value line, and have a linear correlation (reduced $%
\protect\chi^{2}=0.172$). The dispersion is likely to be caused by the
estimation error for $E_{\text{k}}$, since many X-ray afterglow light curves
are not the power law with a slope of $-1$. (b) and (c) The X-ray afterglow
light curves for the bursts (7 bursts) with almost same $E_{\text{ratio}}$
and $E_{\text{iso}}/E_{\text{k}}$ (see Table 5). We find all these light
curves do have the power-law shape with a slope of $\sim$ $-1$. (d) The
distribution of $E_{\text{ratio}}$, $E_{\text{iso}}/E_{\text{k}}$ and $(R_{%
\text{ph}}/R_{s})^{-2/3}$ for the sample (6 bursts, see Table 2) with
detections of peak time of the optical afterglow (to estimate the $\Gamma $
and thus $(R_{\text{ph}}/R_{s})^{-2/3}$). It is found that 1 burst has
the almost same values for these three quantities and the other 5 bursts
have the almost same values for two quantities of them (reduced $\protect\chi%
^{2}=0.089$ for all 12 markers). (e) The consistent efficiency from
observation (with $E_{\text{iso}}/E_{k}$) and the prediction of photosphere
emission model (with $\protect\eta /\Gamma $) for the $\protect\epsilon _{%
\protect\gamma }\gtrsim 50\%$ sample with detections of peak time of the
optical afterglow (7 bursts, see Table 3; reduced $\protect\chi^{2}=0.120$).}
\end{figure*}

When the outflow Lorentz factor $\Gamma$ at the photosphere radius $R_{\text{%
ph}}$ is less than the baryon loading $\eta$ ($\eta =E/Mc^{2}$, where $E$ and $M$ are the injected energy and the baryon mass at the outflow base, respectively),
the photosphere emission is in the unsaturated acceleration case (or $R_{%
\text{ph}}<R_{s}$; where $R_{s}=\eta r_{0}$, and $r_{0}$ is the initial
acceleration radius). In this case, $\Gamma =R_{\text{ph}}/r_{0}$.

For the unsaturated acceleration case, the observed temperature $T_{\text{ob}%
}$ $=$ $D\cdot $ $T_{\text{comoving}}=$ $\Gamma \cdot $ ($T_{0}/\Gamma
)=T_{0}$. Here, $D$ is the Doppler factor, $T_{\text{comoving}}$ is the
photon temperature in the outflow comoving frame, and $T_{0}$ is the
temperature at the outflow base $r_{0}$. Thus, $E_{\text{iso}}=E=\eta Mc^{2}$. Also,
because all the thermal energy is released at the photosphere radius (where there
is no remaining energy to accelerate the jet), the Lorentz factor in the afterglow
phase will remain as $\Gamma$, namely $E_{\text{k}}=\Gamma Mc^{2}$. So
we should have $E_{\text{iso}}/E_{\text{k}}=\eta Mc^{2}/\Gamma Mc^{2}=\eta
/\Gamma$ $\gtrsim 1$ (see Figure 1(b) and 2(e)), corresponding to $\epsilon
_{\gamma}\gtrsim 50\%$.

\subsubsection{$E_{\text{p}}\propto (E_{\text{iso}})^{1/4}$ for $\protect%
\epsilon _{\protect\gamma}\gtrsim 50\%$}

\begin{figure*}[th]
\label{Fig_3} \centering\includegraphics[angle=0,height=2.4in]{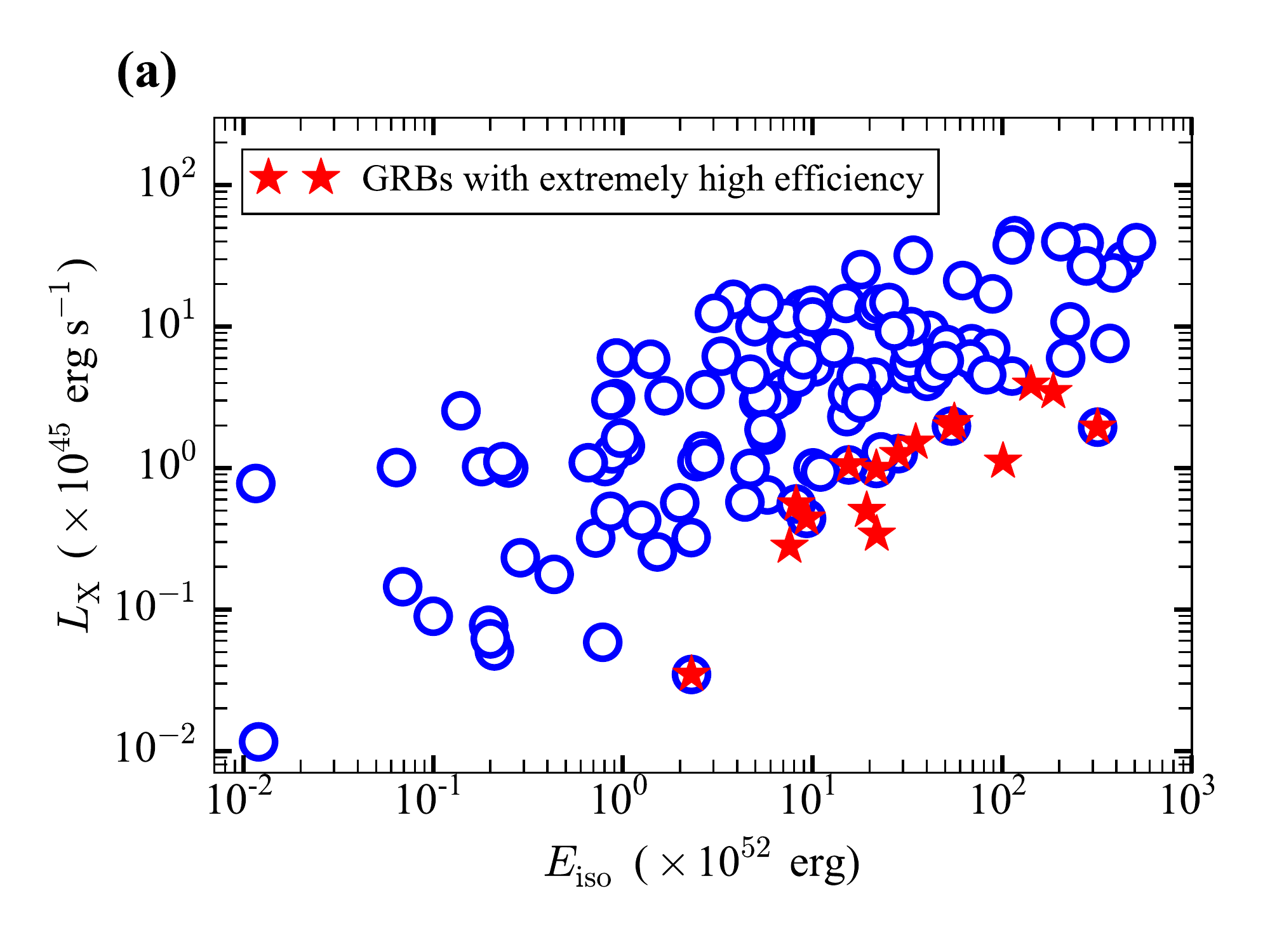} %
\centering\includegraphics[angle=0,height=2.4in]{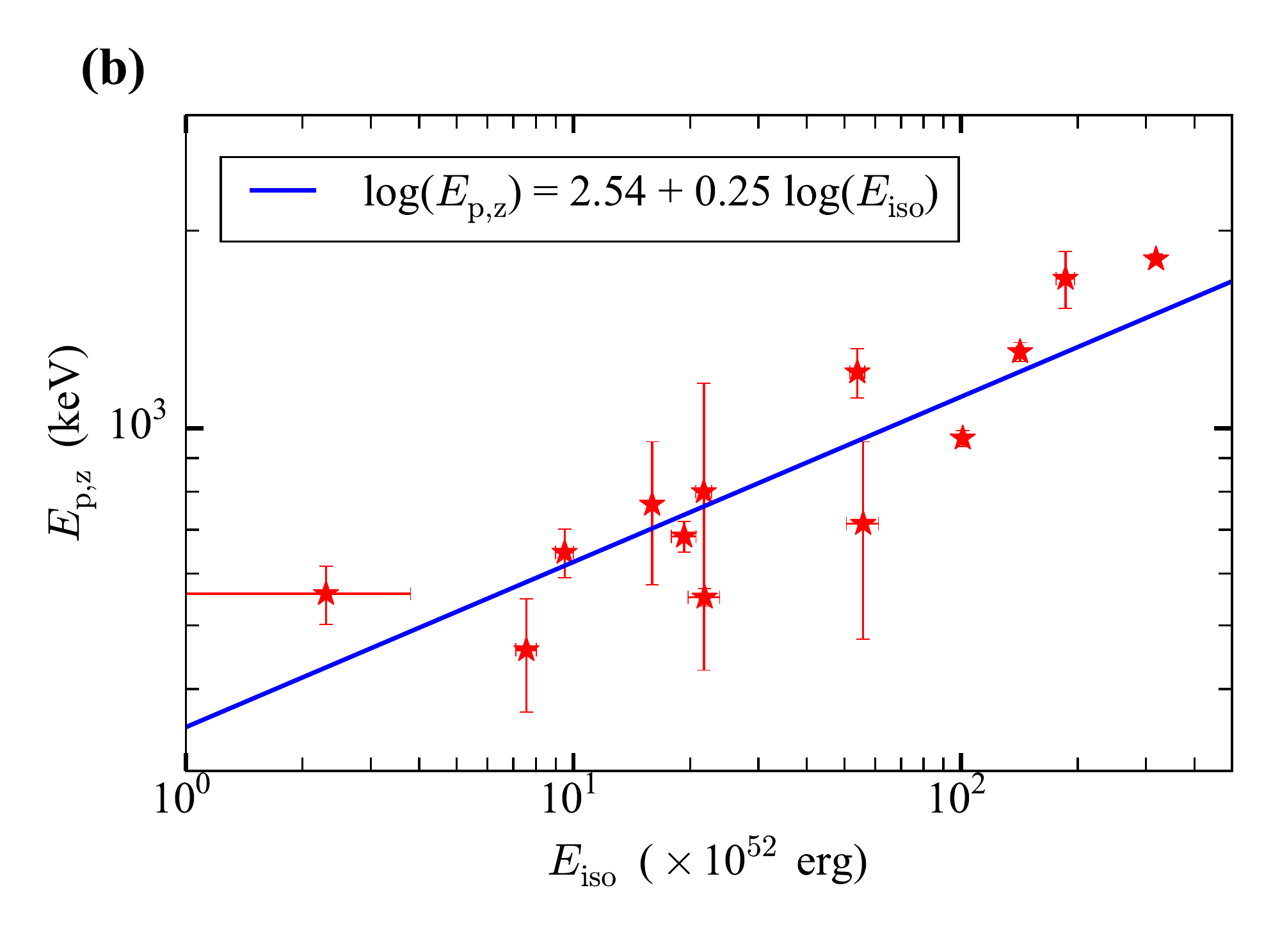} \centering%
\includegraphics[angle=0,height=2.4in]{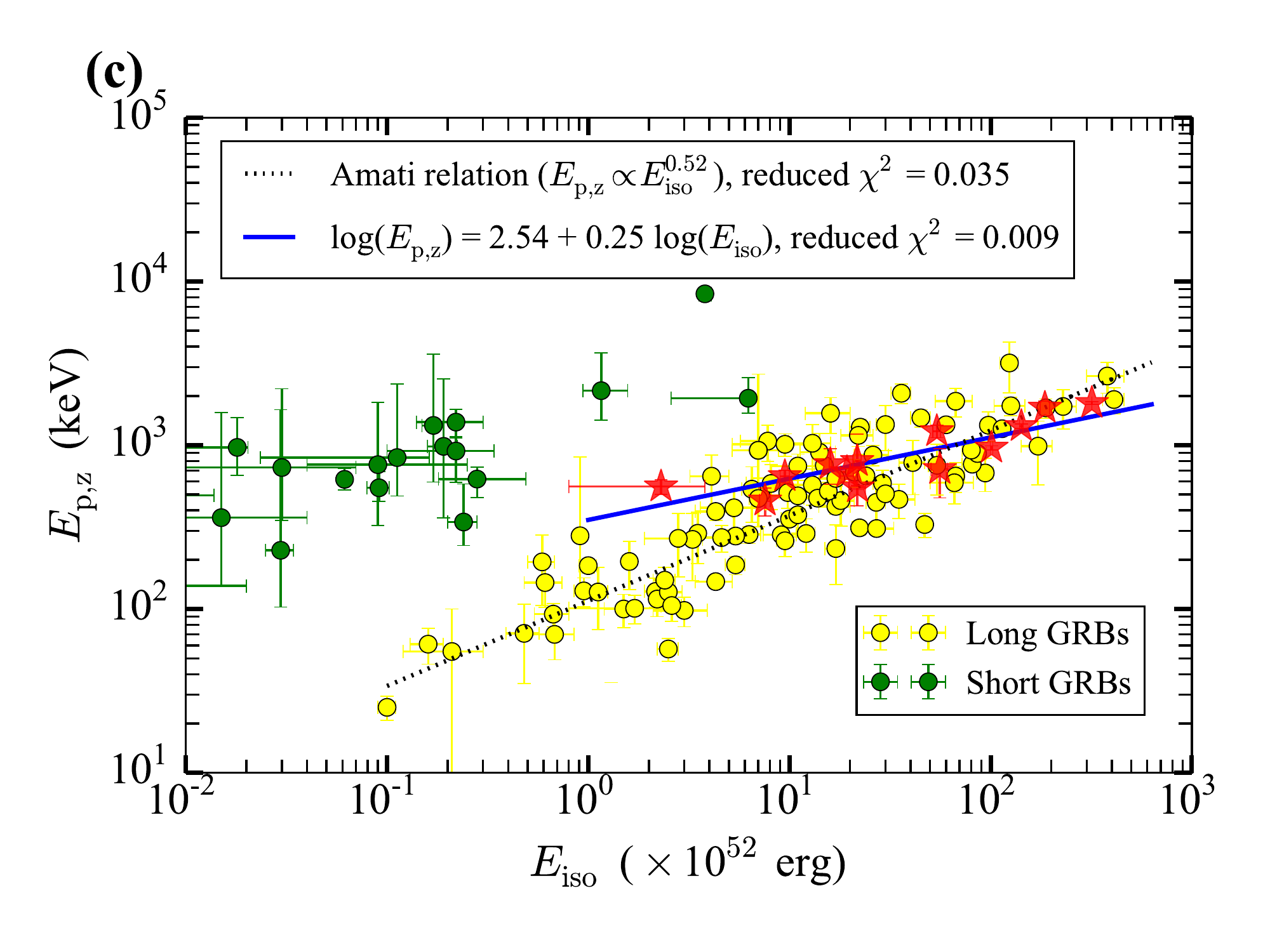} \centering%
\includegraphics[angle=0,height=2.4in]{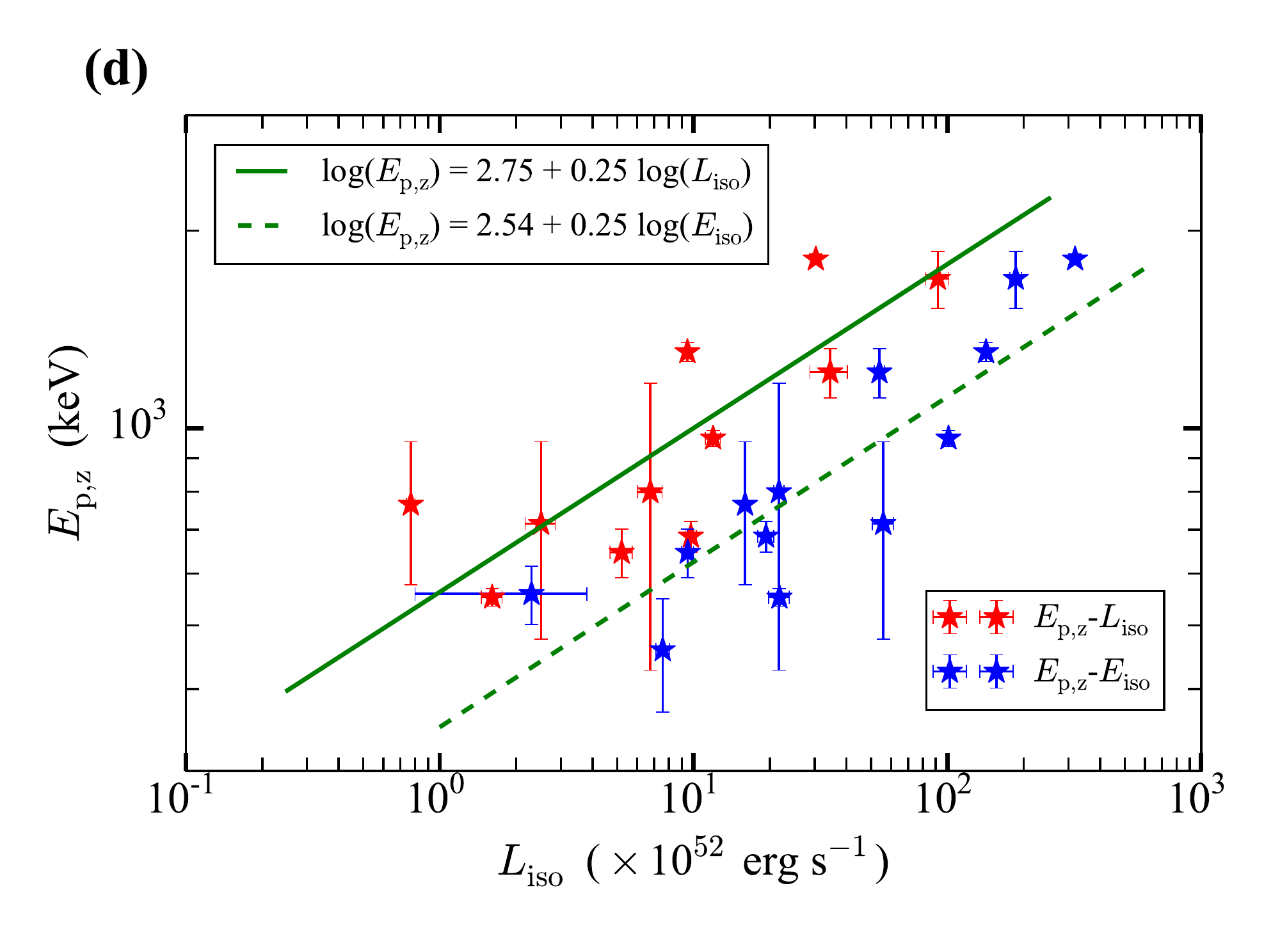}\ \ \ 
\caption{The $E_{\text{p}}\propto (E_{\text{iso}})^{1/4}$ and $E_{\text{p}%
}\propto (L_{\text{iso}})^{1/4}$ relations for the GRBs with the highest
prompt efficiency $\protect\epsilon _{\protect\gamma }$ ($\protect\epsilon _{%
\protect\gamma }\gtrsim 80\%$; see Table 1). (a) The $E_{\text{iso}}$ and
calculated $L_{\text{X,11h}}$ (the late-time X-ray afterglow luminosity at
11 hours) distribution for the whole GRB sample used (117 bursts, after GRB
110213A) with redshift. The red stars with the smallest $L_{\text{X,11h}}$
represent the bursts with the highest prompt efficiency. (b) The $E_{\text{p}%
}$ and $E_{\text{iso}}$ distribution for the selected 15 long GRBs. The
best-fit result is log ($E_{\text{p}}$) $=2.54+0.25\log $ ($E_{\text{iso}}$%
), quite consistent with the $E_{\text{p}}\propto (E_{\text{iso}})^{1/4}$
relation predicted by the photosphere (thermal) emission model. (c)
Comparison of the $E_{\text{p}}$ and $E_{\text{iso}}$ distributions for the
selected GRBs (red stars) and the large sample of long GRBs (see Figure 3 in 
\citet{ZhangBB18b}) (yellow circles). Obviously, the dispersion for the
selected GRBs is quite small (reduced $\protect\chi^{2}=0.009$) relative to
that for the large sample (reduced $\protect\chi^{2}=0.035$). The black dotted line shows the Amati relation \citep{Amati2002} for the large sample. (d) Comparison
of the $E_{\text{p}}$ $- $ $L_{\text{iso}}$ distribution (red stars) and the 
$E_{\text{p}}$ $-$ $E_{\text{iso}}$ distribution (blue stars) for the
selected GRBs.\ Likewise, the $E_{\text{p}}\propto (L_{\text{iso}})^{1/4}$
relation exists. The best-fit result is log ($E_{\text{p}}$) $=2.75+0.23\log 
$ ($L_{\text{iso}}$). And the dispersion (reduced $\protect\chi^{2}=0.012$)
is found to be similar to that of $E_{\text{p}}\propto (E_{\text{iso}%
})^{1/4} $.}
\end{figure*}

\begin{figure*}[th]
\label{Fig_4} \centering\includegraphics[angle=0,height=2.4in]{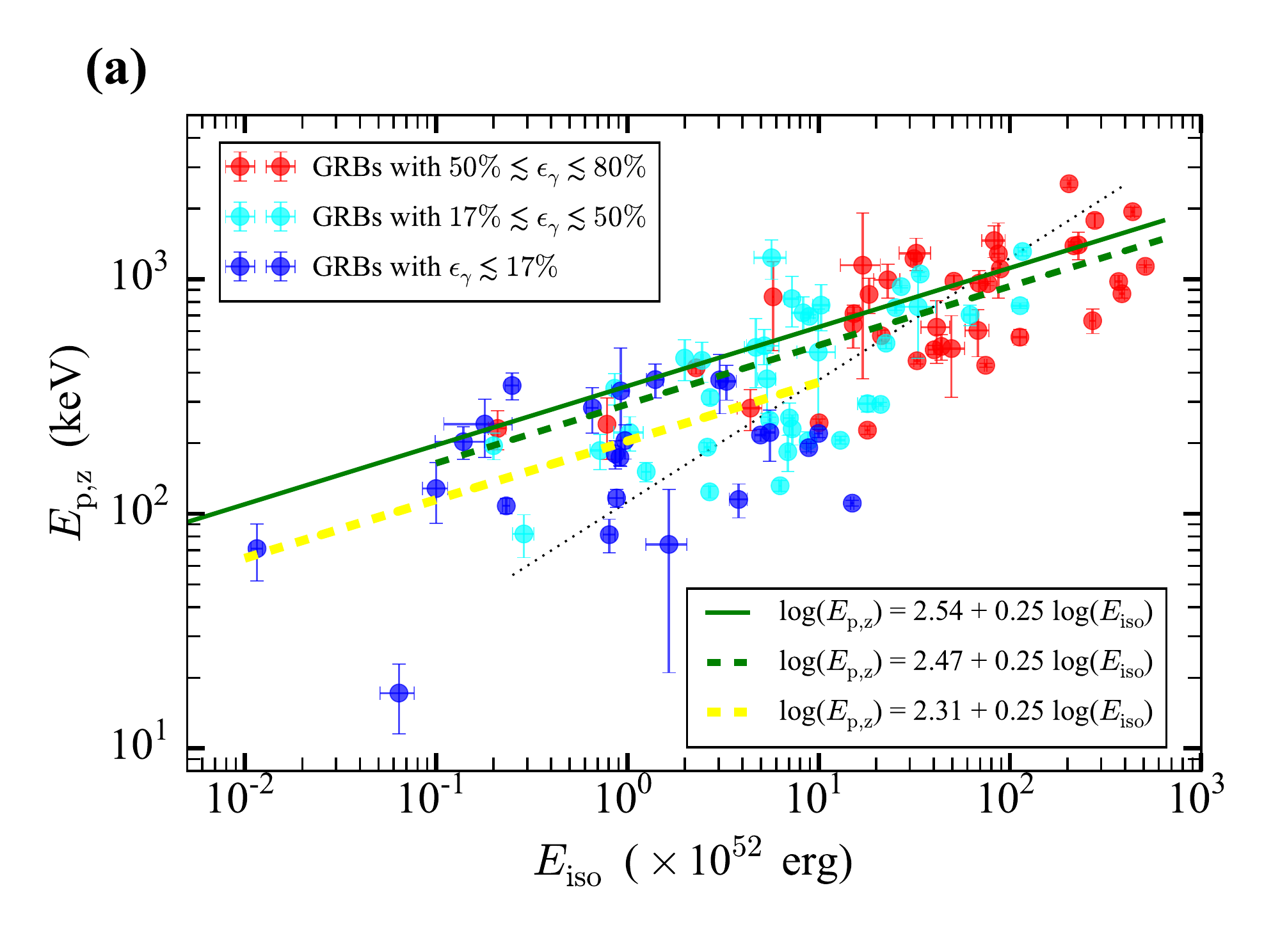} %
\centering\includegraphics[angle=0,height=2.4in]{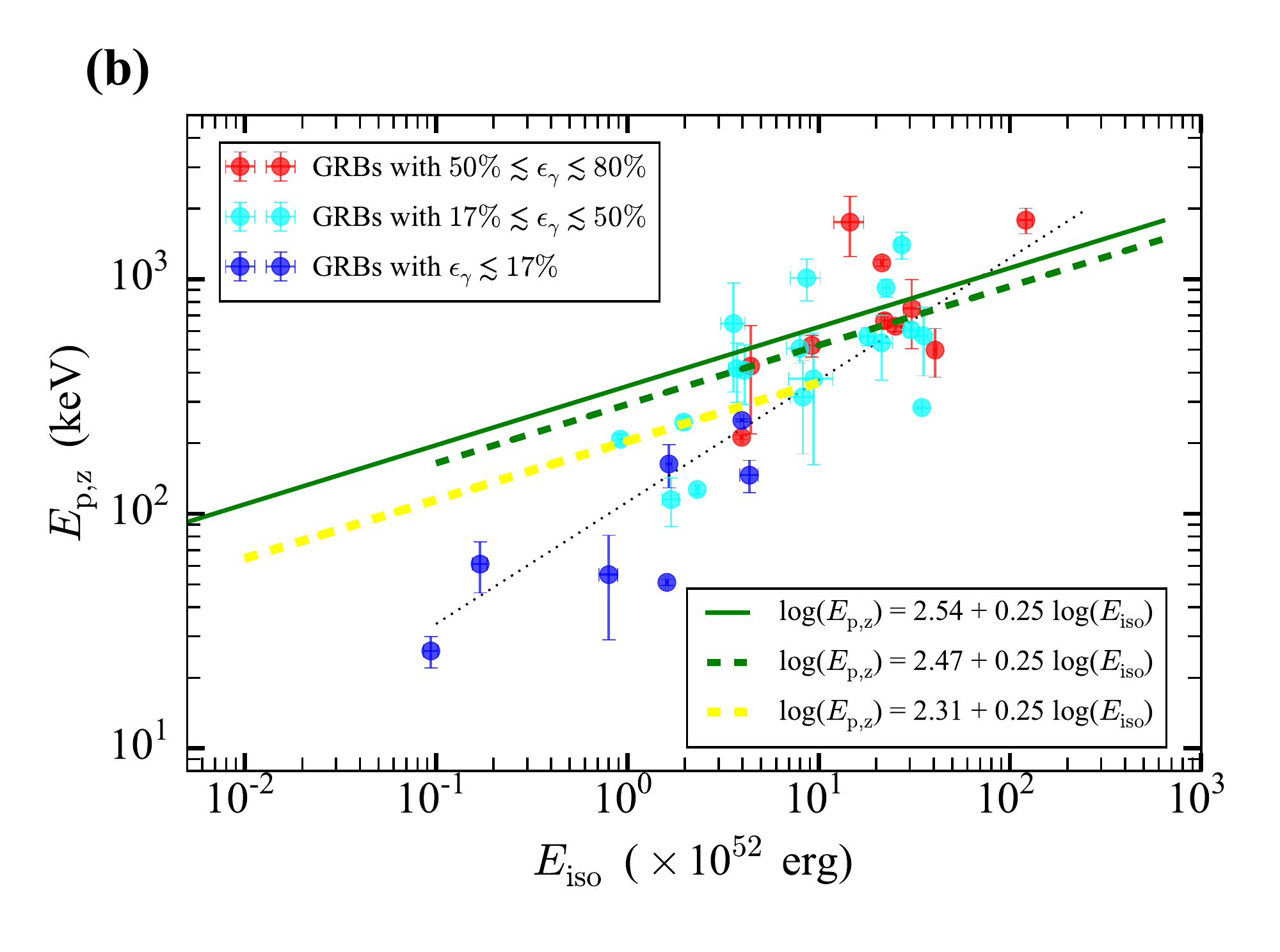} \ \ 
\caption{The different distributions for $E_{\text{p}}$ and $E_{\text{iso}}$
of the two distinguished samples ($\protect\epsilon _{\protect\gamma %
}\gtrsim 50\%$ and $\protect\epsilon _{\protect\gamma }\lesssim 50\%$; see
Tables 2 and 3) for the long GRBs. (a) The $E_{\text{p}}$ and $E_{\text{%
iso}}$ distribution for the bursts after GRB 110213A (117 bursts, with $L_{%
\text{X,11h}}$ derived in this work). For the $\protect\epsilon _{\protect%
\gamma }\gtrsim 50\%$ (red circles) sample, the best-fit result is log ($E_{%
\text{p}}$) $=2.47+0.25\log $ ($E_{\text{iso}}$) (reduced $\protect\chi%
^{2}=0.039$; for the black dotted line of the typical Amati relation, reduced $%
\protect\chi^{2}=0.079$), well consistent with the predicted $E_{\text{p}%
}\propto (E_{\text{iso}})^{1/4}$ by the photosphere emission model. For the $%
\protect\epsilon _{\protect\gamma }\lesssim 50\%$ (blue and cyan circles)
sample, the up-most distribution is found well around $\log $ ($E_{\text{p}}$%
) $=2.54+0.25\log $ ($E_{\text{iso}}$), and the best-fit result is log ($E_{%
\text{p}}$) $=2.31+0.26\log $ ($E_{\text{iso}}$) (reduced $\protect\chi%
^{2}=0.076$; for the typical Amati relation, reduced $\protect\chi^{2}=0.113$%
), quite below that. (b) The $E_{\text{p}}$ and $E_{\text{iso}}$
distribution for the bursts before GRB 110213A (46 bursts, with $L_{\text{%
X,11h}}$ given in \citealt{Avan2012}). The different distributions for $E_{%
\text{p}}$ and $E_{\text{iso}}$ of the two distinguished samples ($\protect%
\epsilon _{\protect\gamma }\gtrsim 50\%$ and $\protect\epsilon _{\protect%
\gamma }\lesssim 50\%$) are also found, similar to the
distributions for the bursts after GRB 110213A.}
\end{figure*}

\begin{figure*}[th]
\label{Fig_5} \centering\includegraphics[angle=0,height=2.4in]{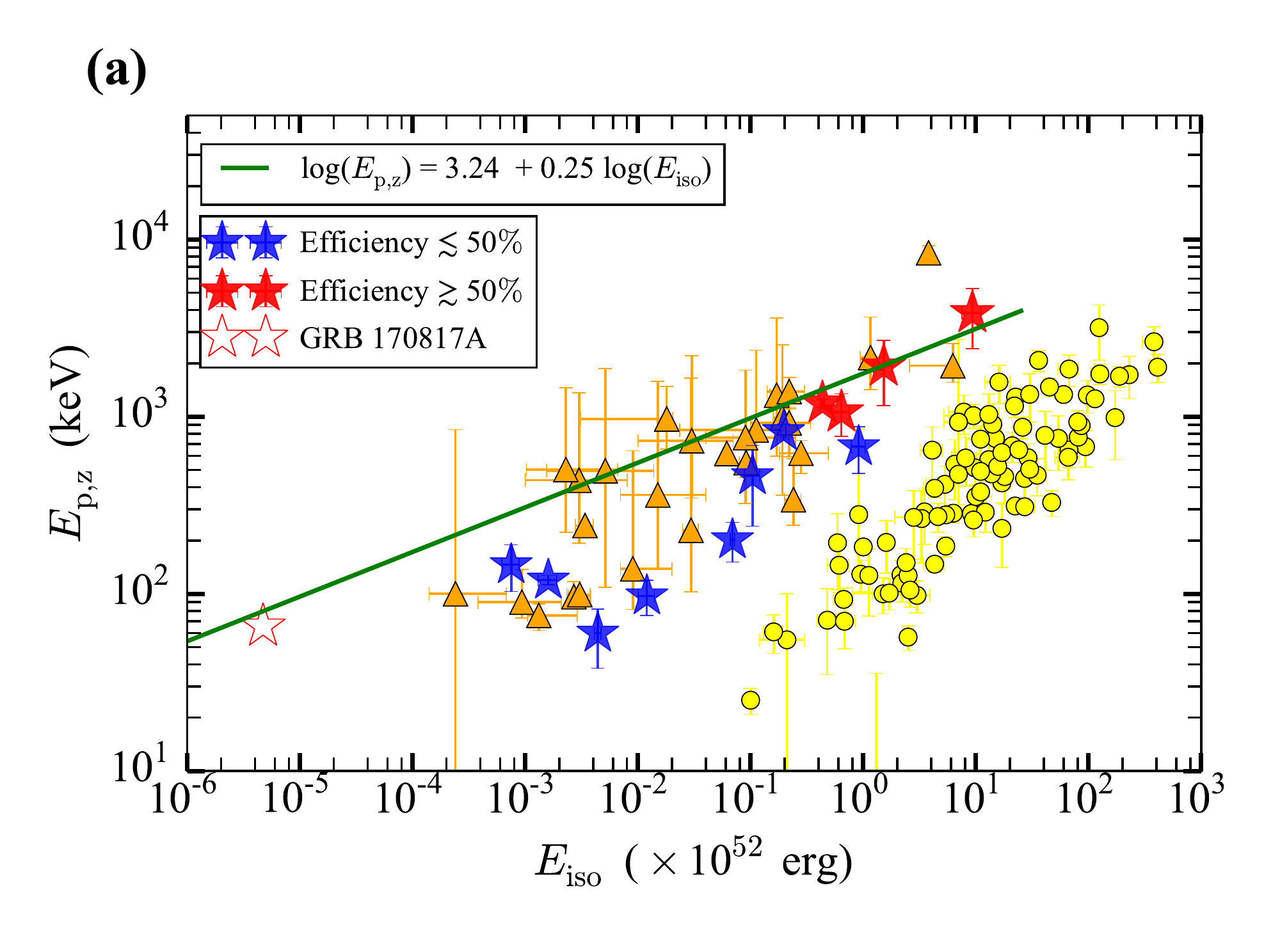} %
\centering\includegraphics[angle=0,height=2.4in]{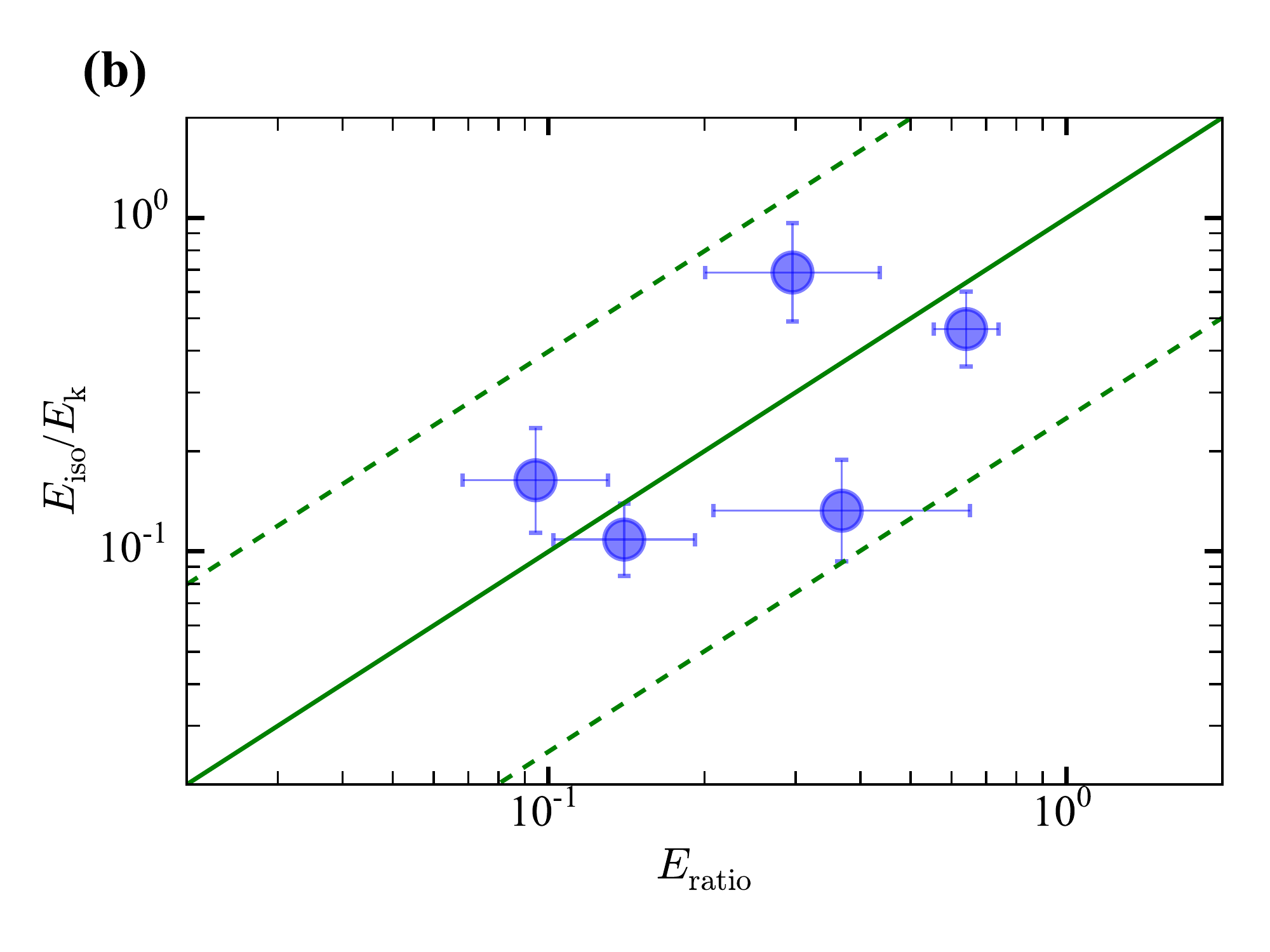} \centering%
\includegraphics[angle=0,height=2.4in]{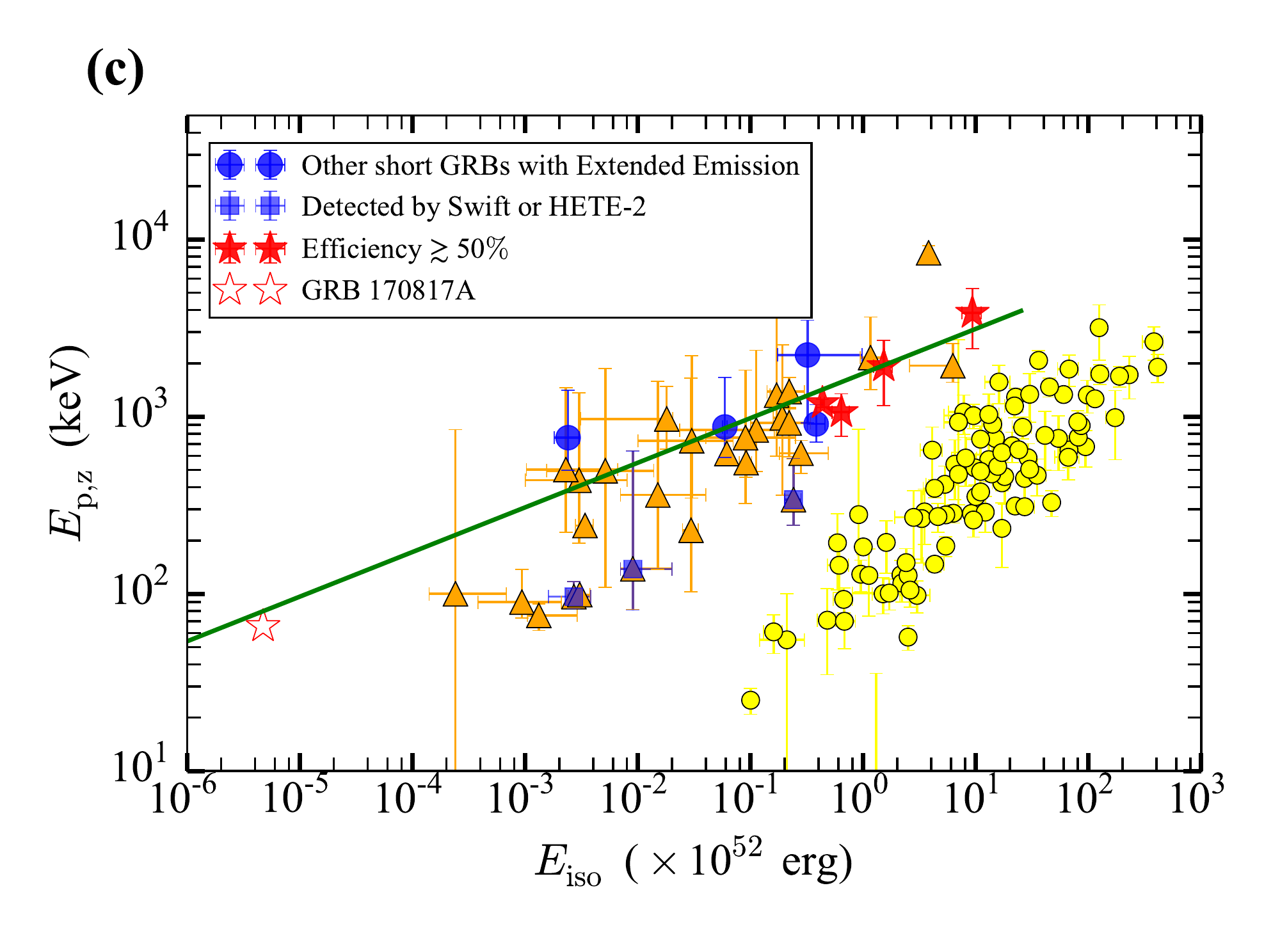} \centering%
\includegraphics[angle=0,height=2.4in]{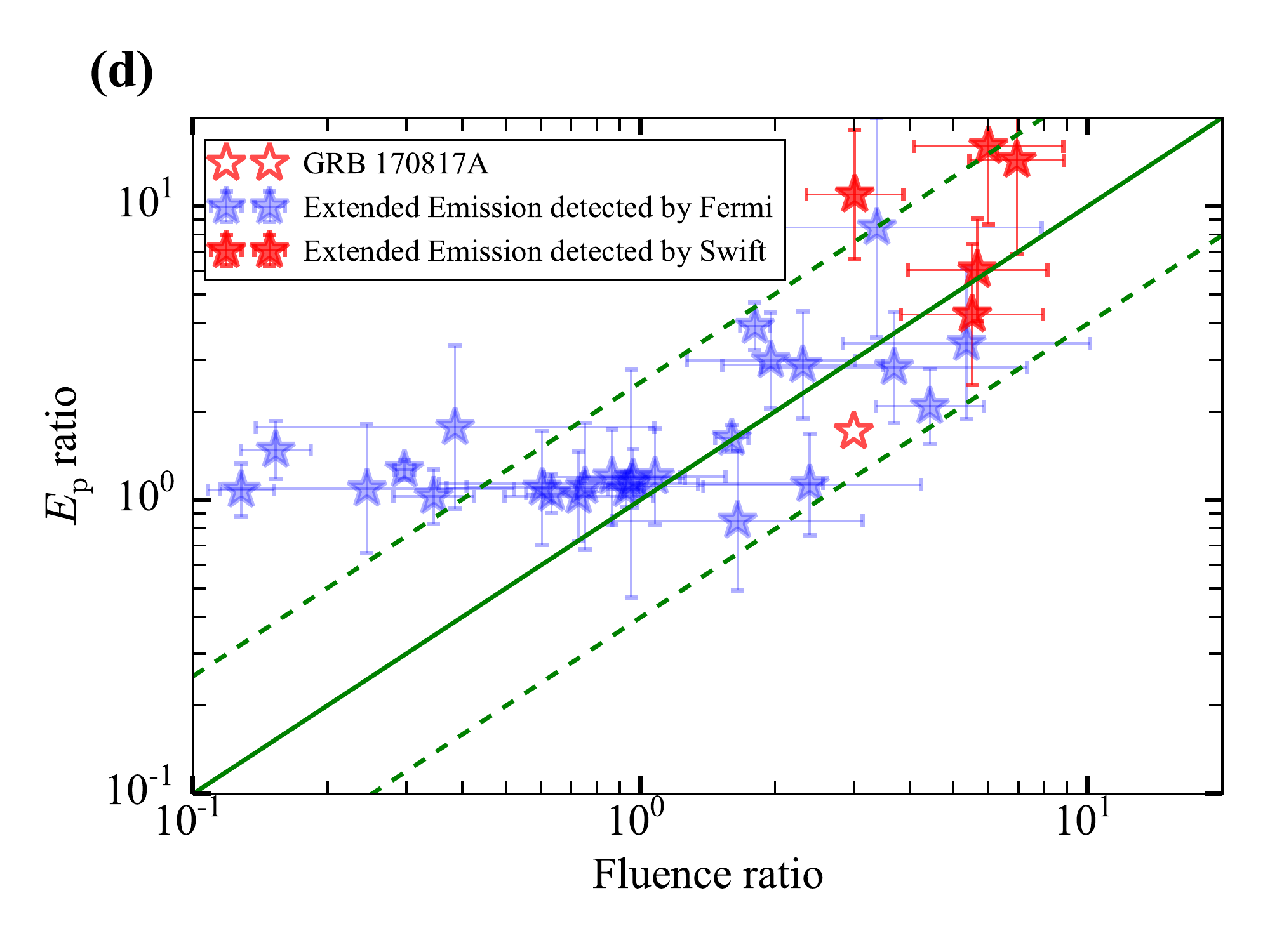} \ \ 
\caption{Evidence from the short GRBs (see Table 8). (a) The different distributions of $E_{%
\text{p}}$ and $E_{\text{iso}}$ for the two distinguished samples: $\protect%
\epsilon _{\protect\gamma }\gtrsim 50\%$ (red stars) and $\protect\epsilon _{%
\protect\gamma }\lesssim 50\%$ (blue stars). (b) The distribution of $E_{%
\text{ratio}}$ and $E_{\text{iso}}/E_{\text{k}}$ for the $\protect\epsilon _{%
\protect\gamma }\lesssim 50\%$ sample (reduced $\chi^{2}=0.140$). (c) The $E_{\text{p}}$ and $E_{\text{%
iso}}$ distribution of the main pulse for other 7 bursts (blue circles and
boxes), which have extended emission (see \citealt{Minaev2019}) and lack
efficiency. Note that all the 5 bursts of the $\protect\epsilon _{\protect%
\gamma }\gtrsim 50\%$ sample (red stars) have extended emission, including
GRB 170817A. (d) The comparison of the ratios of the $E_{\text{p}}$ and the
fluence (or $E_{\text{iso}}$) for the main pulse and the extended emission
of a large extended emission sample (reduced $\chi^{2}=0.196$), including the bursts without redshift
(blue stars). }
\end{figure*}

For the photosphere emission model, the peak energy of the observed spectrum 
$E_{\text{p}}$ corresponds to the temperature of the observed blackbody $T_{%
\text{ob}}$. In the unsaturated regime ($R_{\text{ph}}<R_{s}$, $\epsilon
_{\gamma }\gtrsim 50\%$), $T_{\text{ob}}=T_{0}$. Since $T_{0}\propto E^{1/4}\propto (E_{\text{iso}})^{1/4}$, we should have $E_{\text{p}}\propto (E_{\text{iso}})^{1/4}$ (see
Figures 3–5; for $\epsilon _{\gamma }\gtrsim 80\%$ and $\epsilon_{\gamma }\gtrsim 50\%$ of long GRBs, and short GRBs).

\subsubsection{For $\protect\epsilon _{\protect\gamma}\lesssim 50\%$, $E_{%
\text{iso}}=E\cdot (R_{\text{ph}}/R_{s})^{-2/3}$ and $E_{\text{p}}\propto
E^{1/4}\cdot (R_{\text{ph}}/R_{s})^{-2/3}$}

For $\epsilon _{\gamma}\lesssim 50\%$ ($R_{\text{ph}}>R_{s}$), the outflow
performs adiabatic expansion at $r>R_{s}$. Thus, the comoving temperature
decreased as $T_{\text{comoving}}=(T_{0}/\Gamma)\cdot(r/R_{s})^{-2/3}$. The
escaped photons at $R_{\text{ph}}$ have $E_{\text{iso}}=E\cdot (R_{\text{ph}%
}/R_{s})^{-2/3}$ and $E_{\text{p}}=2.7kT_{0}\cdot (R_{\text{ph}%
}/R_{s})^{-2/3}\propto E^{1/4}\cdot (R_{\text{ph}}/R_{s})^{-2/3}$. So, $E_{%
\text{iso}}$ and $E_{\text{p}}$ should both decrease by the same factor of $%
(R_{\text{ph}}/R_{s})^{-2/3}$, compared with the $E_{\text{p}}\propto (E_{%
\text{iso}})^{1/4}$ correlation for $\epsilon _{\gamma}\gtrsim 50\%$ (see
Figures 4 and 5).

\subsubsection{For $\protect\epsilon _{\protect\gamma}\lesssim 50\%$, $E_{%
\text{iso}}/E_{\text{k}}=(R_{\text{ph}}/R_{s})^{-2/3}=[(E_{\text{p}%
}/2.7k)^{4}\ast (4\protect\pi r_{1}^{2}ac)^{{}}/E_{\text{iso}}]^{1/3}$}

\begin{figure*}[th]
\label{Fig_6} \centering\includegraphics[angle=0,height=2.4in]{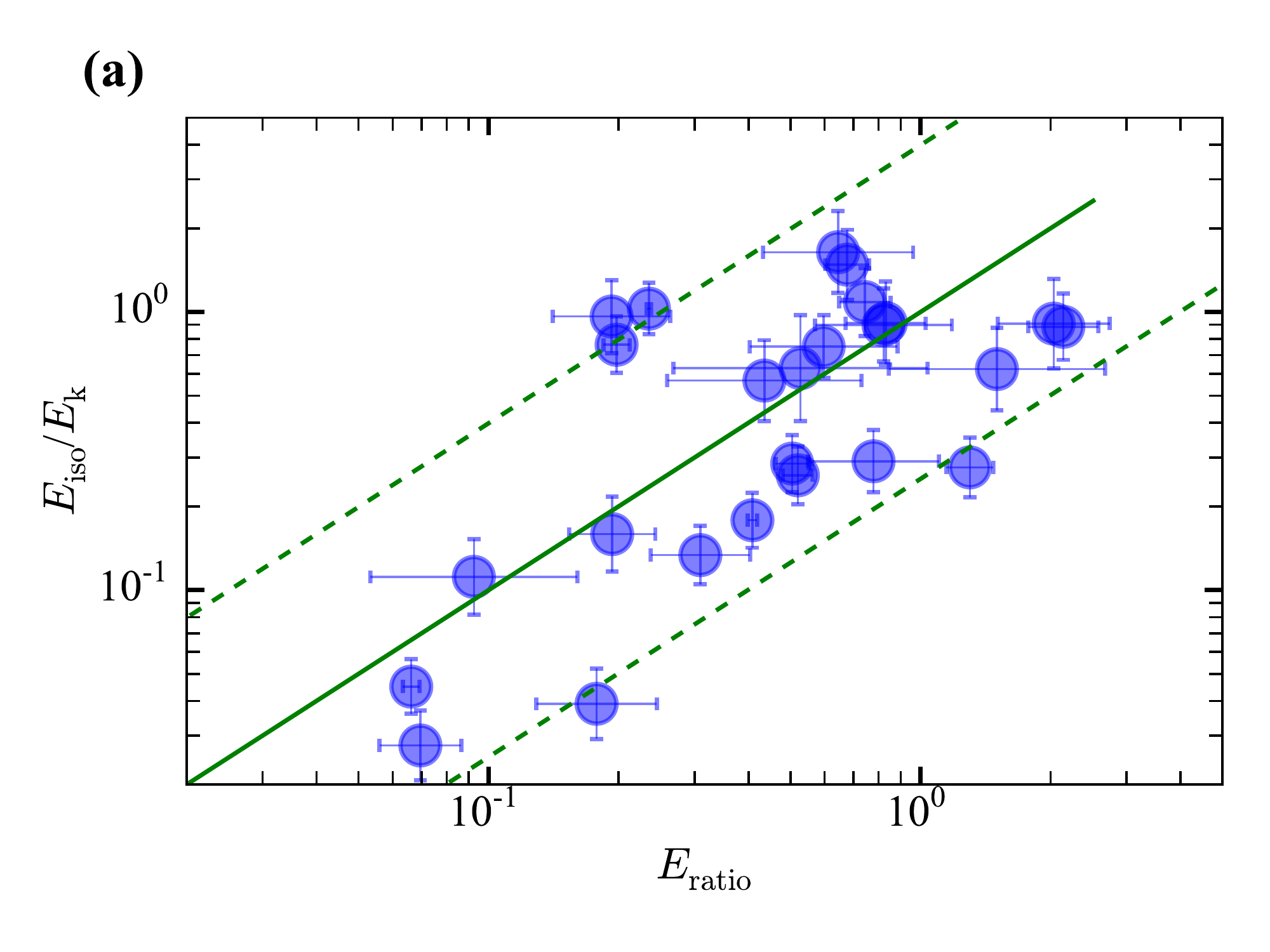} %
\centering\includegraphics[angle=0,height=2.4in]{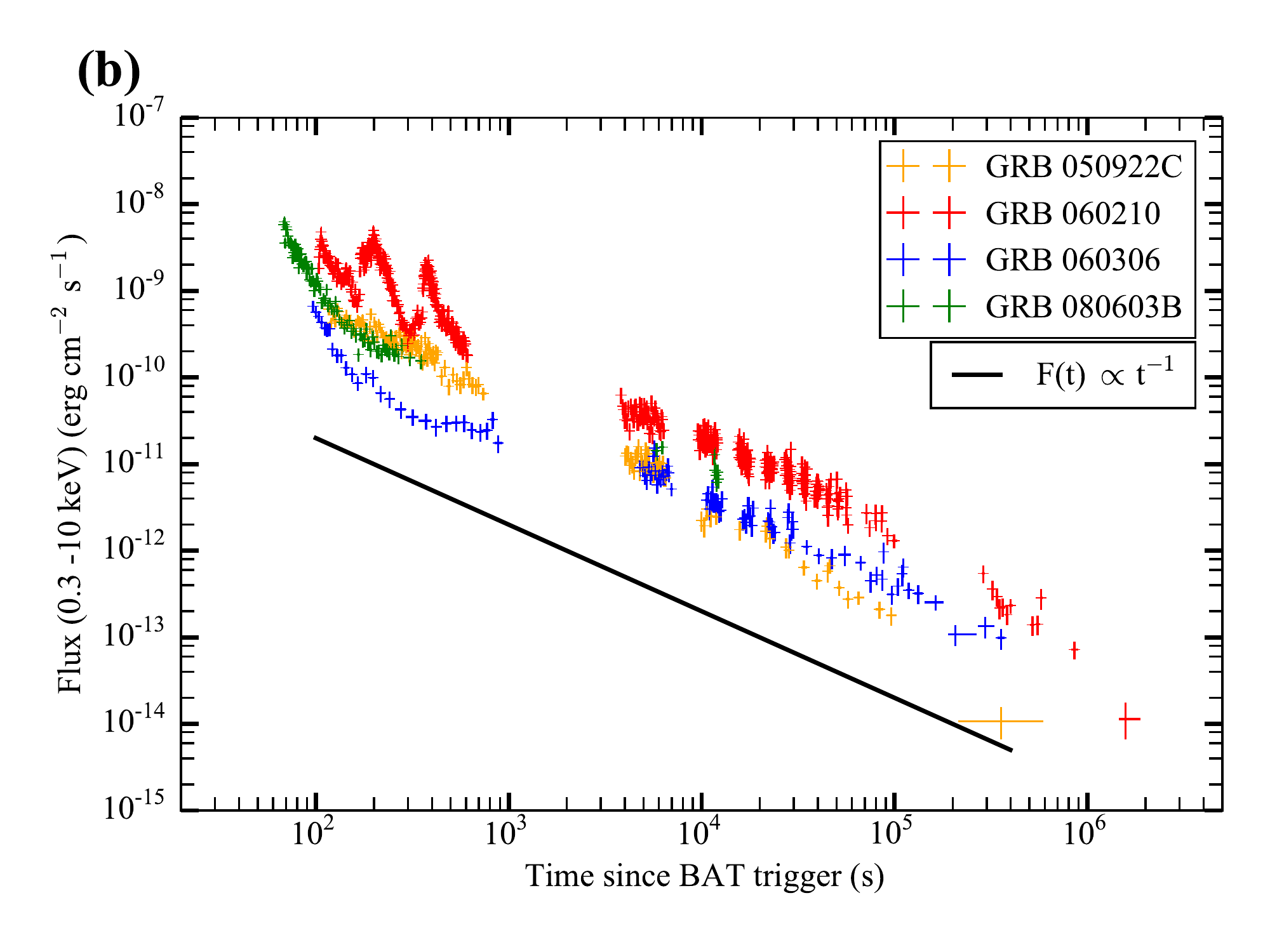} \centering%
\includegraphics[angle=0,height=2.4in]{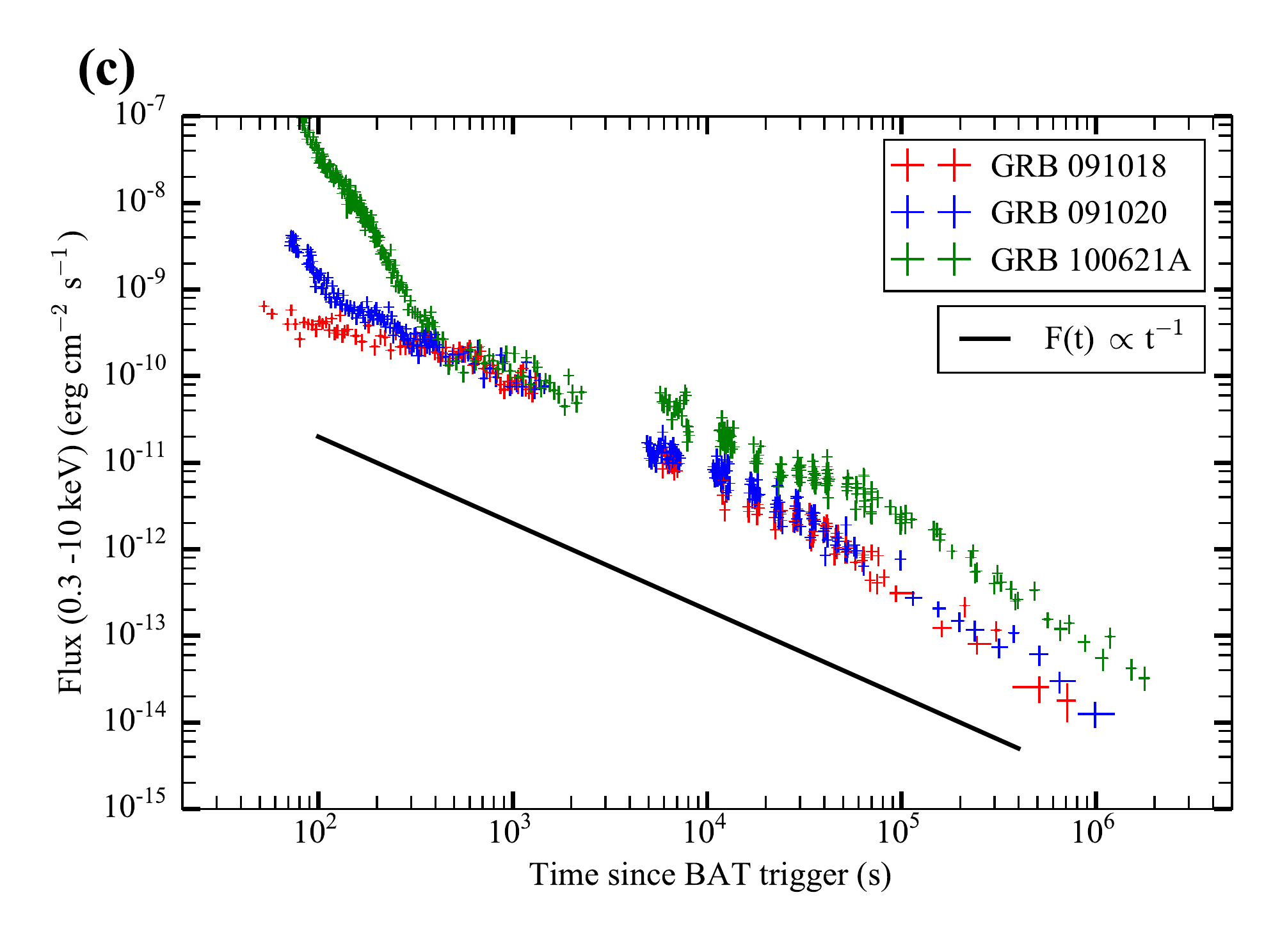}
\caption{The similar observed efficiency (with $E_{\text{iso}}/E_{\text{k}}$
from the X-ray afterglow) and theoretically predicted efficiency (from the
prompt emission) for the bursts before GRB 110213A (with $L_{\text{X,11h}}$
given in \citealt{Avan2012}). (a) The distribution of $E_{\text{ratio}}$ and 
$E_{\text{iso}}/E_{\text{k}}$ for the $\protect\epsilon _{\protect\gamma %
}\lesssim 50\%$ sample, which is found to be well centered around the
equal-value line and have a linear correlation (reduced $\protect\chi%
^{2}=0.160$). (b), (c) The X-ray afterglow light curves for the bursts (7
bursts) with almost same $E_{\text{ratio}}$ and $E_{\text{iso}}/E_{\text{k}}$
(see Table 5). We find that all these light curves do have a power-law shape
with a slope of $\sim$ $-1$. }
\end{figure*}

For the $\epsilon _{\gamma }\lesssim 50\%$ sample (saturated acceleration, $E_{\text{k}}=E$),
we should have
\begin{equation}
E_{\text{iso}}=E\cdot (R_{\text{ph}}/R_{s})^{-2/3}  \label{a}
\end{equation}%
and 
\begin{eqnarray}
E_{\text{p}} &=&2.7kT_{0}\cdot (R_{\text{ph}}/R_{s})^{-2/3}  \notag \\
&=&2.7k(L/4\pi r_{0}^{2}ac)^{1/4}\cdot (R_{\text{ph}}/R_{s})^{-2/3}  \notag
\\
&=&2.7k(E/4\pi r_{1}^{2}ac)^{1/4}\cdot (R_{\text{ph}}/R_{s})^{-2/3}  \notag
\\
&=&2.7k(4\pi r_{1}^{2}ac)^{-1/4}\cdot E^{1/4}\cdot (R_{\text{ph}%
}/R_{s})^{-2/3},  \label{b}
\end{eqnarray}%
here $r_{1}^{2}\simeq r_{0}^{2}\ast T_{90\text{.}}$ From the relation of $%
\log $ ($E_{\text{p}}$) $=2.54+0.25\log $ ($E_{\text{iso}}$) in Figure 3(b),
we obtain $r_{1}^{{}}=8.45\times 10^{8}$ cm. Then, from Equation (\ref{a})
and Equation (\ref{b}), we get $\ \ \ \ \ \ \ \ \ \ \ \ \ \ \ \ \ $%
\begin{equation}
(E_{\text{p}})^{4}/E_{\text{iso}}=(2.7k)^{4}\ast (4\pi
r_{1}^{2}ac)^{-1}\cdot (R_{\text{ph}}/R_{s})^{-2}\text{,}
\end{equation}%
thus $\ \ \ \ \ \ \ \ \ \ \ \ \ \ \ \ \ \ \ \ \ \ \ \ \ \ $%
\begin{equation}
\ (R_{\text{ph}}/R_{s})^{-2/3}=[(E_{\text{p}}/2.7k)^{4}\ast (4\pi
r_{1}^{2}ac)^{{}}/E_{\text{iso}}]^{1/3}.
\end{equation}

Combined with the abovementioned $E_{\text{iso}}/E_{\text{k}}=(R_{\text{ph}%
}/R_{s})^{-2/3}$, $E_{\text{ratio}}=E_{\text{iso}}/E_{\text{k}}=(R_{\text{ph}%
}/R_{s})^{-2/3}$ is predicted ($E_{\text{ratio}}=[(E_{\text{p}%
}/2.7k)^{4}\ast (4\pi r_{1}^{2}ac)^{{}}/E_{\text{iso}}]^{1/3}$; see Figures 2, and 5–7).

\subsubsection{$\Gamma \propto E_{\text{iso}}^{1/8}E_{\text{p}%
}^{1/2}/(T_{90})^{1/4}$ correlation}

\begin{figure*}[th]
\label{Fig_7} \centering\includegraphics[angle=0,height=2.4in]{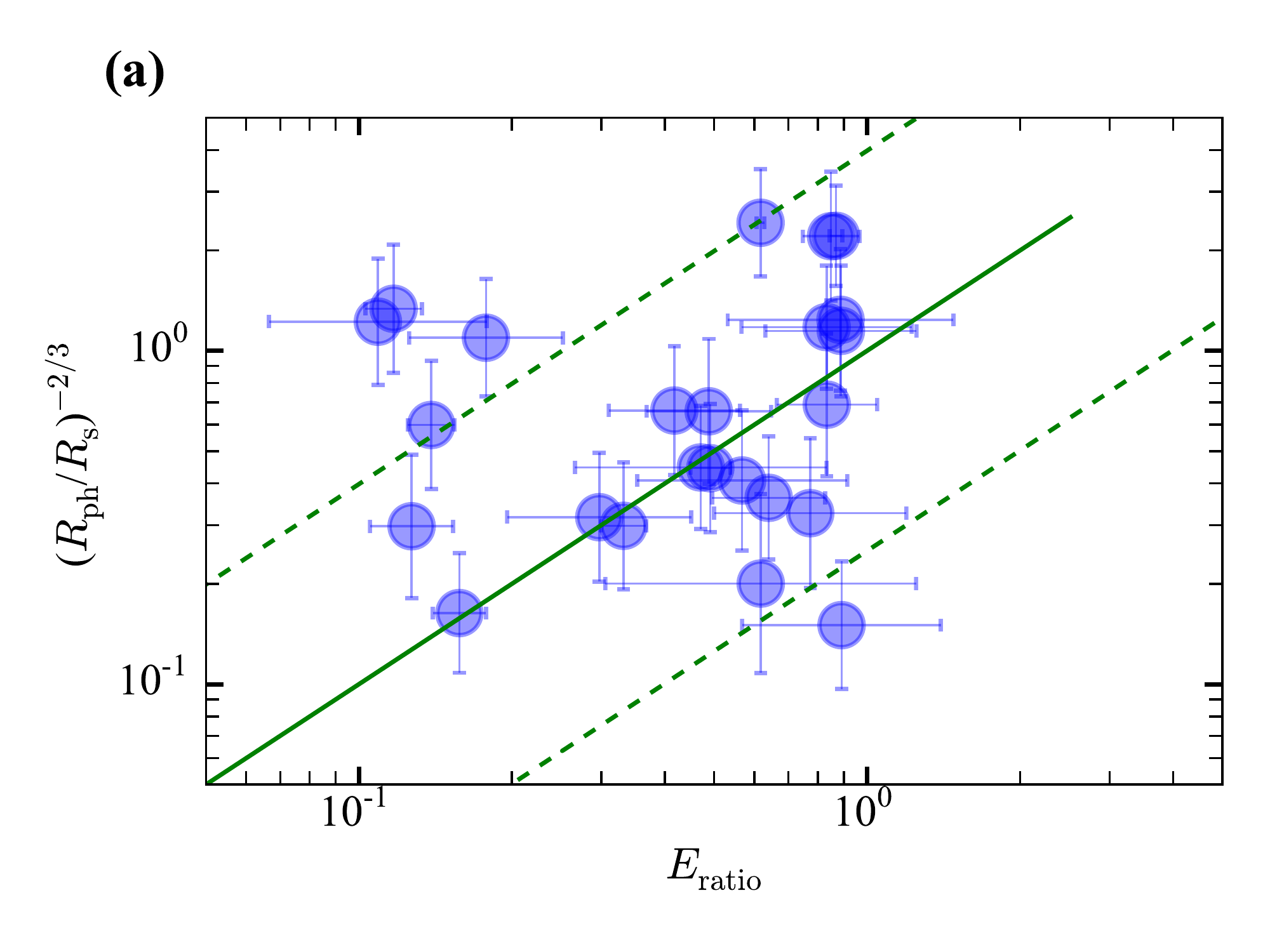} %
\centering\includegraphics[angle=0,height=2.4in]{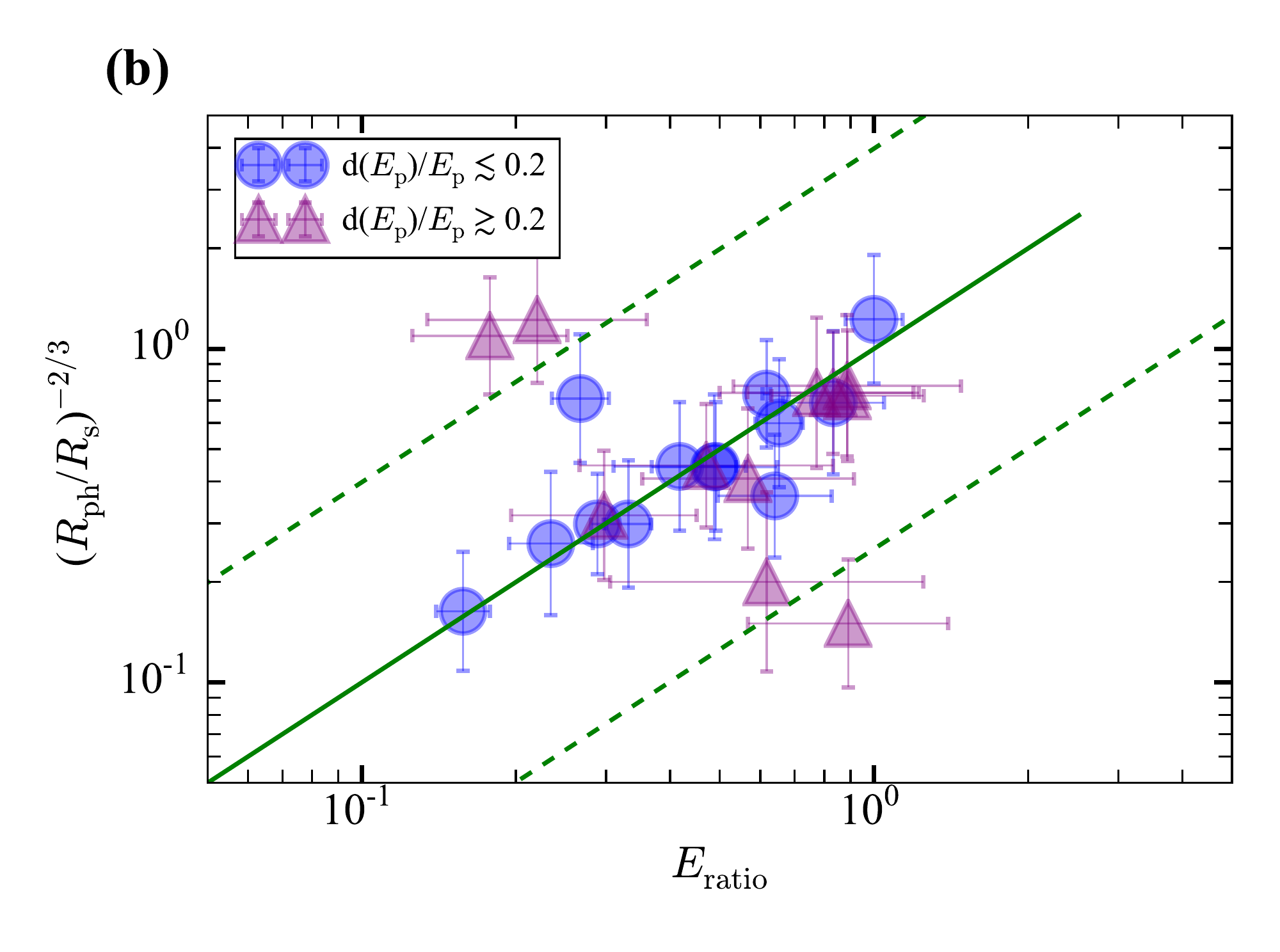} \centering%
\includegraphics[angle=0,height=2.4in]{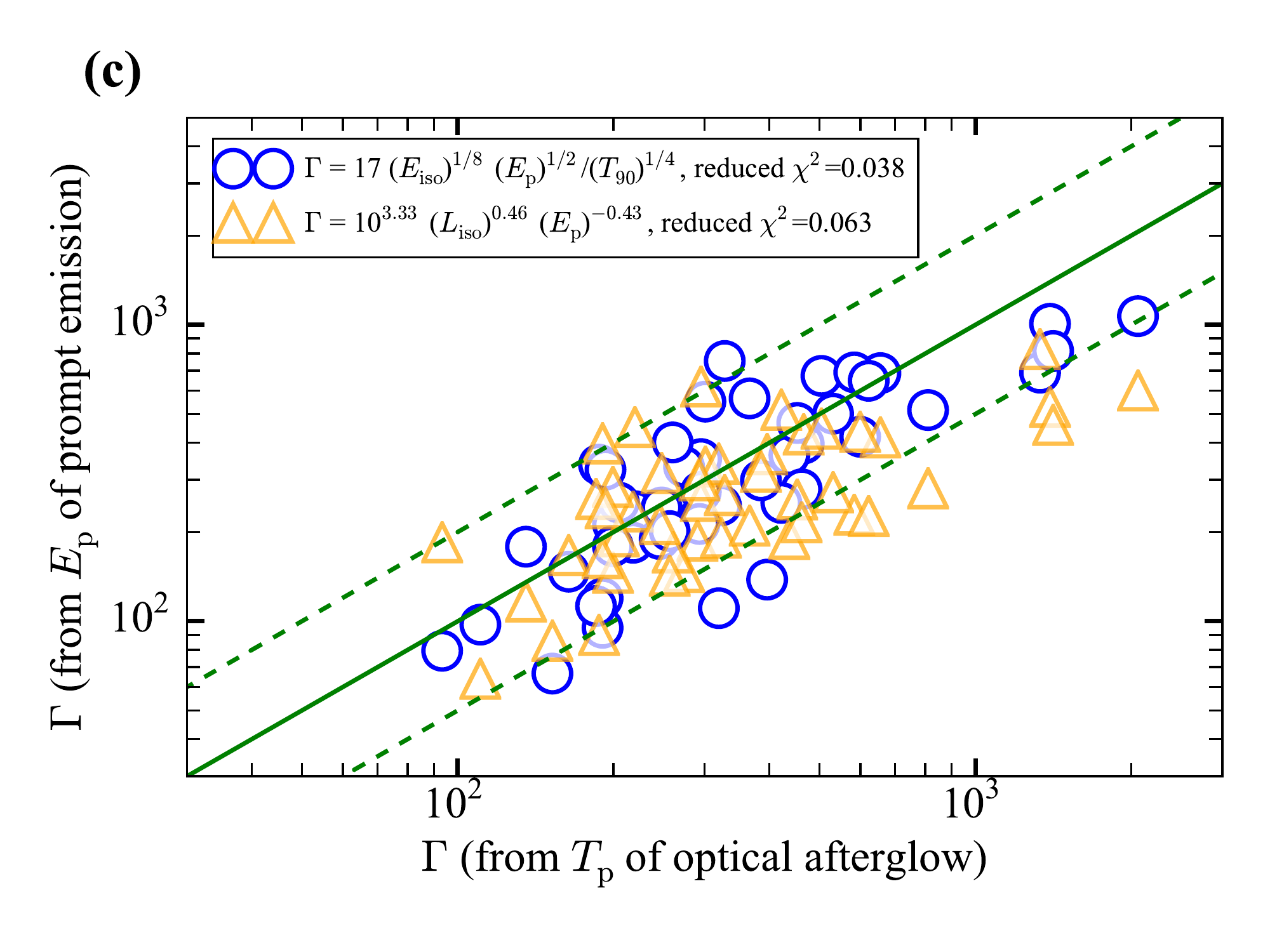} \centering%
\includegraphics[angle=0,height=2.4in]{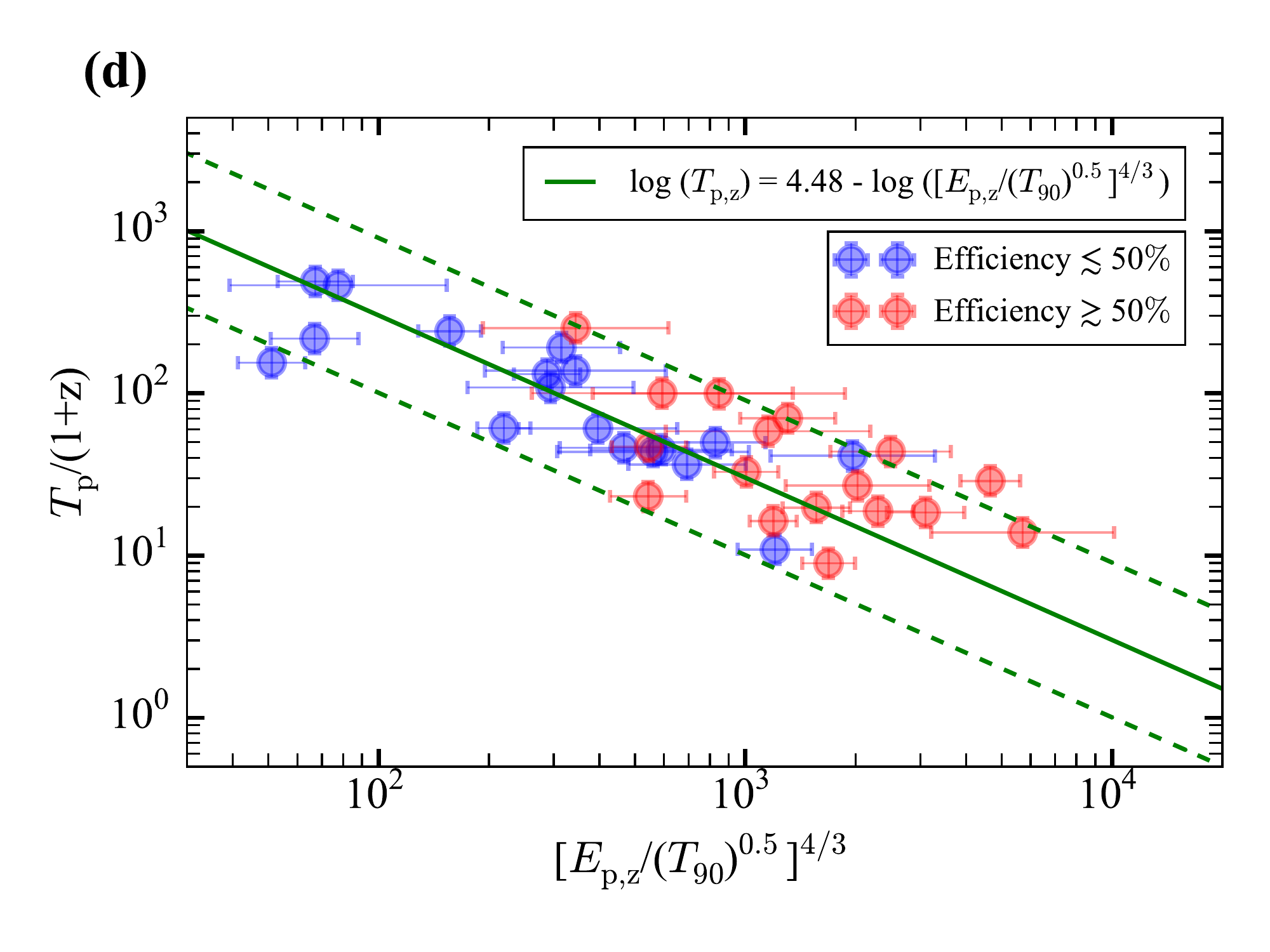} \centering%
\includegraphics[angle=0,height=2.4in]{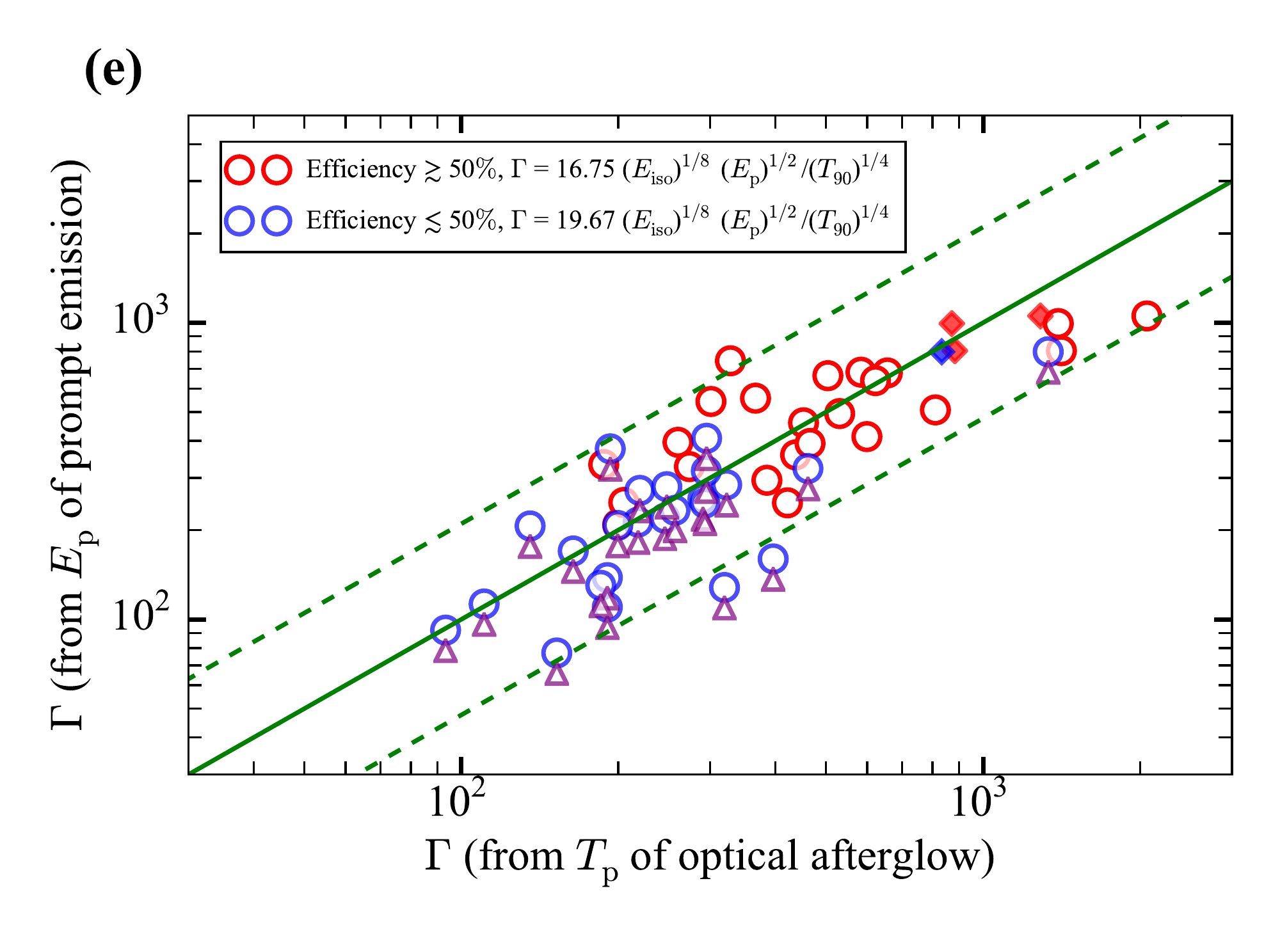} \ \ 
\caption{The tight correlation of $\Gamma \propto E_{\text{iso}}^{1/8}E_{%
\text{p}}^{1/2}/(T_{90})^{1/4}$ to estimate $\Gamma $, we derived based on
the same efficiency of $E_{\text{ratio}}$ and $(R_{\text{ph}}/R_{s})^{-2/3}$
in the $\protect\epsilon _{\protect\gamma }\lesssim 50\%$ case (see Table
6). (a) The distribution of $E_{\text{ratio}}$ and $(R_{\text{ph}%
}/R_{s})^{-2/3}$ for the selected $\protect\epsilon _{\protect\gamma %
}\lesssim 50\%$ sample (based on $E_{\text{ratio}}\leq 0.9$; 24 bursts;
reduced $\protect\chi^{2}=0.239$). (b) The distributions of $E_{\text{ratio}%
} $ and $(R_{\text{ph}}/R_{s})^{-2/3}$ for the two sub-samples with smaller $%
E_{\text{p}}$ errors ($dE_{\text{p}}/E_{\text{p}}\leq 0.2$; blue circles;
reduced $\protect\chi^{2}=0.025$) and larger $E_{\text{p}}$ errors ($dE_{%
\text{p}}/E_{\text{p}}\geq 0.2$; purple triangles; reduced $\protect\chi%
^{2}=0.228$). (c) A comparison of the $\Gamma $ obtained from the optical
afterglow (for 47 bursts in \citealt{Ghirlan2018}) and the $\Gamma $
obtained from the prompt emission (orange triangles for $\Gamma =10^{3.33}L_{%
\text{iso}}^{0.46}E_{\text{p}}^{-0.43}$ and blue circles for $\Gamma
=17\cdot E_{\text{iso}}^{1/8}E_{\text{p}}^{1/2}/(T_{90})^{1/4}$). (d) The
distribution of $[E_{\text{p}}/(T_{90})^{1/2}]^{4/3}$ and $T_{p}^{{}}/(1+z)$
for 35 bursts (reduced $\protect\chi^{2}=0.100$). (e) The slightly different
constants for the $\protect\epsilon _{\protect\gamma }\lesssim 50\%$ case
(Const=19.67, reduced $\protect\chi^{2}=0.034$; Const=16.75, purple
triangles, reduced $\protect\chi^{2}=0.048$) and the $\protect\epsilon _{%
\protect\gamma }\gtrsim 50\% $ case (Const = 16.75). }
\end{figure*}

\begin{figure*}[th]
\label{Fig_8} \centering\includegraphics[angle=0,height=2.4in]{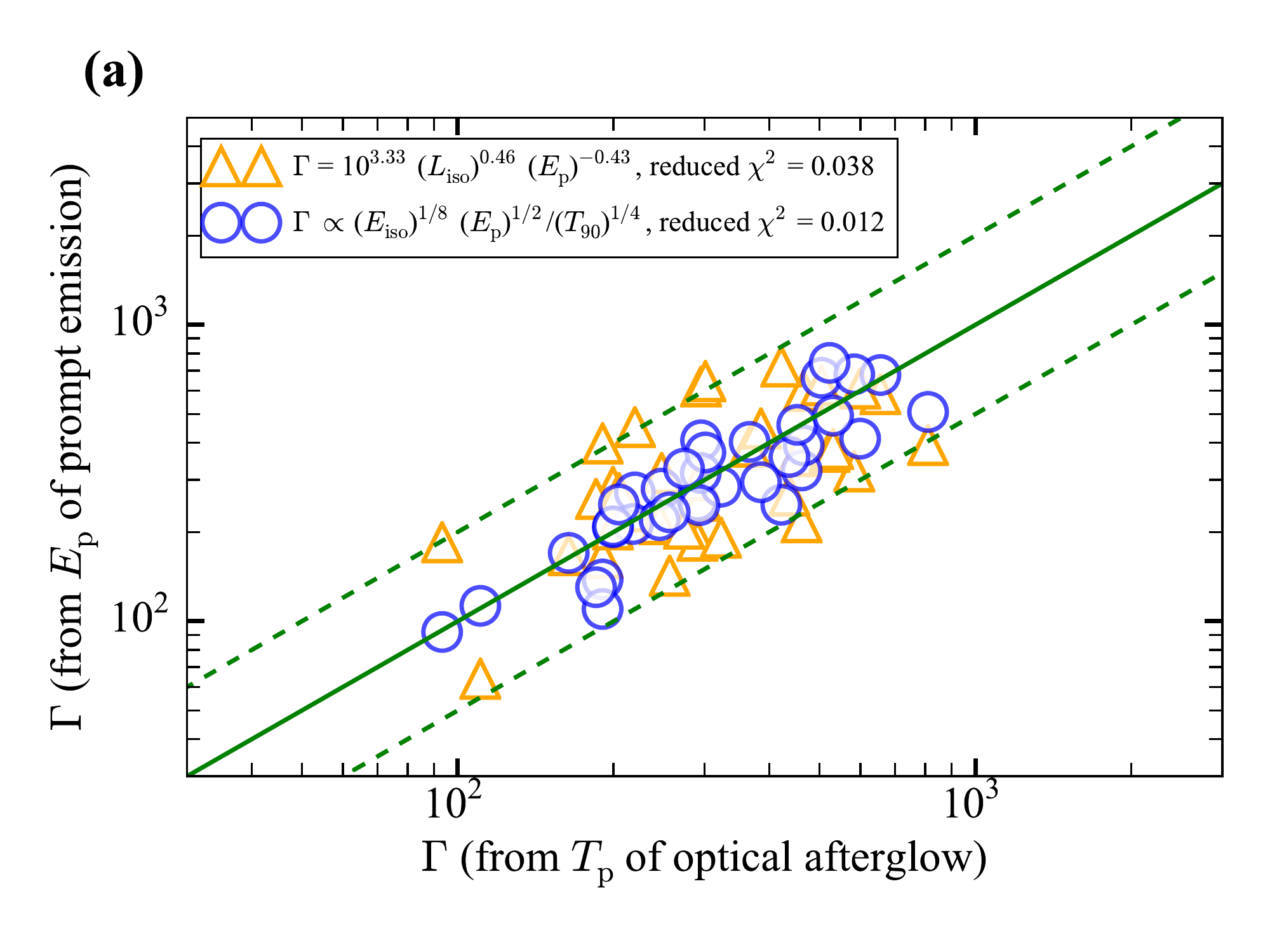} %
\centering\includegraphics[angle=0,height=2.4in]{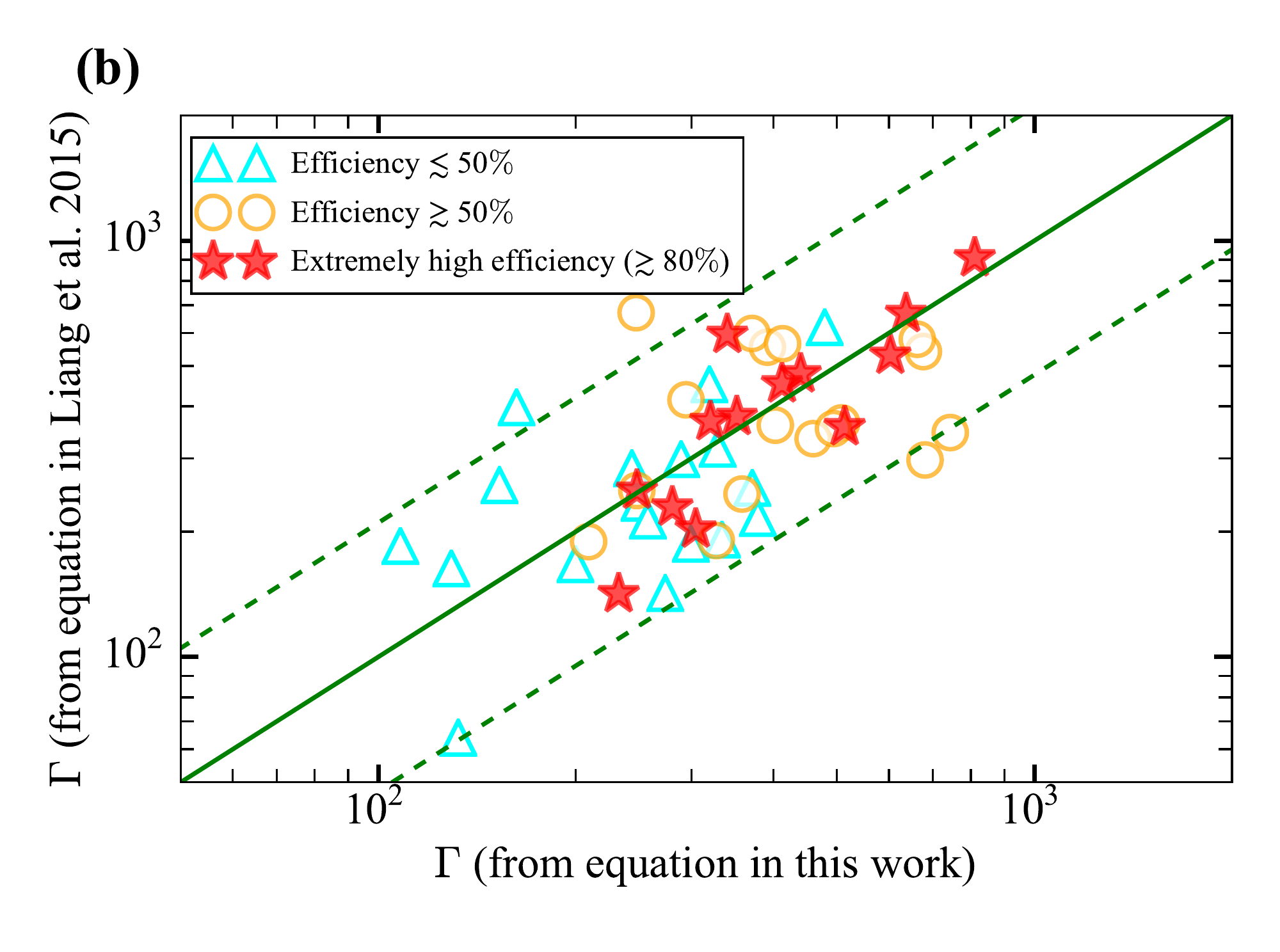} \centering%
\includegraphics[angle=0,height=2.4in]{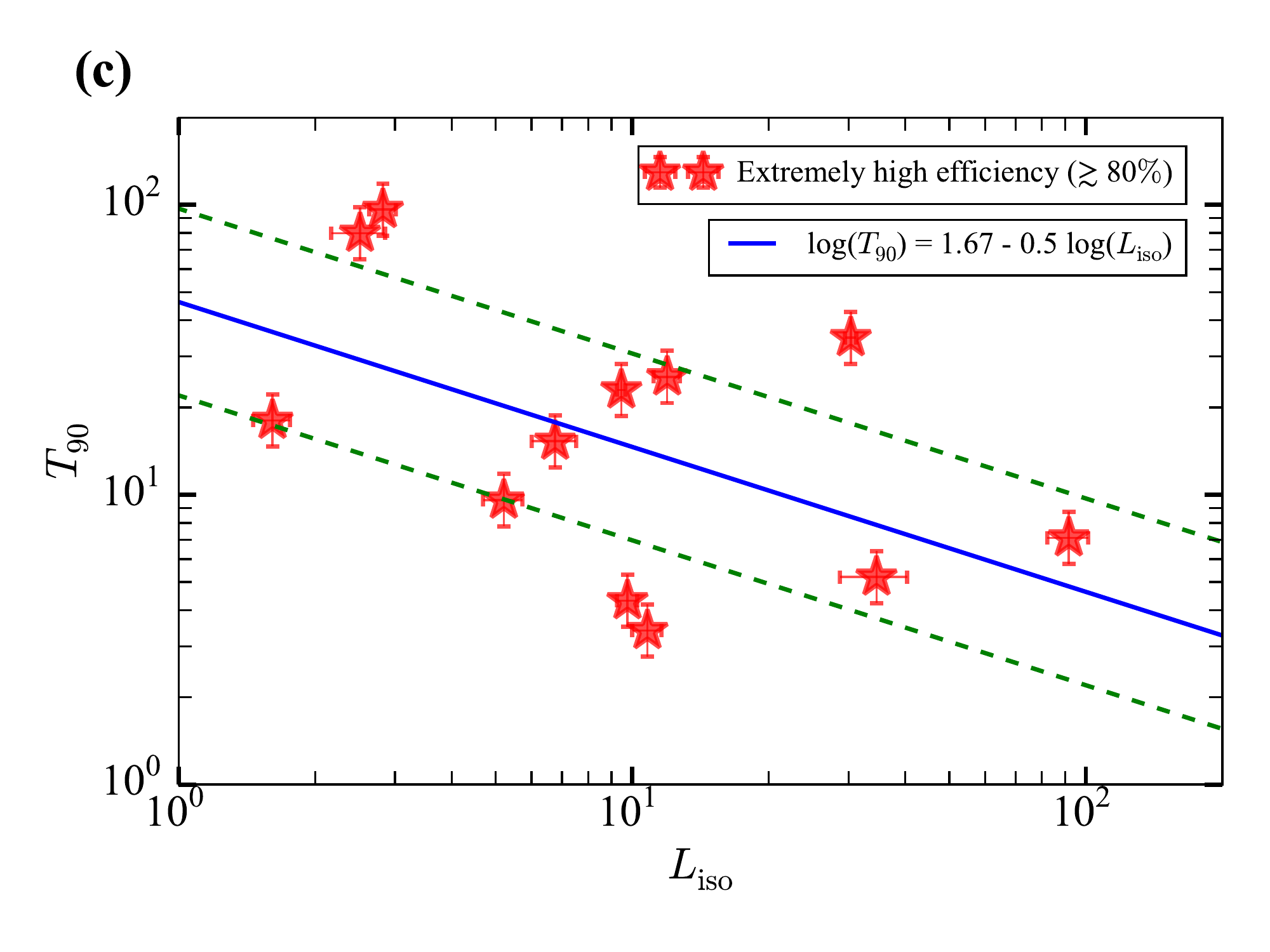} \ \ 
\caption{Comparison of the $\Gamma $ obtained from $\Gamma \propto E_{\text{%
iso}}^{1/8}E_{\text{p}}^{1/2}/(T_{90})^{1/4}$ and $\Gamma =10^{3.33}L_{\text{%
iso}}^{0.46}E_{\text{p}}^{-0.43}$ (see Tables 6 and 1). (a) The comparison of the $\Gamma $
obtained from the optical afterglow and the $\Gamma $ obtained from the prompt
emission (orange triangles for $\Gamma =10^{3.33}L_{\text{iso}}^{0.46}E_{%
\text{p}}^{-0.43}$ and blue circles for $\Gamma \propto E_{\text{iso}%
}^{1/8}E_{\text{p}}^{1/2}/(T_{90})^{1/4}$)\ for the better sample (35
bursts) in Figure 7(d). The different constants are used for the $\protect%
\epsilon _{\protect\gamma }\gtrsim 50\%$ and $\protect\epsilon _{\protect%
\gamma }\lesssim 50\%$ sub-samples. (b) Comparison of
the $\Gamma $ obtained from $\Gamma =10^{3.33}L_{\text{iso}}^{0.46}E_{\text{p}%
}^{-0.43}$ and $\Gamma \propto E_{\text{iso}}^{1/8}E_{\text{p}%
}^{1/2}/(T_{90})^{1/4}$ for the $\protect\epsilon _{\protect\gamma }\lesssim 50\%$ (cyan triangles; reduced $\chi^{2}=0.043$), the 
$\protect\epsilon _{\protect\gamma }\gtrsim 50\%$ (orange circles; reduced $\chi^{2}=0.046$) and the
high-efficiency ($\protect\epsilon _{\protect\gamma }\gtrsim 80\%$; red
stars; reduced $\chi^{2}=0.017$) sub-samples. Obviously, these two estimations are consistent with each
other, since they can be transferred with $E_{\text{p}}\propto (E_{\text{iso}%
})^{1/4}$. (c) The $T_{90}\propto
(L_{\text{iso}})^{-0.5}$ correlation (reduced $\chi^{2}=0.174$), found for the high-efficiency
sub-sample ($\protect\epsilon _{\protect\gamma }\gtrsim 80\%$).}
\end{figure*}

\begin{figure*}[th]
\label{Fig_9} \centering\includegraphics[angle=0,height=2.4in]{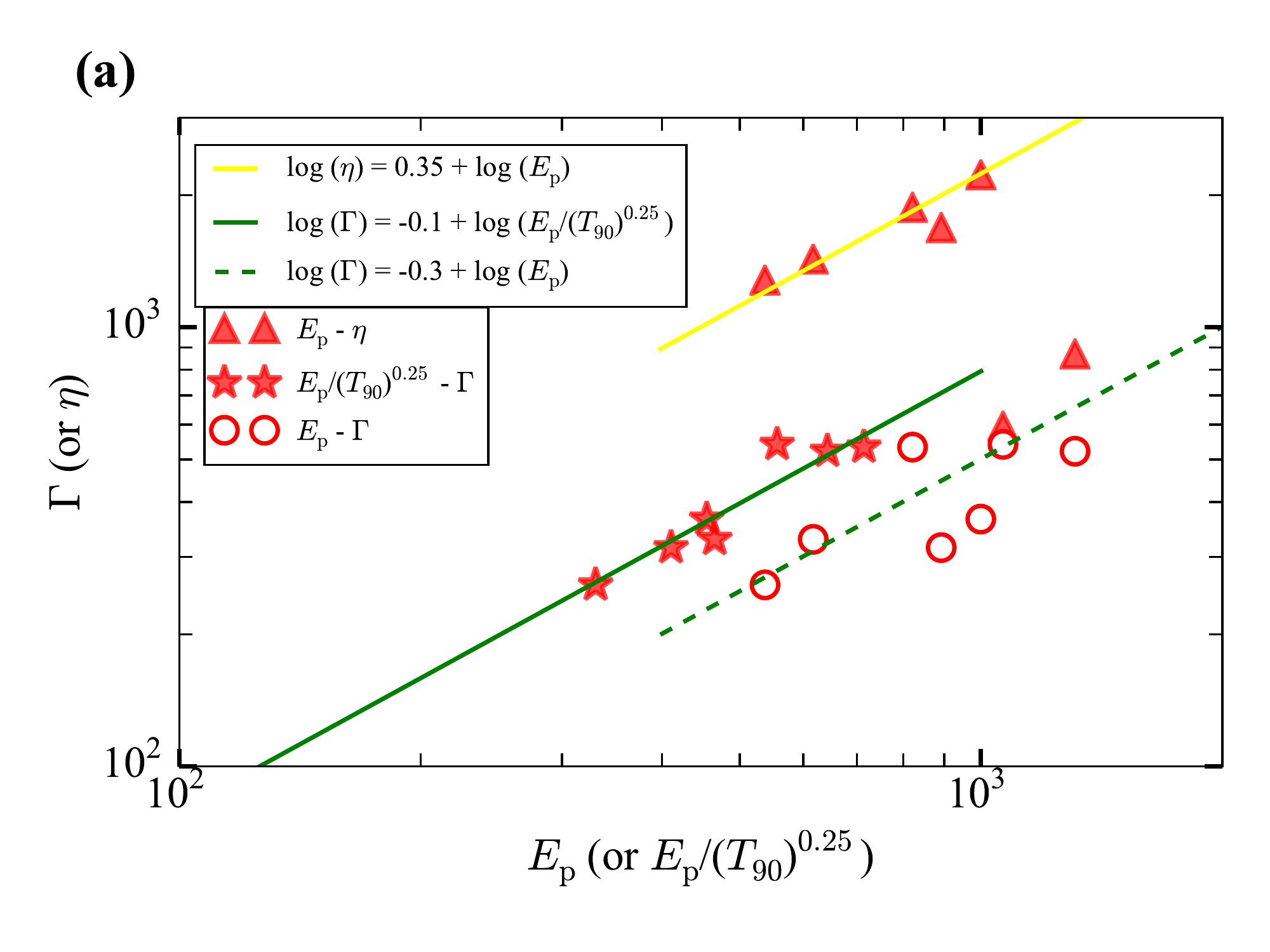} %
\centering\includegraphics[angle=0,height=2.4in]{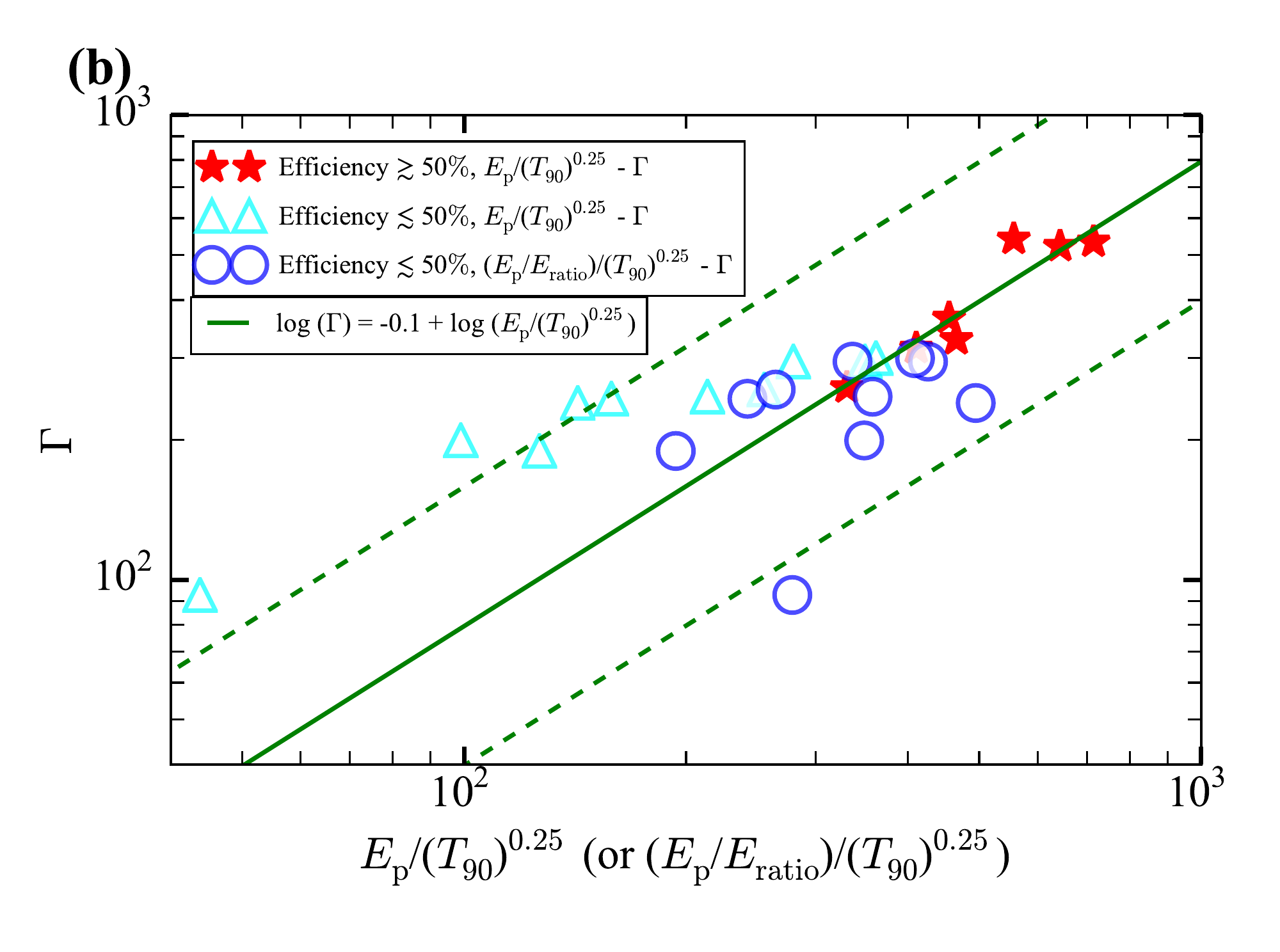} \centering%
\includegraphics[angle=0,height=2.4in]{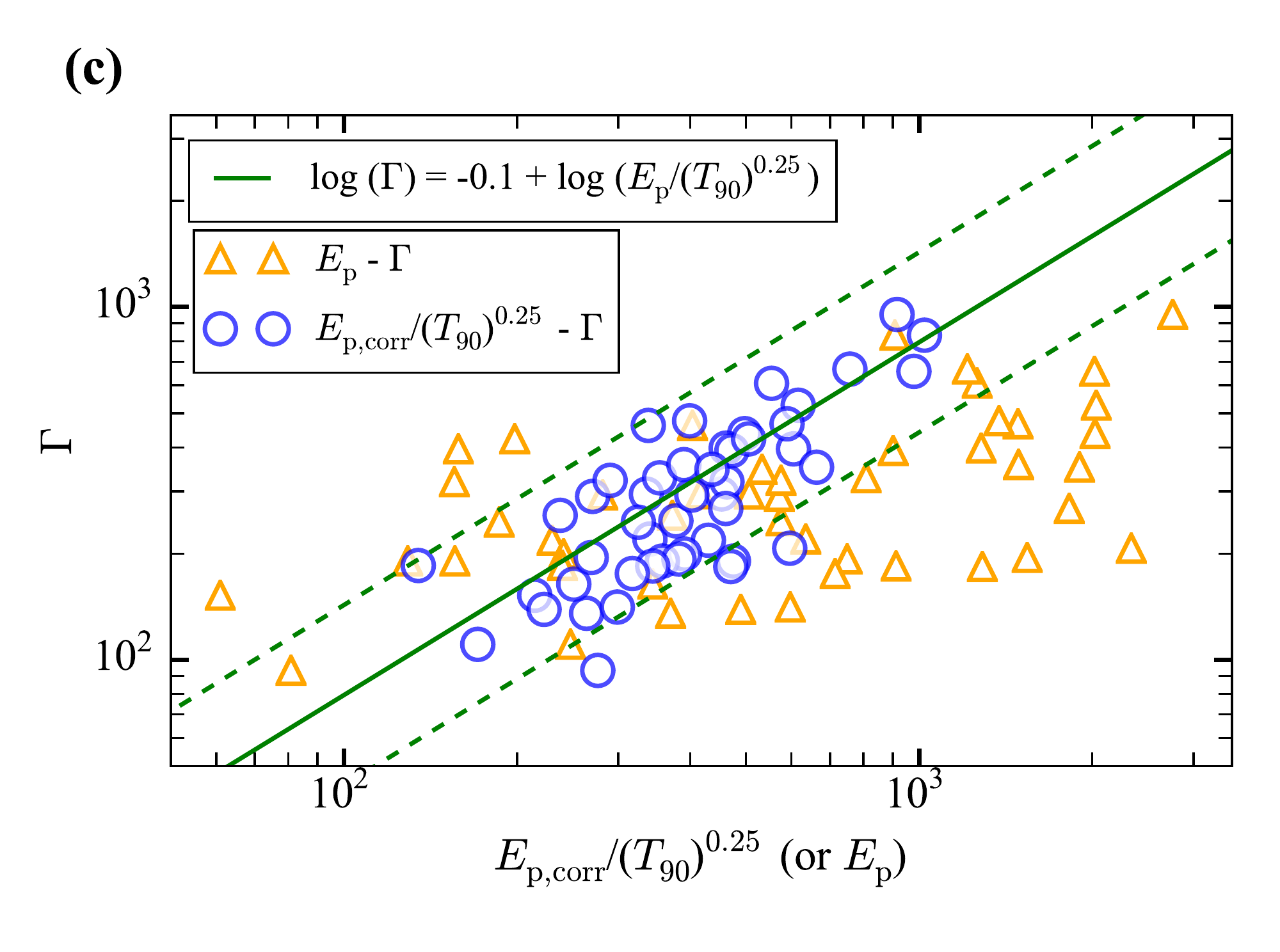} \ \ 
\caption{The tight correlation of $\Gamma \propto $ $E_{\text{p}%
}^{{}}/(T_{90})^{1/4}$, predicted by $\Gamma \propto E_{\text{iso}}^{1/8}E_{%
\text{p}}^{1/2}/(T_{90})^{1/4}$ and $E_{\text{p}}\propto (E_{\text{iso}%
})^{1/4}$. (a) The distributions of $E_{\text{p}}-\Gamma $ (red circles;
reduced $\protect\chi^{2}=0.013$), $E_{\text{p}}-\protect\eta $ (red
triangles; reduced $\protect\chi^{2}=0.002$) and $E_{\text{p}%
}^{{}}/(T_{90})^{1/4}-\Gamma $ (red stars; reduced $\protect\chi^{2}=0.002$)
for the selected $\protect\epsilon _{\protect\gamma }\gtrsim 50\%$ sample
with $T_{p}^{{}}$ detection (7 bursts in Figure 2(e); see Table 3). It is
obvious that, the tight correlations of $\Gamma =10^{-0.1}\cdot E_{\text{p}%
}^{{}}/(T_{90})^{1/4}$ and $E_{\text{p}}\propto \protect\eta $ (for the 5
bursts with higher efficiency, thus $\protect\eta \propto (E_{\text{iso}%
})^{1/4}$) are found. (b) The distributions of $(E_{\text{p}}/E_{\text{ratio}%
})/(T_{90})^{1/4}-\Gamma $ (blue circles; reduced $\protect\chi^{2}=0.031$)
and $E_{\text{p}}^{{}}/(T_{90})^{1/4}-\Gamma $ (cyan triangles) for the
selected $\protect\epsilon _{\protect\gamma }\lesssim 50\%$ sample with $%
T_{p}^{{}}$ detection (6 bursts in Figure 2(d); see Table 2). The tight
correlation of $(E_{\text{p}}/E_{\text{ratio}})/(T_{90})^{1/4}\propto \Gamma 
$ is found, which is in line with the $E_{\text{p}}^{{}}/(T_{90})^{1/4}%
\propto \Gamma $ correlation for the $\protect\epsilon _{\protect\gamma %
}\gtrsim 50\%$ case. (c) The distribution of $\Gamma -$ $E_{\text{p}%
}^{{}}/(T_{90})^{1/4}$ (blue circles; reduced $\protect\chi^{2}=0.025$) for
the large sample with $\Gamma $ (47 bursts; see Table 6) in 
\citet{Ghirlan2018}. For the $\protect\epsilon _{\protect\gamma }\gtrsim
50\% $ case the $E_{\text{p}}^{{}}$ is re-derived from the $\log $ ($E_{%
\text{p}}$) $=2.54+0.25\log $ ($E_{\text{iso}}$) correlation (using the $E_{%
\text{iso}} $), and for the $\protect\epsilon _{\protect\gamma }\lesssim
50\% $ case the $E_{\text{p}}^{{}}$ is re-derived from $E_{\text{p}}/E_{%
\text{ratio}}$. Obviously, we find the distribution of $\Gamma $ and $E_{%
\text{p}}^{{}}/(T_{90})^{1/4}$ is well centered around $\Gamma
=10^{-0.1}\cdot E_{\text{p}}^{{}}/(T_{90})^{1/4}$ and have a linear
correlation.}
\end{figure*}

\begin{figure*}[h]
\label{Fig_10} \centering\includegraphics[angle=0,height=2.1in]{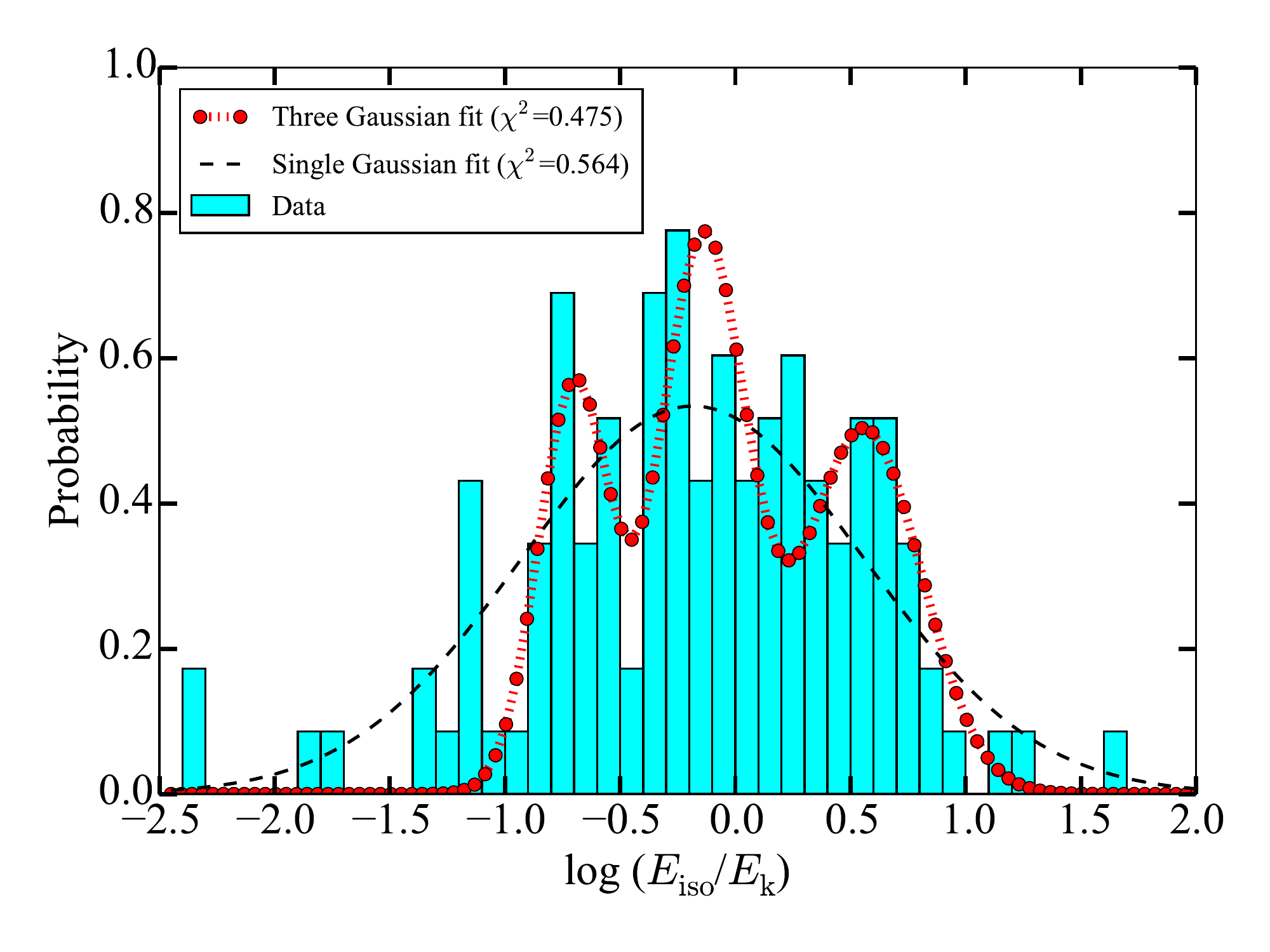}
\caption{The distribution of the $E_{\text{iso}}/E_{\text{k}}$ for our whole
sample (117 bursts, after GRB 110213A; see Tables 1-3). The mean value is around $\sim $ $%
10^{-0.2}$ to $10^{-0.3}$, thus indicating the average efficiency of $%
\protect\epsilon _{\protect\gamma }$ $\sim $ $33\%$ to $40\%$. Also, the
distribution seems to consist of three Gaussian distributions (smaller $\chi^{2}$ for $E_{\text{iso}}/E_{\text{k}} \gtrsim 10^{-1.0}$). }
\end{figure*}

As in the previous statement and shown in Figure 7(b), for $\epsilon _{\gamma
}\lesssim 50\%$, we have $E_{\text{ratio}}=(R_{\text{ph}}/R_{s})^{-2/3} $.
Considering
\begin{equation}
R_{\text{ph}}=\frac{\sigma _{T}}{8\pi m_{p}c^{3}}\frac{L_{{}}}{\Gamma ^{3}},
\label{a1}
\end{equation}%
and $R_{s}$ = $\Gamma \cdot r_{0}$, we have
\begin{equation}
\ [(E_{\text{p}}/2.7k)^{4}\ast (4\pi r_{1}^{2}ac)^{{}}/E_{\text{iso}%
}]^{1/3}=(\frac{\sigma _{T}}{8\pi m_{p}c^{3}}\frac{E_{\text{iso}}/T_{90}}{%
\Gamma ^{4}\cdot r_{0}})^{-2/3}.  \label{e}
\end{equation}

Thus, we can use the quantities of the prompt emission ($E_{\text{iso}}$, $E_{%
\text{p}}$ and $T_{90}$) to estimate the Lorentz factor $\Gamma$, just as
the obtained $\Gamma =10^{3.33}L_{\text{iso}}^{0.46}E_{\text{p}}^{-0.43}$
from the statistic fitting in \citet{Liang2015} (see Figures 7(c) and 8).
Based on Equation (\ref{e}), we derive
\begin{equation}
\Gamma \propto E_{\text{iso}}^{1/8}E_{\text{p}}^{1/2}/(T_{90})^{1/4}.
\label{ee}
\end{equation}%
Considering the constants, we obtain  
\begin{equation}
\Gamma =19.67\cdot E_{\text{iso}}^{1/8}E_{\text{p}}^{1/2}/(T_{90})^{1/4}%
\text{.}  \label{abc}
\end{equation}

For the $\epsilon _{\gamma }\gtrsim 50\%$ case, from
\begin{equation}
R_{\text{ph}}=\left[ \frac{\sigma _{T}}{6m_{p}c}\frac{L_{\text{iso}}}{4\pi
c^{2}\eta }r_{0}^{2}\right] ^{1/3}\text{,}  \label{ab}
\end{equation}%
we have 
\begin{eqnarray}
\Gamma &=&R_{\text{ph}}/r_{0}^{{}}=\left[ \frac{\sigma _{T}}{6m_{p}c}\frac{%
L_{\text{iso}}}{4\pi c^{2}\eta }r_{0}^{2}\right] ^{1/3}/r_{0}^{{}}  \notag \\
&=&\left[ \frac{\sigma _{T}}{6m_{p}c}\frac{L_{\text{iso}}}{4\pi c^{2}\eta
r_{0}^{{}}}\right] ^{1/3}\text{.}  \label{de}
\end{eqnarray}%
Thus, $L_{\text{iso}}$ $\propto $ $E_{\text{iso}}^{{}}/(T_{90})$ $\propto $ $%
\eta \cdot \Gamma ^{3}$. Combined with $E_{\text{iso}}/E_{\text{k}}=\eta
Mc^{2}/\Gamma Mc^{2}=\eta /\Gamma $, we have $1/T_{90}$ $\propto $ $(1/E_{%
\text{k}})\cdot \Gamma ^{4}$, namely $E_{\text{k}}\propto $ $\Gamma
^{4}\cdot T_{90}$. Putting this into $\Gamma \propto (E_{\text{k}%
}^{{}})^{1/8}\cdot \lbrack T_{p}^{{}}/(1+z)]^{-3/8}$, we obtain $\Gamma
\propto \Gamma ^{1/2}\cdot (T_{90})^{1/8}\cdot \lbrack
T_{p}^{{}}/(1+z)]^{-3/8}$, namely $\Gamma ^{1/2}\cdot (T_{90})^{-1/8}\propto
\lbrack T_{p}^{{}}/(1+z)]^{-3/8}$. From Figure 9(a) (discussed in Section
2.2.7), we have $\Gamma \propto $ $E_{\text{p}}^{{}}/(T_{90})^{1/4}$. So,
as in the $\epsilon _{\gamma }\lesssim 50\%$ case, $E_{\text{p}%
}^{1/2}/(T_{90})^{1/4}\propto \lbrack T_{p}^{{}}/(1+z)]^{-3/8}$ is obtained
(as discussed in Section 2.2.6). This means that the Equation (\ref{ee}) is
also available for the $\epsilon _{\gamma }\gtrsim 50\%$ case. After
considering the constants, we get
\begin{equation}
\Gamma =16.75\cdot E_{\text{iso}}^{1/8}E_{\text{p}}^{1/2}/(T_{90})^{1/4}%
\text{.}  \label{abc2}
\end{equation}

Note that different constants for the $\epsilon _{\gamma }\lesssim 50\%$
case and the $\epsilon _{\gamma }\gtrsim 50\%$ case are predicted (see Figure
7(d) and (e)). Also, the calculation of $\Gamma $ in Figure 9(a), using
the optical afterglow, has assumed an efficiency of $\sim$ $20\%$ (namely, $%
E_{\text{k}}=5\cdot E_{\text{iso}}$). In fact, when calculating $\Gamma$,
we should modify $E_{\text{k}}$ based on the real efficiency ($\eta
/\Gamma $). For the $\epsilon _{\gamma }\gtrsim 50\%$ case, we typically
have $\eta /\Gamma $ $\sim$ $3.5$ (see Figure 9(a) and Figure 10). Thus,
the constant $\sim$ $16.75$ should be $\sim$ $1.4$ times smaller.

\subsubsection{$T_{p}/(1+z)\propto \lbrack E_{\text{p}%
}/(T_{90})^{1/2}]^{-4/3}$ correlation}

Note that when using the optical afterglow (the peak
time $T_{p}$) to obtain $\Gamma $, we have
\begin{eqnarray}
\Gamma &\propto &(E_{\text{k}}^{{}})^{1/8}\cdot \lbrack
T_{p}^{{}}/(1+z)]^{-3/8}  \notag \\
&\propto &(E_{\text{iso}}^{{}}/\epsilon _{\gamma })^{1/8}\cdot \lbrack
T_{p}^{{}}/(1+z)]^{-3/8}\text{,}  \label{eee}
\end{eqnarray}%
where $\epsilon _{\gamma }$ is normally taken as $0.2$. Then, if the Equation (\ref{ee}) is
correct, combining Equation (\ref{ee}) and Equation (\ref{eee})
should give (see Figure 7(d))
\begin{equation}
T_{p}^{{}}/(1+z)\propto \lbrack E_{\text{p}}/(T_{90})^{1/2}]^{-4/3}\text{.}
\end{equation}

\subsubsection{$\Gamma \propto E_{\text{p}}^{{}}/(T_{90})^{1/4}$ correlation}

\begin{figure*}[th]
\label{Fig_11} \centering\includegraphics[angle=0,height=2.4in]{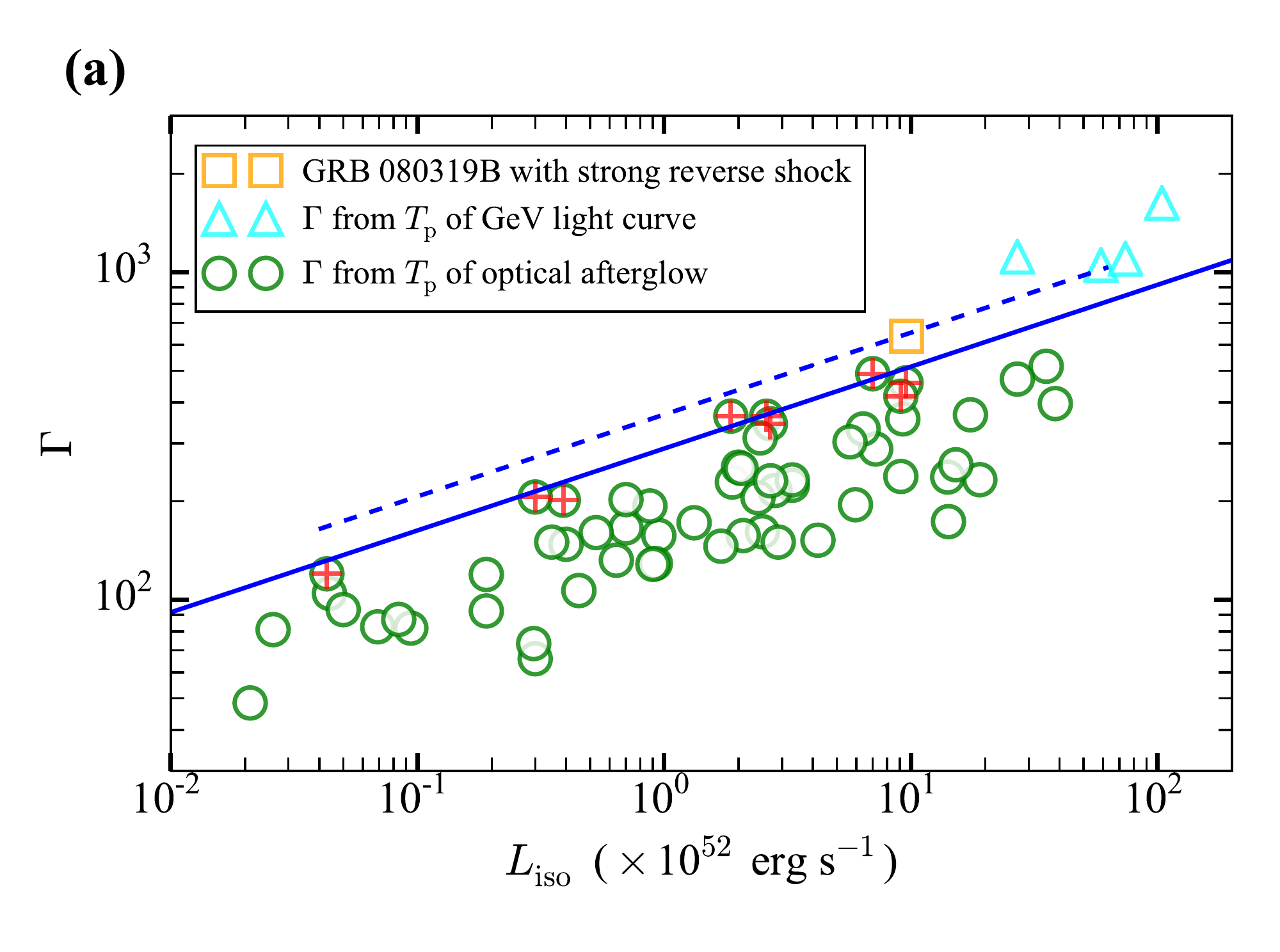} %
\centering\includegraphics[angle=0,height=2.4in]{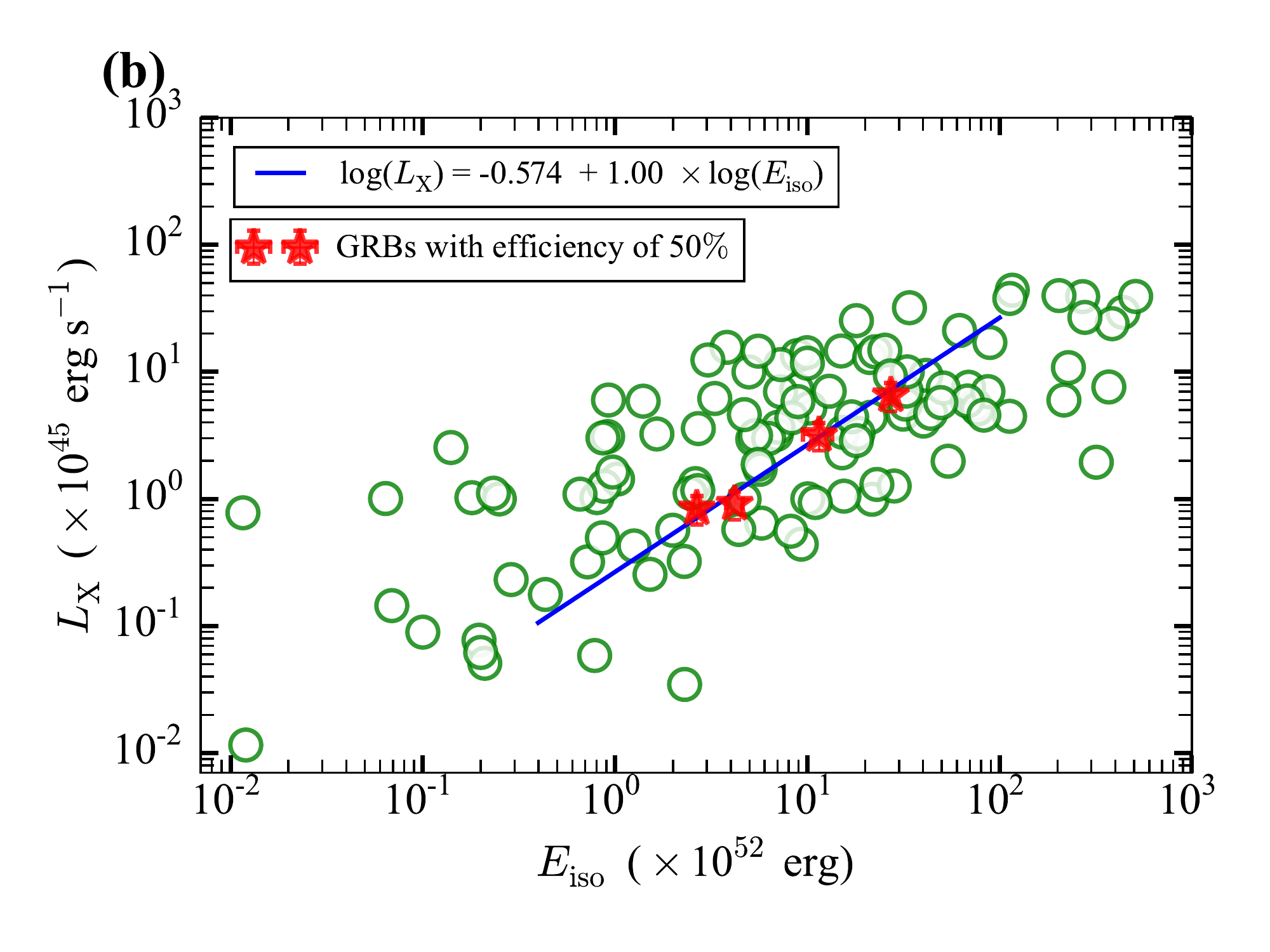} \ \ 
\caption{The $L_{\text{iso}}-\Gamma $ and $E_{\text{iso}}-L_{\text{X,45}}$
distributions for the $\protect\epsilon _{\protect\gamma }=50\%$ sample (see
Table 7). (a) The distribution of $L_{\text{iso}}$ and $\Gamma $ for the
complete sample (62 bursts) with the detection of $T_{p}$. We select the
sample (9 bursts, marked by the red plus) with the maximum $\Gamma $ (for
fixed $L_{\text{iso}} $), to check whether their efficiency is $50\%$ as
predicted by the photosphere emission model. (b) The distribution of $E_{%
\text{iso}}$ and $L_{\text{X,45}}$ for the selected sample with the maximum $%
\Gamma $ (4 bursts with $L_{\text{X,45}}$ detection, red stars). It is found
that all these bursts have almost the same efficiency (with $E_{\text{iso}%
}\propto L_{\text{X}}\propto E_{\text{k}}$, reduced $\protect\chi^{2}=0.008$%
). Thus, we think that the efficiency $\protect\epsilon _{\protect\gamma }$
for these bursts is likely to be $50\%$ ($E_{\text{iso,52}}=E_{\text{k,52}%
}\simeq 3.7\ast L_{\text{X,45}}$). }
\end{figure*}

As shown in Figure 4(a), we have $E_{\text{p}}\propto (E_{\text{iso}})^{1/4}$
for both the $\epsilon _{\gamma }\gtrsim 50\%$ case (smaller dispersion) and
the $\epsilon _{\gamma }\lesssim 50\%$ case (larger dispersion). Then, with
the Equation (\ref{ee}), we should obtain $\Gamma \propto $ $E_{\text{p}%
}^{{}}/(T_{90})^{1/4}$. On the other hand, from Figure 1(a) and Figure 11(a)
we have approximately $\Gamma \propto (L_{\text{iso}})^{1/4}$, thus $\Gamma
\propto $ $E_{\text{p}}^{{}}/(T_{90})^{1/4}$ is also likely to be obtained (see
Figure 9).

\subsection{Tightness of the Scaling Relations and Data Errors}

To quantitatively measure the tightness of the scaling relations, we
calculate the statistical value
\begin{equation}
\chi ^{2}=\sum_{i=1}^{N}(\lg y_{i}-a\lg x_{i}-b)^{2},
\end{equation}
for each relation of $\lg y=a\lg x-b$ (for example, $\lg E_{\text{p}%
}=0.25\lg E_{\text{iso}}+2.54$). The reduced $\chi ^{2}=\chi ^{2}/$degrees of
freedom (dof $=N-2$) for each relation is given in the caption for the corresponding
figure. Note that $\sqrt{\chi ^{2}/\text{dof}}$ is approximate to the
typical dispersion measure $\sigma $ (both in units of dex). In Figure 10,
for the Gaussian fit, $\chi ^{2}=\sum(y_{i}-\bar{y}_{i,\text{%
model}})^{2}$ is adopted.

To show how well the data follows the predicted relations, the missing errors
for the considered quantities are estimated. For the compound quantities (such
as $E_{\text{iso}}/E_{\text{k}}$), the errors are estimated by error
propagation: $\Delta \lbrack \lg (E_{\text{iso}}/E_{\text{k}})]=\Delta
\lbrack \lg (E_{\text{iso}})-\lg (E_{\text{k}})]=\Delta \lbrack \lg (E_{%
\text{iso}})]+\Delta \lbrack \lg (E_{\text{k}})].$ For the $L_{\text{X,11h}}$
derived in this work, $T_{p,op}$, and $T_{90,i}$, the errors at the $90\%$
confidence level ($\sim $ $0.1$ dex) are taken, just as $\Delta (L_{\text{%
X,11h}})$ given in \citet{Avan2012}.

\section{EVIDENCE FROM LONG GRBS WITH EXTREMELY HIGH
EFFICIENCY ($\protect\epsilon _{\protect\gamma }\gtrsim 80\%$)}

\label{sec:extreme}

There is much controversy about the spectral differences between the
photosphere emission model and the synchrotron emission model, after
considering the more natural and complicated physical conditions (jet
structure, decaying magnetic field, and so on;
\citealt{Uhm2014,Geng18,Meng2018,Meng2019,Meng2021}). Nevertheless, a crucial
difference between these two models is that the photosphere emission model
predicts much higher radiation efficiency $\epsilon _{\gamma }$. The
synchrotron emission models mainly include the internal shock model (for a
matter-dominated fireball; \citealt{Rees1994}) and the ICMART model
(internal-collision-induced magnetic reconnection and turbulence, for a
Poynting flux-dominated outflow; \citealt{ZhangYan2011}). For the internal
shock model, since only the relative kinetic energy between different shells
can be released, the radiation efficiency is rather low ($\sim 10\%$; %
\citealt{Koba1997}). For the ICMART model, the radiation efficiency can be
much higher ($\sim 50\%$), and it reaches $\sim 80\%$ in the extreme case. However, the
extremely high $\epsilon _{\gamma }$ ($\epsilon _{\gamma }\gtrsim 80\%$) is
unlikely to be achieved, because the magnetic reconnection requires some
conditions to be triggered, so much magnetic energy is
left. For the photosphere emission model, if only the acceleration is in the
unsaturated regime ($R_{\text{ph}}<R_{s}$), the radiation efficiency can be
close to $100\%$.  Thus, in this work, we select
the GRBs with extremely high $\epsilon _{\gamma }$ ($\epsilon _{\gamma
}\gtrsim 80\%$; see Figure 3(a) and Table 1). Note
that the two bursts (GRB 990705 and GRB 000210) that are claimed to have extremely
high $\epsilon _{\gamma}$ in \citet{Lloy2004} are also included. We then
analyze the prompt\footnote{%
The \textit{Fermi}/GBM data are publicly available at %
\url{https://heasarc.gsfc.nasa.gov/W3Browse/fermi/fermigbrst.html}. The
Konus-Wind data are publicly available at %
\url{https://vizier.cds.unistra.fr/viz-bin/VizieR?-source=J/ApJ/850/161}.}
and afterglow\footnote{%
The X-ray afterglow data are publicly available at %
\url{https://www.swift.ac.uk/xrt_products/}. The optical afterglow data are
taken from \citet{Li2012,Li2018,Liang2013}.} properties of these 15 long
GRBs, to confirm the photosphere emission origin.

\subsection{Characteristics of Prompt Emission}

\begin{figure*}[th]
\label{Fig_12} \centering\includegraphics[angle=0,height=2.1in]{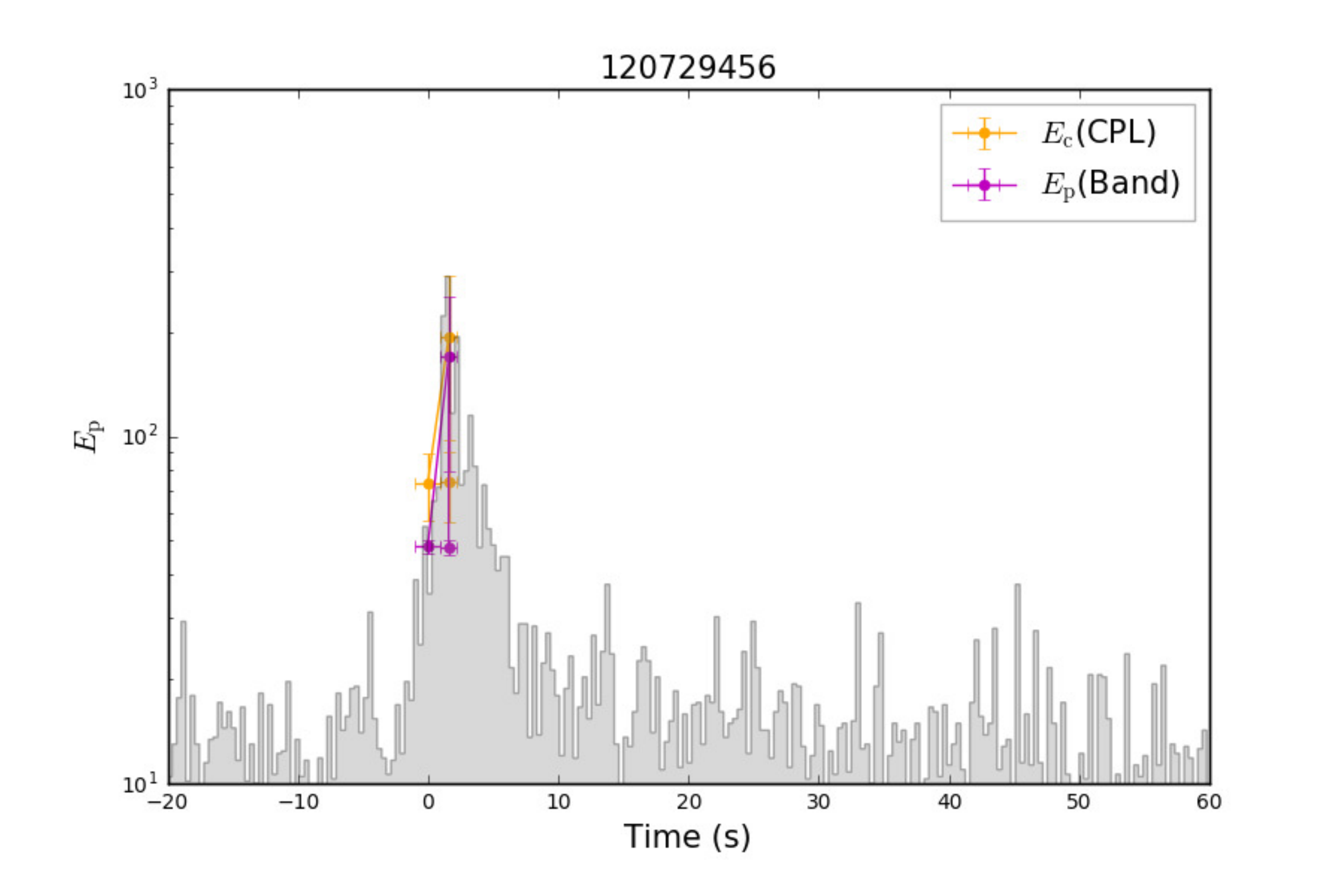} %
\centering\includegraphics[angle=0,height=2.1in]{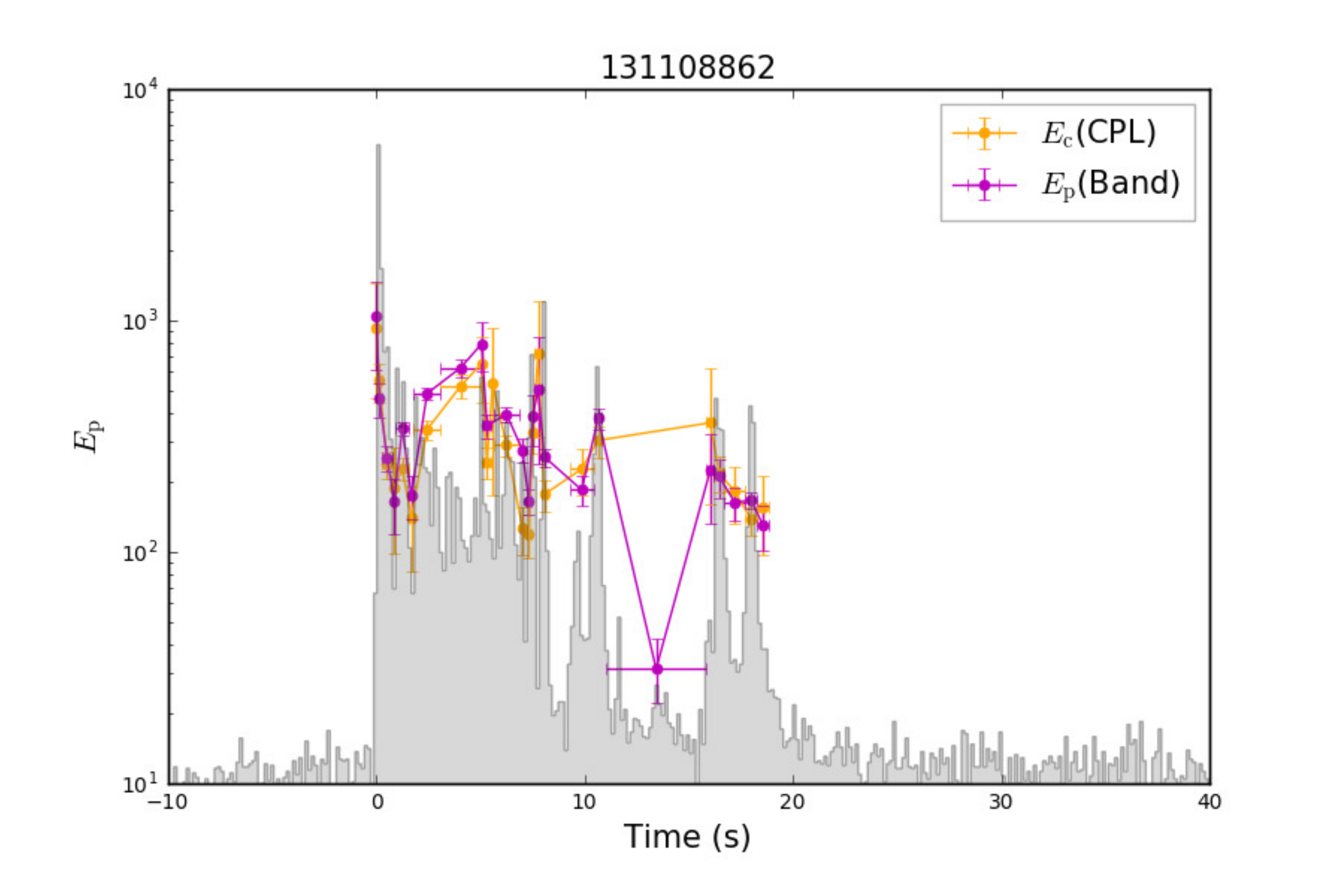} %
\includegraphics[angle=0,height=2.1in]{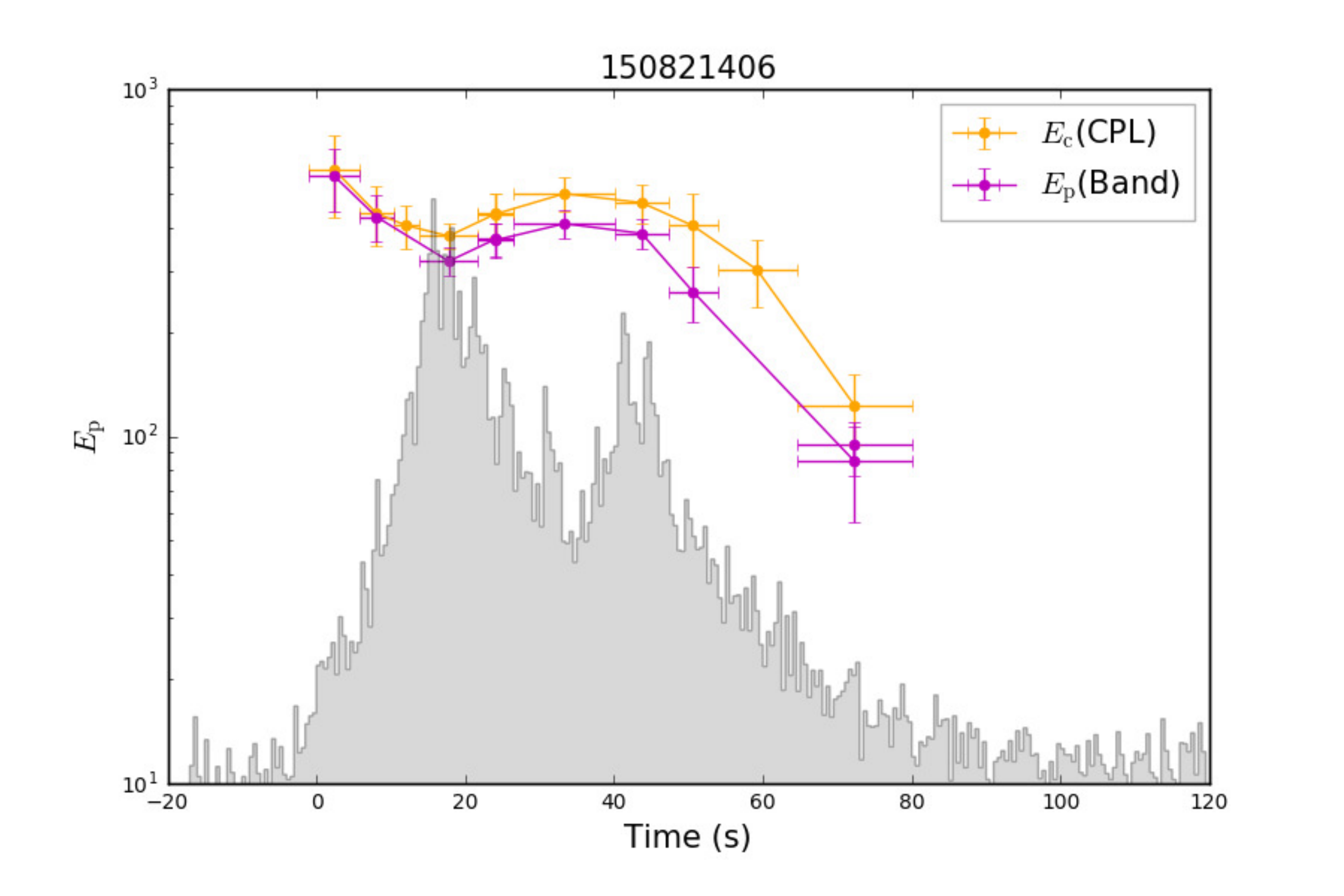} %
\includegraphics[angle=0,height=2.1in]{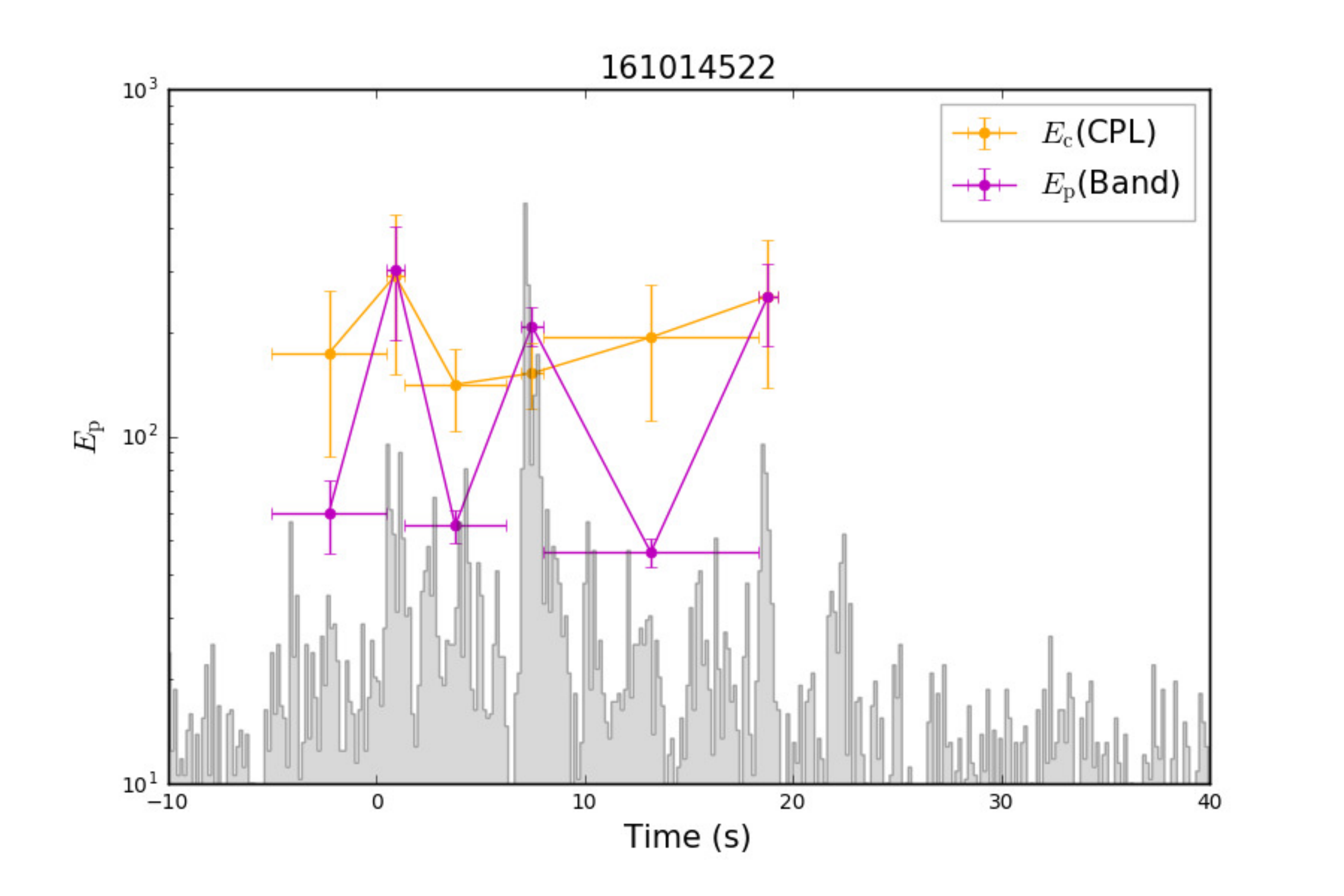} %
\includegraphics[angle=0,height=2.1in]{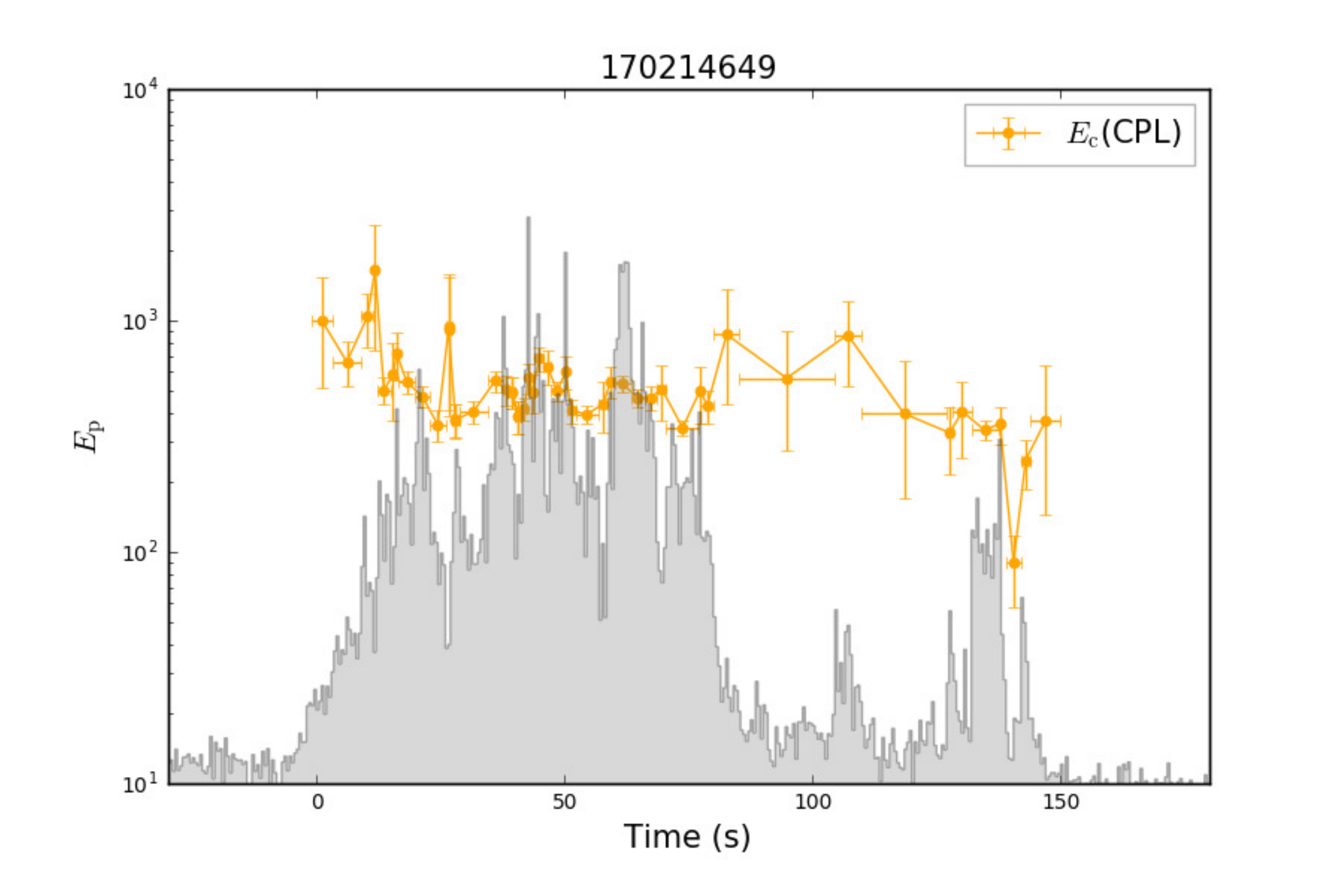}
\caption{The $E_{\text{p}}$ evolutions of the time-resolved spectra for 5
extremely high-efficiency GRBs ($\protect\epsilon _{\protect\gamma }\gtrsim 80\%$) detected by \textit{Fermi}/GBM. The $E_{\text{p}}$ evolutions almost follow the evolution of the flux $F$, consistent with the
predicted $E_{\text{p}}\propto F^{1/4}$ by the photosphere (thermal)
emission in the unsaturated acceleration regime.}
\end{figure*}

In Figure 3(b), we plot the $E_{\text{p}}$ and $E_{\text{iso}}$
distributions of the selected GRBs (omitting GRB 081203A and GRB 130606A,
because of the large $E_{\text{p}}$ error), and we find that they follow the
predicted $E_{\text{p}}\propto (E_{\text{iso}})^{1/4}$ relation (see Section 2.2.2) quite well.
The best-fit result is log ($E_{\text{p}}$) $=2.54+0.25\log $ ($E_{\text{iso}%
}$). In Figure 3(c), we compare the $E_{\text{p}}$ and $E_{\text{iso}}$
distributions of the selected GRBs with those of the large sample of long
GRBs, and find that the dispersion is quite small relative to that of the
large sample.

In Figure 3(d), we plot the $E_{\text{p}}$ and $L_{\text{iso}}$ distributions of the selected GRBs, and find that they also follow the $E_{\text{p}}\propto (L_{\text{iso}})^{1/4}$ relation well.
The best-fit result is log ($E_{\text{p}}$) $=2.75+0.23\log $ ($L_{\text{iso}%
}$). And the dispersion is found to be similar to that of $E_{\text{p}%
}\propto (E_{\text{iso}})^{1/4}$. Furthermore, based on the best-fit $E_{%
\text{p}}\sim 10^{2.75}\cdot (L_{\text{iso}})^{1/4}$ and $E_{\text{p}%
}=2.7kT_{0}=2.7k(L_{\text{iso}}/4\pi r_{0}^{2}ac)^{1/4}$, we obtain the
initial acceleration radius $r_{0}\sim 3.21\times 10^{8}$ cm, well
consistent with the quite high mean value $\left\langle r_{0}\right\rangle
\sim 10^{8.5} $ cm deduced in \citet{Pe2015}.

In Figure 12, we show the $E_{\text{p}}$ evolutions of the time-resolved
spectra for 5 GRBs detected by Fermi/GBM. The $E_{\text{p}}$ evolutions are
found to follow the evolution of the flux $F$ quite well (intensity tracking
pattern; \citealt{Lia1996}). This positive correlation is consistent with the abovementioned unsaturated acceleration condition of the photosphere emission. From Tables 1 and 4, we can see that the best-fit spectral model of the time-integrated spectra is the CPL model, or that the high-energy spectral index $\beta$ (using the BAND function to fit) is minimal. Thus, the photosphere emission model can better explain the high-energy spectra of these high-efficiency GRBs.

\subsection{Characteristics of Afterglows}

\begin{figure*}[th]
\label{Fig_13} \centering\includegraphics[angle=0,height=2.4in]{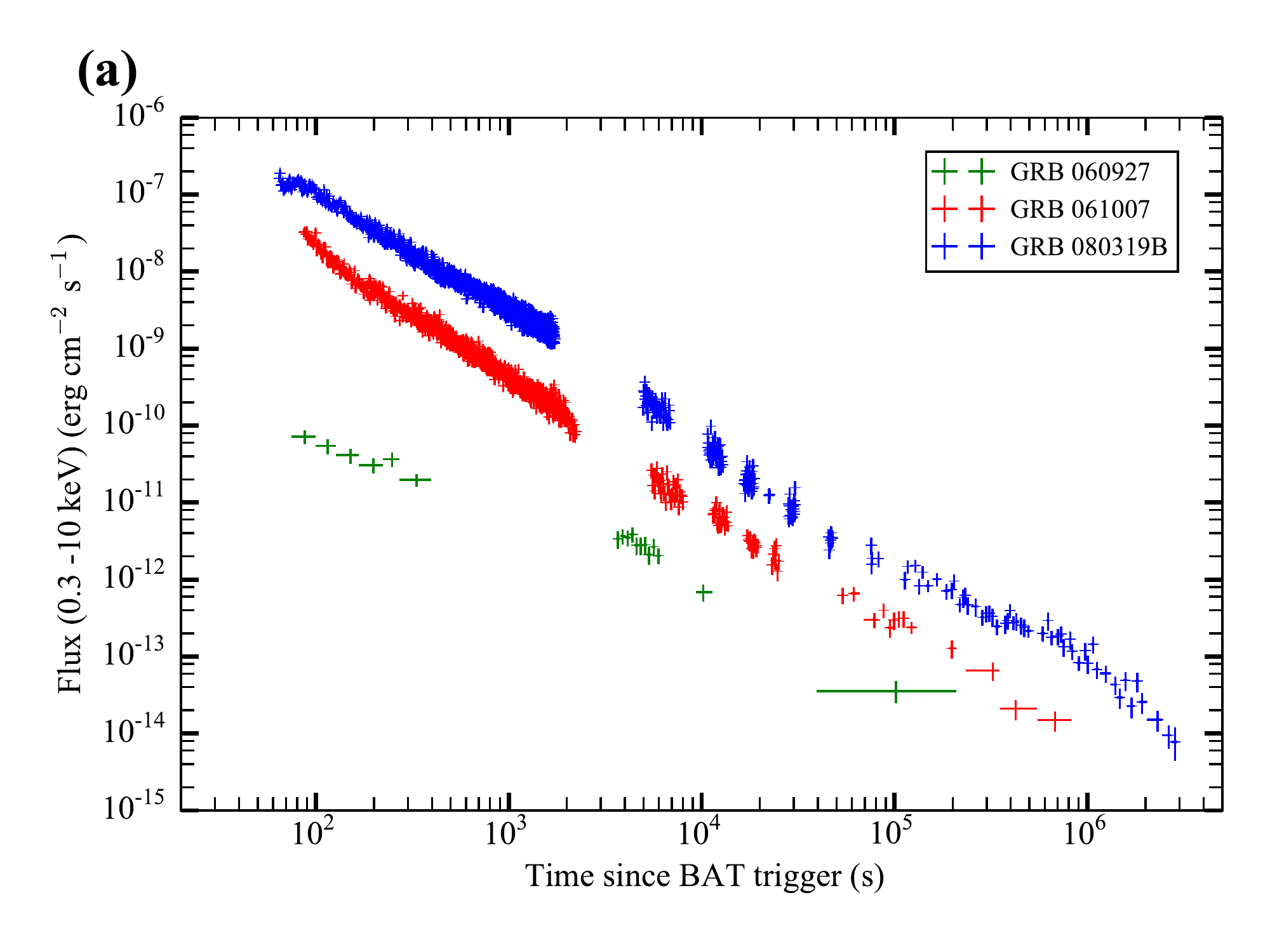} %
\centering\includegraphics[angle=0,height=2.4in]{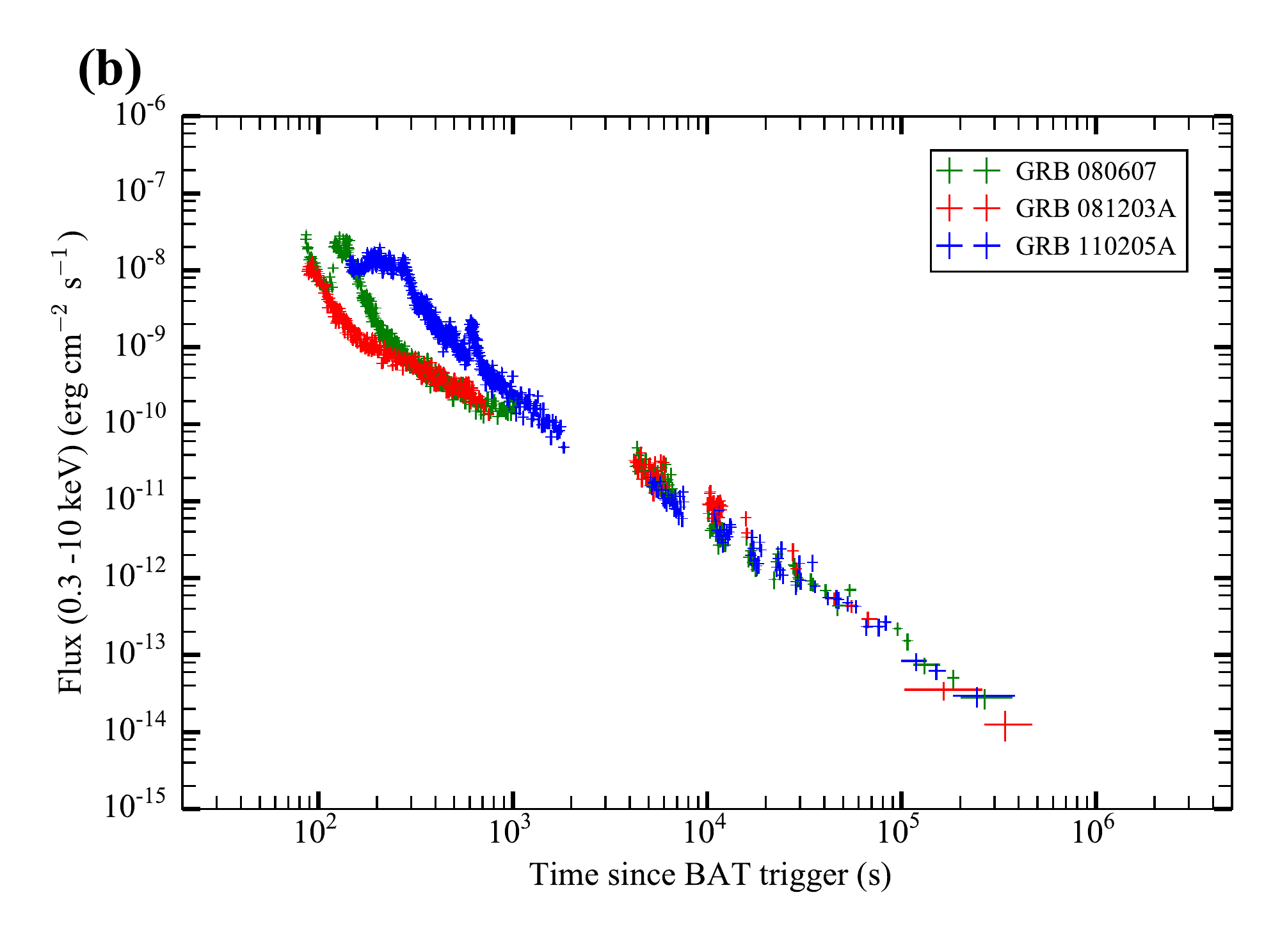} \centering%
\includegraphics[angle=0,height=2.4in]{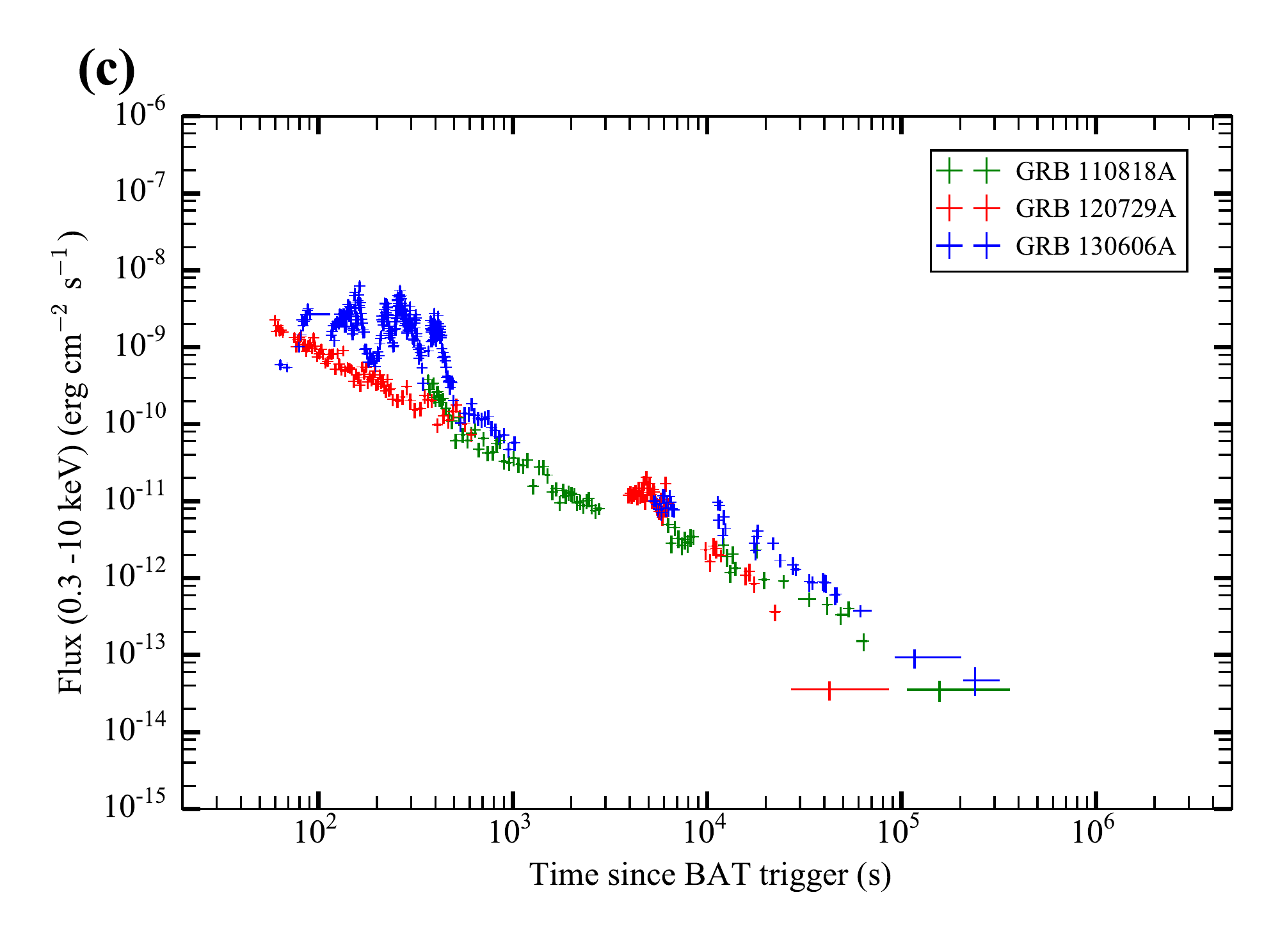} \centering%
\includegraphics[angle=0,height=2.4in]{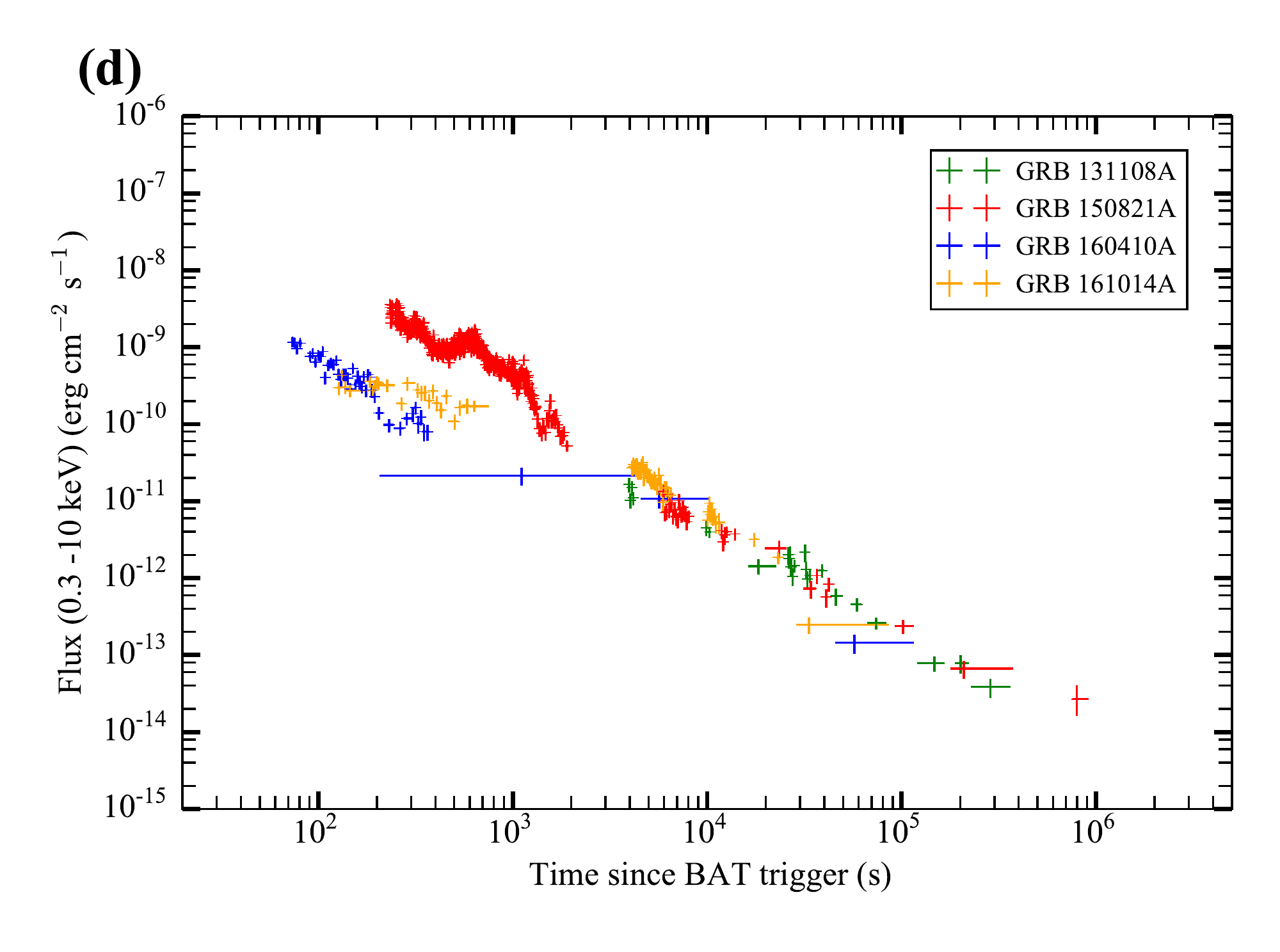}\ \ \ 
\caption{The X-ray afterglow light curves for the selected extremely high-efficiency
GRBs (except for GRB 990705 and GRB 000210). All these light curves appear
as a simple power-law shape, without any plateau, steep decay, or significant
flare (with weak flare in the early times).}
\end{figure*}

\begin{figure*}[th]
\label{Fig_14} \centering\includegraphics[angle=0,height=2.4in]{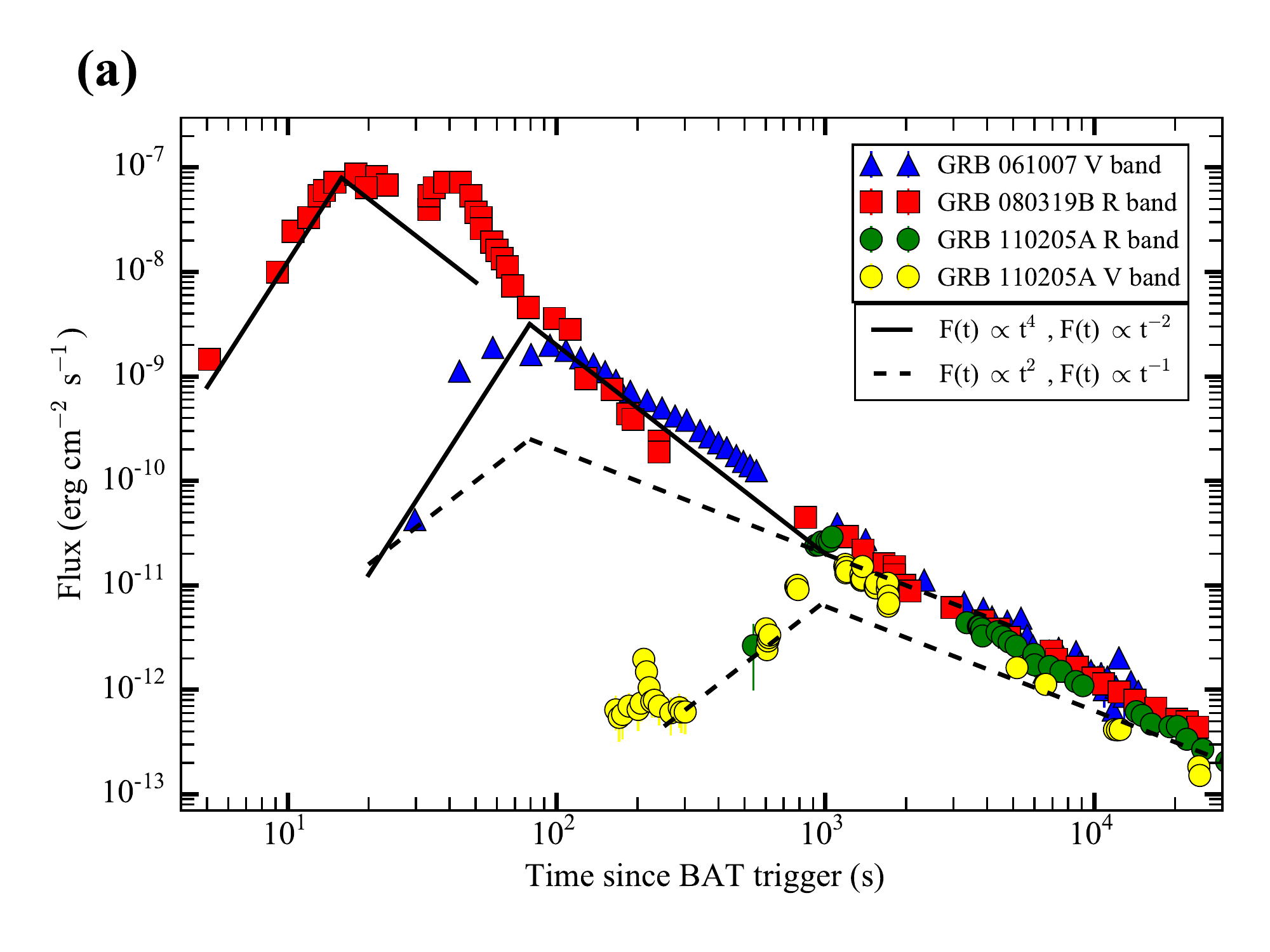} %
\centering\includegraphics[angle=0,height=2.4in]{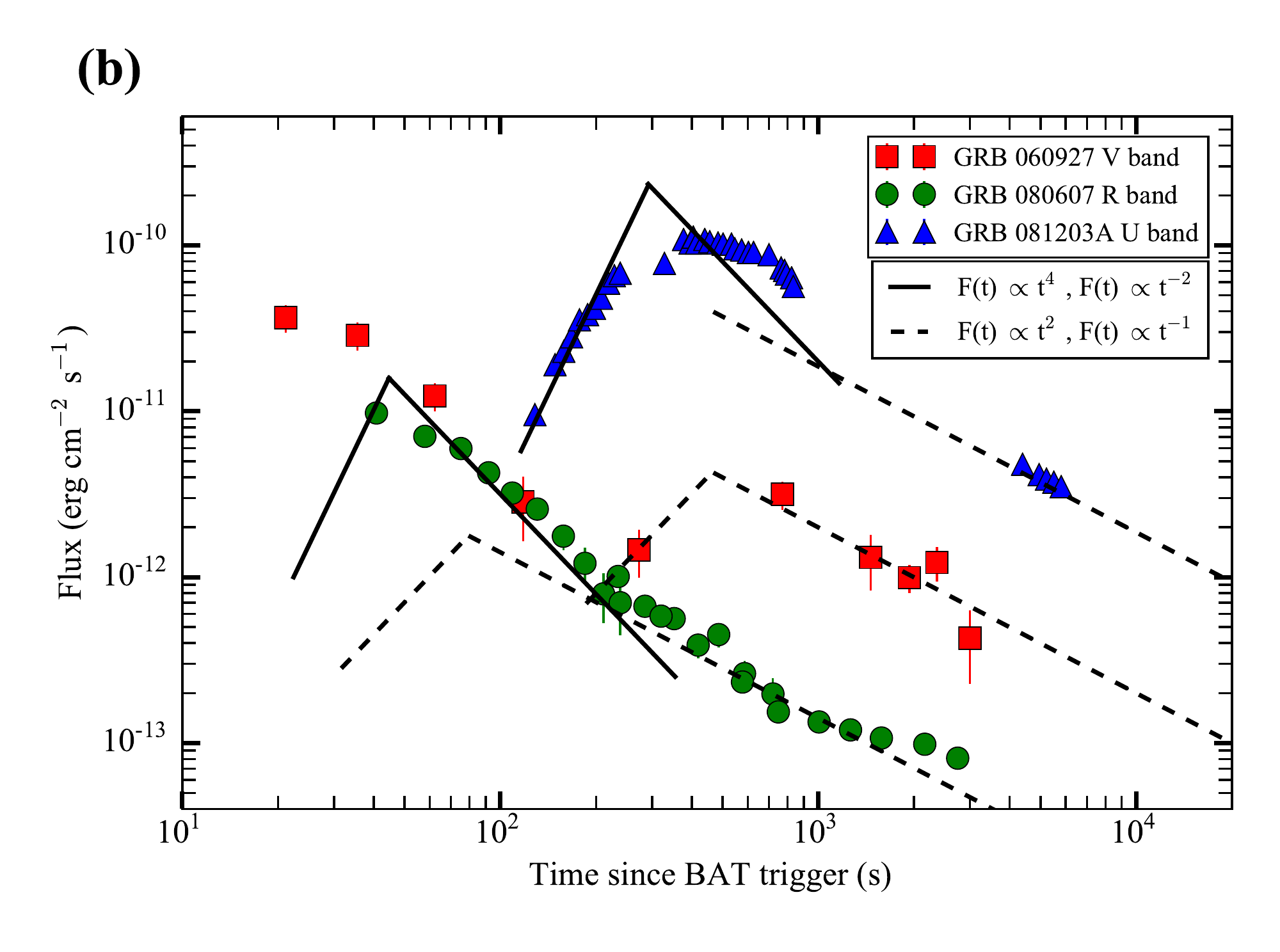} \ \ 
\caption{The optical afterglow light curves for 6 extremely high-efficiency GRBs ($\protect\epsilon _{\protect\gamma }\gtrsim 80\%$) whose early peaks can be detected. All these light curves show significant reverse
shock signals. (a) 3 GRBs considered also to have reverse shock signals
(with detections of both the rapid rise $f$ $\gtrsim $ $t_{{}}^{3}$ and fall 
$f$ $\sim $ $t_{{}}^{-2}$) in other works (\citealt{Gao2015,Yi2020}). (b) 3
GRBs with detection of only the rapid rise $f $ $\gtrsim $ $t_{{}}^{3}$ or
the rapid fall $f$ $\sim $ $t_{{}}^{-2}$.}
\end{figure*}

Figure 13 shows the X-ray afterglow light curves of the selected GRBs
(except for GRB 990705 and GRB 000210). We find that all the X-ray afterglow
light curves appear as simple power-law shapes\footnote{Interestingly, a similar X-ray afterglow characteristic has been found for GeV-/TeV-detected GRBs \citep{Yama2020}.}, without any plateau, steep
decay \citep{Zhang2006}, or significant flare (with weak flares in the early
times). In Figure 14, we show the optical afterglow light curves of 6 GRBs
whose early peaks can be detected. All the optical afterglow light
curves show significant reverse shock signals. The power-law shape of the
X-ray afterglow and the reverse shock in the optical afterglow are the basic
predictions \citep{Pac1993,Mes1997,Sari1999} of the classical hot fireball
model of GRBs (see Appendices B.2 and B.3). Thus, the jets of these
high-efficiency GRBs are likely to be thermal-dominated, and the radiation
mechanism of the prompt emission is unlikely to be the ICMART model (for
Poynting flux-dominated outflow; \citealt{ZhangYan2011}). Also, considering
the high efficiency ($\epsilon _{\gamma }\gtrsim 80\%$), the internal shock
model ($\epsilon _{\gamma }$ $\sim 10\%$; \citealt{Rees1994,Koba1997}) is
unlikely. The prompt emission of these GRBs is likely to be produced
by the photosphere emission, then.

In Figure 1(a), we show the correlation of $L_{\text{iso}}$ and $\Gamma $ for the selected GRBs. The $\Gamma $ is obtained by the tight $L_{\text{iso}%
}-E_{\text{p}}-\Gamma $ correlation (\citealt{Liang2015}; the values are taken from \citealt{Xue2019}). In Section 4.3, we find that this estimation is likely to be quite accurate for these
high-efficiency GRBs with $E_{\text{p}}\propto (L_{\text{iso}})^{1/4}$.
Though 4 GRBs have detections of the peak time of the optical afterglow, we
do not use them to estimate the $\Gamma$, because of the significant reverse
shock signals. We find that $\Gamma $ is tightly correlated with $L_{\text{%
iso}}$, $\Gamma \propto (L_{\text{iso}})^{0.29}$. This is well consistent
with the prediction of the neutrino annihilation from the hyperaccretion
disk \citep{Lv2012}, $\Gamma \propto (L_{\text{iso}})^{7/27}=(L_{\text{iso}%
})^{0.26}$. This therefore also supports the jets of these GRBs being
thermal-dominated.

In Figure 1(b), we show the correlation of $E_{\text{iso}}/E_{\text{k}}$ and 
$\eta /\Gamma $ (see Section 2.2.1) for the selected GRBs. According to Equation (\ref{de}), along with $r_{0}\sim 3.21\times 10^{8}$ cm, as derived above,
and $L_{\text{iso}}$, we can use $\Gamma $ to obtain the $\eta $ for each
burst. We find the obvious linear correlations for $E_{%
\text{iso}}/E_{\text{k}}$ and $\eta /\Gamma $, and they are almost the same
(aside from 3 bursts: note that we obtain 4 bursts that are almost the same when
we take $r_{0}=3.21\times 10^{8}$ cm, and then use the offset from the
best-fit $E_{\text{p}}\propto (L_{\text{iso}})^{1/4}$ relation to slightly
modify $r_{0}$ for the other bursts, to obtain 3 other bursts that are almost the
same) when we take $E_{\text{k,52}}=5\ast L_{\text{X,45}}$ ($E_{\text{k,52}%
}=E_{\text{k}}/10^{52}$, $L_{\text{X,45}}=L_{\text{X}}/10^{45}$; this is
quite close to the derivation of $E_{\text{k,52}}=3.7\ast L_{\text{X,45}}$
described in Section 4.1, and the slight difference is likely to result from
the slight error of $\Gamma $ estimated by the $L_{\text{iso}}-E_{\text{p}%
}-\Gamma $ correlation). Again, this result strongly supports the photosphere
emission origin in the unsaturated acceleration regime for these
high-efficiency GRBs.

\subsection{Discussion of the Probability Photosphere Model and the
Dissipative Photosphere Model}

According to the above statements, the prompt emission of the selected
high-efficiency GRBs is likely to be produced by the photosphere emission in
the unsaturated acceleration regime. But noteworthily, from Table 1, we can
see that the low-energy spectral index $\alpha $ is quite typical (around $%
-1 $), rather than very hard. This strongly supports that the photosphere
emission model having the capacity to produce the observed typical soft low-energy spectrum.
Theoretically, the probability photosphere model (with geometric broadening)
and the dissipative photosphere model (with subphotospheric energy
dissipation) can both achieve this. But for the dissipative photosphere
model, the $E_{\text{p}}\propto (E_{\text{iso}})^{1/4}$ relation should be
violated, since the inverse Compton scattering below the photosphere radius
will change the photon energy (namely $T_{\text{ob}}$ $\neq T_{0}$; see
Appendix A). Also, the high-energy spectrum for this model should be a power
law, rather than the exponential cutoff. So the characteristics of the
selected high-efficiency GRBs favor the probability photosphere model.

\label{sec:result}

\section{EVIDENCE FROM LONG GRBS WITH $\protect\epsilon %
_{\protect\gamma }\gtrsim 50\%$ AND $\protect\epsilon _{\protect\gamma %
}\lesssim 50\%$}

\label{sec:long}

\subsection{\protect\bigskip $\protect\epsilon _{\protect\gamma }=50\%$ and
Maximum $\Gamma $}

For the $\epsilon _{\gamma }=50\%$ ($R_{\text{ph}}=R_{s}$) case, with a
fixed $L_{\text{iso}}$ the observed Lorentz factor $\Gamma $ in the
afterglow phase should be the maximum, because of the following reason. To obtain $%
\epsilon _{\gamma }<50\%$ $(R_{\text{ph}}>R_{s}$, $R_{\text{ph}}\propto L_{%
\text{iso}}/\Gamma ^{3}$ and $R_{s}=\Gamma \cdot r_{0}$), $\Gamma $ ($\Gamma
=\eta $) should be smaller. Conversely, for $\epsilon _{\gamma }>50\%$, $%
\eta $ should be larger. And in this case, from Equation (\ref{de}), we
have $\Gamma $ $\propto (L_{\text{iso}}/\eta )^{1/3} $, thus $\Gamma $
should also be smaller. Note that this maximum $\Gamma $ exists for the hot
fireball, while the corresponding $\epsilon _{\gamma }=50\%$ is the
prediction of the photosphere emission origin for the prompt emission. The
maximum $\Gamma $ is given as (see also Equation 16 in \citealt{Ghirlan2018})

\begin{equation}
\Gamma _{\max }=\left[ \frac{L_{\text{iso}}\sigma _{T}}{8\pi m_{p}c^{3}r_{0}}%
\right] ^{1/4}.^{{}}  \label{e5}
\end{equation}

In Figure 11(a), we show the distribution of $L_{\text{iso}}$ and $\Gamma $
for the complete sample (62 bursts), with the detection of the peak time of the
early optical afterglow (\citealt{Ghirlan2018}; obtaining $\Gamma $).
Obviously, with the exception of GRB 080319B (with a strong reverse shock signal) and 4
bursts (peak time is obtained from the \textit{Fermi}/LAT light curve, and
the decay slope of $\sim$ $1.5$ implies that it is likely to be produced by
the radiative fireball and the $\Gamma $ should be smaller by a factor of $%
\sim$ $1.6$, see \citet{Ghise2010} and Appendix B.4), the distribution of the
maximum $\Gamma $ well follows the predicted $L_{\text{iso}}^{1/4}{}^{{}}$
correlation, and only has the difference of a constant $\sim$ $10^{0.1}$ ($%
1.27$) from the prediction of Equation (\ref{e5}) (the dashed line, $%
r_{0}\sim 3.21\times 10^{8}$ cm is used based on Figure 3). Note that though
the equation for calculating $\Gamma$ is confirmed to act as $\Gamma \propto
(E_{\text{k}}^{{}})^{1/8}\cdot \lbrack T_{p}^{{}}/(1+z)]^{-3/8}$, its
constant is highly uncertain (see Table 2 in \citet{Ghirlan2018}). The
constant that is given in other works (with different methods) can be 1.7 (or 0.5)
times that used in \citet{Ghirlan2018}. So the above difference ($1.27$)
obtained by our work is reasonable, and may be more accurate (since it does
not strongly depend on the model assumption; if $r_{0}$ is accurate, then it is
likely to be accurate).

Then, we select the sample (9 bursts) with the maximum $\Gamma $ (see Table
7) to check their efficiency properties. In Figure 11(b), we show the
distribution of $E_{\text{iso}}$ and $L_{\text{X,45}}$ for this sample (4
bursts with $L_{\text{X,45}}$ detection). Note that we exclude GRB 081007
due to the too small $E_{\text{iso}}$ and GRB 080310 due to the plateau in
the early optical afterglow. As expected from the photosphere emission model,
all these bursts have almost the same efficiency (with $E_{\text{iso}%
}\propto L_{\text{X}}\propto E_{\text{k}}$). Thus, we think that the
efficiency $\epsilon _{\gamma }$ for these bursts is likely to be $50\%$ ($%
E_{\text{iso,52}}=E_{\text{k,52}}\simeq 3.7\ast L_{\text{X,45}}$). Note
that, based on this, the average derived efficiency ($\epsilon _{\gamma }$ $%
\sim$ $33\%$ to $40\%$; see Figure 10) for the whole sample (117 bursts) is
almost consistent with that given in other works %
\citep{Lloy2004,Fan2006,ZhaB2007,Avan2012,Wygo2016}.

\subsection{$E_{\text{p}}-E_{\text{iso}}$ Distributions and Consistent
Efficiency}

Based on the derivation of $E_{\text{k,52}}=3.7\ast L_{\text{X,45}}$
described above, and separated by $E_{\text{iso,52}}=E_{\text{k,52}}=3.7\ast
L_{\text{X,45}}$ ($\epsilon _{\gamma }=50\%$) for the distribution of $E_{%
\text{iso,52}}$ and $L_{\text{X,45}}$ in Figure 11, we obtain two distinguished
long-GRB samples ($\epsilon _{\gamma }\lesssim 50\%$, see Table 2; and $%
\epsilon _{\gamma }\gtrsim 50\%$, see Table 3). Note that we exclude the
above high-efficiency sample ($\epsilon _{\gamma }\gtrsim 80\%$) and the
sample with $\epsilon _{\gamma }=50\%$.

In Figure 4(a), we show that the best-fit result for the $\epsilon _{\gamma
}\gtrsim 50\%$ sample is log ($E_{\text{p}}$) $=2.47+0.25\log $ ($E_{\text{%
iso}}$), which is consistent with the prediction of the photosphere emission
model. Also, this result is almost the same as that for the above
high-efficiency sample. The offset from the best-fit result is likely to be
caused by a distribution of $r_{0}$ (as for the constrained results in %
\citealt{Pe2015}). For the $\epsilon _{\gamma }\lesssim 50\%$ sample, $E_{%
\text{iso}}$ and $E_{\text{p}}$ should both decrease by the same factor of $%
(R_{\text{ph}}/R_{s})^{-2/3}$ (see Section 2.2.3), compared with the distribution of $\log $ ($%
E_{\text{p}}$) $=2.54+0.25\log $ ($E_{\text{iso}}$) for the above
high-efficiency sample. In Figure 4(a), we show that the upmost
distribution for the $\epsilon _{\gamma }\lesssim 50\%$ sample is well
around $\log $ ($E_{\text{p}}$) $=2.54+0.25\log $ ($E_{\text{iso}}$), and
that the best-fit result (much smaller) is log ($E_{\text{p}}$) $=2.31+0.26\log $
($E_{\text{iso}}$). The decreases of $E_{\text{iso}}$ and $E_{\text{p}}$ are
more obvious when we divide the $\epsilon_{\gamma }\lesssim 50\%$ sample
into two subsamples ($\epsilon_{\gamma }\lesssim 17\%$ and $%
\epsilon_{\gamma }\gtrsim 17\%$).

In Figure 2(a), we show the distributions of $E_{\text{ratio}}$ ($E_{\text{%
ratio}}=[(E_{\text{p}}/2.7k)^{4}\ast (4\pi r_{1}^{2}ac)^{{}}/E_{\text{iso}%
}]^{1/3}$) and $E_{\text{iso}}/E_{\text{k}}$ (implicitly, $E_{\text{k,52}%
}=3.7\ast L_{\text{X,45}}/((1+z)/2)$ is adopted) for the $\epsilon
_{\gamma}\lesssim 50\%$ sample. They are well centered around
the equal-value line and have a linear correlation. This is well
consistent with the prediction from the photosphere emission model (see Section 2.2.4). The
dispersion is likely to be caused by the estimation error for $E_{\text{k}}$
(we have excluded the bursts with large $E_{\text{p}}$ errors of $dE_{\text{p}%
}/E_{\text{p}}\geq 0.2$), since many X-ray afterglow light curves are quite
complex (with plateaus, steep decays, or significant flares). The method of
using $L_{\text{X,11h}}$ to estimate $E_{\text{k}}$ should only be
completely correct for X-ray afterglows with power-law shapes and slopes of 
$-1$. To check the origin of the dispersion, we select the bursts with
almost the same $E_{\text{ratio}}$ and $E_{\text{iso}}/E_{\text{k}}$ (see Table
5). As expected, we find that all these bursts (7 bursts) have a power-law X-ray
afterglow light curve with a slope of $\sim $ $-1$ (shown in Figures 2(b) and
(c)).

Also, according to Section 2.2.4, for the $\epsilon _{\gamma }\lesssim 50\%$ sample,
we should have $E_{\text{ratio}}=E_{\text{iso}}/E_{\text{k}}=(R_{\text{ph}%
}/R_{s})^{-2/3}$. To check this, we select the bursts with
detections of the peak time of the optical afterglow (using them to estimate the $%
\Gamma $ and thus $(R_{\text{ph}}/R_{s})^{-2/3}$). Note that since $E_{%
\text{ratio}}$ and $E_{\text{iso}}/E_{\text{k}}$ are the average results for
the whole duration, and $(R_{\text{ph}}/R_{s})^{-2/3}$ $\propto $ $L^{-2/3}$,
we use $L=$ $E_{\text{iso}}/T_{90}$ (rather than $L_{\text{iso}}$). In
Figure 2(d), we show the distributions of $E_{\text{ratio}}$, $E_{\text{iso}%
}/E_{\text{k}}$ and $(R_{\text{ph}}/R_{s})^{-2/3}$ for the selected sample
(6 bursts). Similar to the above, they are well centered around the
equal-value line and have a linear correlation. Also, 1 burst has almost the same values for these three quantities, and the other 5 bursts have
almost the same values for two quantities. So the predicted $E_{%
\text{ratio}}=E_{\text{iso}}/E_{\text{k}}=(R_{\text{ph}}/R_{s})^{-2/3}$ from
the photosphere emission model can be well reproduced.

For the $\epsilon _{\gamma }\gtrsim 50\%$ sample, similar to Figure 1(b)
(for the high-efficiency GRBs), we should have $E_{\text{iso}}/E_{\text{k}%
}=\eta /\Gamma $. To check this, we also select the bursts with detections of
the peak time of the optical afterglow (using them to estimate $\Gamma $ and
thus $\eta $). Note that since $\Gamma $ $\propto $ $L^{1/3}$ and the
derived $\Gamma $ is likely to correspond to $L_{\text{iso}}$ (the maximum $L$),
we use $L=$ $L_{\text{iso}}$ when calculating $\eta $. In Figure 2(e), we
show the distribution of $E_{\text{iso}}/E_{\text{k }}$and $\eta /\Gamma $
for the selected sample (7 bursts). As expected, they are well
centered around the equal-value line and have a linear correlation.

The analysis results obtained above are for the sample with $L_{\text{X,11h}%
} $ as derived in this work (for GRBs after GRB 110213A). For another sample
with $L_{\text{X,11h}}$ presented in \citet{Avan2012} (for GRBs before GRB
110213A), we perform a similar analysis and obtain similar results. For the $%
\epsilon _{\gamma }\gtrsim 50\%$ sub-sample, the $E_{\text{p}}$ and $E_{%
\text{iso}}$ distribution is also centered around log ($E_{\text{p}}$) $%
=2.47+0.25\log $ ($E_{\text{iso}}$) (see Figure 4(b)). For the $\epsilon
_{\gamma }\lesssim 50\%$ sub-sample, the decreases of $E_{\text{iso}}$ and $%
E_{\text{p}}$ are also obvious. Also, the distributions of $E_{\text{ratio}}$
and $E_{\text{iso}}/E_{\text{k}}$ are well centered around the equal-value
line, and have linear correlations (see Figure 6(a)). For the selected
bursts with almost same $E_{\text{ratio}}$ and $E_{\text{iso}}/E_{\text{k}}$
(see Table 5), all (7 bursts) show a power-law X-ray afterglow light curve
with a slope of $\sim$ $-1 $ (see Figures 6(b) and (c)).

\subsection{The Excellent Derived $\Gamma \propto E_{\text{iso}}^{1/8}E_{%
\text{p}}^{1/2}/(T_{90})^{1/4}$ Correlation}

The small burst number in Figure 2(d) is a result of
obtaining both the X-ray afterglow light curve and the detection of the peak
time of the early optical afterglow. To further check $E_{\text{ratio}%
}=E_{\text{iso}}/E_{\text{k}}=(R_{\text{ph}}/R_{s})^{-2/3}$ for the $%
\epsilon _{\gamma }\lesssim 50\%$ case, we then analyze the complete sample with
detections of the peak time of the early optical afterglow (obtaining $\Gamma 
$ and thus $(R_{\text{ph}}/R_{s})^{-2/3}$; \citealt{Ghirlan2018}; see Table
6). Note that the used constant is $1.27$ times that given in %
\citet{Ghirlan2018} (see Section 4.1). Though lacking of $E_{\text{iso}}/E_{%
\text{k}}$ for most bursts in the sample, considering the different
distributions of $E_{\text{p}}$ and $E_{\text{iso}} $ for $\epsilon _{\gamma
}\lesssim 50\% $ and $\epsilon _{\gamma }\gtrsim 50\% $, we can use the
judgment of ($E_{\text{ratio}}\lesssim $ $0.9$) to roughly select the $%
\epsilon _{\gamma }\lesssim 50\%$ sub-sample. Note that we do not use the
bursts without the $E_{\text{p}}$ value in \citet{Minaev2019} and \citet{Xue2019}
due to the large $E_{\text{p}}$ errors, and we move 4 bursts to the $\epsilon
_{\gamma }\gtrsim 50\%$ sub-sample based on their detections of $E_{\text{iso%
}}/E_{\text{k}}$. In Figure 7(a), we show the distribution of $E_{\text{%
ratio}}$ and $(R_{\text{ph}}/R_{s})^{-2/3}$ for the selected $\epsilon
_{\gamma }\lesssim 50\%$ sub-sample (24 bursts). This distribution is
roughly centered around the equal-value line, and has a linear
correlation. After modifying the $\Gamma $ or $E_{\text{p}}$ based on Figure
2(d) (using $E_{\text{ratio}}=E_{\text{iso}}/E_{\text{k}}=(R_{\text{ph}%
}/R_{s})^{-2/3}$) for the 6 bursts there, in Figure 7(b), we show the
distributions of $E_{\text{ratio}}$ and $(R_{\text{ph}}/R_{s})^{-2/3}$ for
the two sub-samples with smaller $E_{\text{p}}$ errors ($dE_{\text{p}}/E_{%
\text{p}}\leq 0.2$) and larger $E_{\text{p}}$ errors ($dE_{\text{p}}/E_{\text{%
p}}\geq 0.2$). Obviously, for the sub-sample with smaller $E_{\text{p}}$
errors, the values of $E_{\text{ratio}}$ and $(R_{\text{ph}}/R_{s})^{-2/3}$
are almost the same. Note that we decrease the $\Gamma $ of GRB 090926A by a factor of 1.6, since its peak time is obtained from the LAT light curve and
the decay slope of $\sim$ $1.5$ implies that it is likely to be produced by
the radiative fireball \citep{Ghise2010}. And, for the large offset in
Figure 7(b) (the upper offset), we check its (GRB 090618) optical afterglow
light curve, and find that the reverse shock signal is significant, thus
overestimating the $\Gamma $ and $(R_{\text{ph}}/R_{s})^{-2/3}$.

Based on $E_{\text{ratio}}=(R_{\text{ph}}/R_{s})^{-2/3} $, we derive $\Gamma
\propto E_{\text{iso}}^{1/8}E_{\text{p}}^{1/2}/(T_{90})^{1/4}$ (see Equation 
$(\ref{e})$ and Equation (\ref{ee}) in Section 2.2.5). In Figure 7(c), we show a
comparison of the $\Gamma $ obtained from the optical afterglow (for 47 bursts in \citealt{Ghirlan2018}) and the $\Gamma $ obtained from the prompt
emission (orange triangles for $\Gamma =10^{3.33}L_{\text{iso}}^{0.46}E_{%
\text{p}}^{-0.43}$ and blue circles for $\Gamma =17\cdot E_{\text{iso}%
}^{1/8}E_{\text{p}}^{1/2}/(T_{90})^{1/4}$). Obviously, using these two
correlations, we can give an approximate estimation for $\Gamma $, both.
Furthermore, the Equation (\ref{ee}) derived in our work from the
photosphere emission model can give a better estimation for $\Gamma $ (with
a smaller reduced $\chi ^{2}$).

According to Equation (\ref{ee}), two correlations of $T_{p}/(1+z)\propto
\lbrack E_{\text{p}}/(T_{90})^{1/2}]^{-4/3}$ (see Section 2.2.6) and $\Gamma
\propto $ $E_{\text{p}}^{{}}/(T_{90})^{1/4}$ (see Section 2.2.7) are
predicted, which will be tested in the following section.

\subsubsection{$T_{p}/(1+z)\propto \lbrack E_{\text{p}%
}/(T_{90})^{1/2}]^{-4/3}$ test}

In Figure 7(d) we show the distribution of $E_{\text{p}}/(T_{90})^{1/2}$
and $T_{p}^{{}}/(1+z)$ for 35 bursts (we delect 7 bursts with $dE_{\text{p}%
}/E_{\text{p}}\geq 0.2$ and 5 bursts with LAT light curves, and we modify the 
$E_{\text{p}}$ or $T_{p}^{{}}$ for 5 bursts based on Figure 2(d)). Just as
predicted, this distribution shows a linear correlation with a slope of $%
\sim-4/3$, and it is consistent with $T_{p}^{{}}/(1+z)\propto (E_{\text{p}%
})^{-1.25}$ presented in Table 1 of \citet{Ghirlan2018}. So Equation (%
\ref{ee}) is likely to be correct.

Besides, from Figure 7(d), we can see that the $T_{p}^{{}}$ for the
sub-sample of $\epsilon _{\gamma }\gtrsim 50\%$ is a bit larger than that
for $\epsilon _{\gamma }\lesssim 50\%$, though both satisfy the $%
T_{p}^{{}}/(1+z)\propto \lbrack E_{\text{p}}/(T_{90})^{1/2}]^{-4/3}$
correlation. This is well consistent with the slightly different predicted
constants of Equation (\ref{ee}) for the $\epsilon _{\gamma }\lesssim 50\%$
case (19.67, see Equation (\ref{abc})) and the $\epsilon _{\gamma }\gtrsim
50\%$ case (16.75, see Equation (\ref{abc2})).

The smaller constant ($16.75$) for the $\epsilon _{\gamma }\gtrsim 50\%$
case is consistent with the larger $T_{p}^{{}}$ in Figure 7(d). In Figure
7(e), we show the derived $\Gamma $ using the Equation (\ref{abc}) and
the Equation (\ref{abc2}), and find that the distribution for the $%
\epsilon _{\gamma }\lesssim 50\%$ case is better (more symmetric) than that
in Figure 7(c).

\subsubsection{$\Gamma =10^{-0.1}\cdot E_{\text{p}}^{{}}/(T_{90})^{1/4}$
correlation}

In Figure 9(a), we do find the tight correlation of 
\begin{equation}
\Gamma =10^{-0.1}\cdot E_{\text{p}}^{{}}/(T_{90})^{1/4}
\end{equation}%
for the $\epsilon _{\gamma }\gtrsim 50\%$ case (with $L_{\text{X,11h}}$
detection). Note that the $E_{\text{p}}^{{}}$ here is re-derived from the $%
\log $ ($E_{\text{p}}$) $=2.54+0.25\log $ ($E_{\text{iso}}$) correlation
(using the $E_{\text{iso}}$), since the observed $E_{\text{p}}^{{}}$ with an 
offset from the above line is likely to arise from the different $r_{0}^{{}}$
(actually $E_{\text{p}}\propto (E_{\text{iso}}/r_{0}^{2})^{1/4}$) or the
error of $E_{\text{p}}$. Besides, we modify the $\Gamma $ (using $E_{\text{%
iso}}/E_{\text{k}}$, mainly for 3 bursts) based on Figure 2(e). In Figure
9(a), we also show the distributions for $E_{\text{p}}-\Gamma $ and $E_{%
\text{p}}-\eta $, here $\eta $ is obtained by $\eta /\Gamma =E_{\text{iso}%
}/E_{\text{k}}$. It is obvious that there is a tight correlation of $E_{%
\text{p}}\propto \eta $ for the sub-sample (5 bursts) with higher
efficiency. This means that $\eta \propto (E_{\text{iso}})^{1/4}$, which is well
consistent with the prediction of the neutrino annihilation from the
hyperaccretion disk (for the hot fireball).

For the $\epsilon _{\gamma }\lesssim 50\%$ case, since $(E_{\text{p}}/E_{%
\text{ratio}})\propto (E_{\text{iso}}/E_{\text{ratio}})^{1/4}$ and $\Gamma
\propto (L_{\text{iso}}/E_{\text{ratio}})^{1/4}$ we should have $\Gamma
\propto $ $(E_{\text{p}}/E_{\text{ratio}})/(T_{90})^{1/4}$. From Figure
9(b) we do find this correlation, which is also in line with that for the $%
\epsilon _{\gamma }\gtrsim 50\%$ case. Note that we modify the $E_{\text{p}}$
or $\Gamma $ based on Figure 2(d). In Figure 9(c) we show the distribution
of $\Gamma -$ $E_{\text{p}}^{{}}/(T_{90})^{1/4}$ for the large $\Gamma $
sample (47 bursts) in \citet{Ghirlan2018}. Note that for the $\epsilon
_{\gamma }\gtrsim 50\%$ case, the $E_{\text{p}}^{{}}$ is re-derived from the $%
\log $ ($E_{\text{p}}$) $=2.54+0.25\log $ ($E_{\text{iso}}$) correlation
(using the $E_{\text{iso}}$), and for the $\epsilon _{\gamma }\lesssim 50\%$
case, the $E_{\text{p}}^{{}}$ is re-derived from $E_{\text{p}}/E_{\text{ratio}%
}$. For the $\epsilon _{\gamma }\gtrsim 50\%$ case, when calculating $\Gamma 
$, we modify the $E_{\text{k}}$ based on the original $\eta /\Gamma $.
Obviously, we find that the distribution of $\Gamma $ and $E_{\text{p}%
}^{{}}/(T_{90})^{1/4}$ is well centered around $\Gamma =10^{-0.1}\cdot E_{%
\text{p}}^{{}}/(T_{90})^{1/4}$ and shows a linear correlation. Note that the $\Gamma -$ $E_{\text{p}}$ correlation is also found in \citet{Ghirlan2012}.

\subsubsection{The consistency of our $\Gamma \propto E_{\text{iso}}^{1/8}E_{%
\text{p}}^{1/2}/(T_{90})^{1/4}$ correlation and the $\Gamma =10^{3.33}L_{%
\text{iso}}^{0.46}E_{\text{p}}^{-0.43}$ correlation}

For the better sample (35 bursts) in Figure 7(d), in Figure 8(a), we replot
the comparison of the $\Gamma $ obtained from the optical afterglow and the $%
\Gamma $ obtained from the prompt emission (orange triangles for $\Gamma
=10^{3.33}L_{\text{iso}}^{0.46}E_{\text{p}}^{-0.43}$ and blue circles for $%
\Gamma \propto E_{\text{iso}}^{1/8}E_{\text{p}}^{1/2}/(T_{90})^{1/4}$). For
the $\epsilon _{\gamma }\gtrsim 50\%$ sub-sample (17 bursts) and the $%
\epsilon _{\gamma }\lesssim 50\%$ sub-sample (18 bursts), we use the different derived
constants. Obviously, the Equation (\ref{ee}) derived in our
work gives a much better estimation of $\Gamma $ (with a much smaller reduced 
$\chi ^{2}$, compared with Figure 7(c)). Noteworthily, the $\Gamma
=10^{3.33}L_{\text{iso}}^{0.46}E_{\text{p}}^{-0.43}$ correlation obtained
from the statistical fitting is actually consistent with our $\Gamma \propto
E_{\text{iso}}^{1/8}E_{\text{p}}^{1/2}/(T_{90})^{1/4}$ correlation derived
from the photosphere emission model. Because, along with $E_{\text{p}}\propto
(E_{\text{iso}})^{1/4}$ (or $E_{\text{p}}\propto (L_{\text{iso}})^{1/4}$;
see Figures 3 and 4), they can be transferred to each other, as shown
in the following:

\begin{eqnarray}
\Gamma &\propto &E_{\text{iso}}^{1/8}E_{\text{p}}^{1/2}/(T_{90})^{1/4} 
\notag \\
&\propto &(E_{\text{iso}}/T_{90})^{1/8}\cdot E_{\text{p}%
}^{1/2}/(T_{90})^{1/8},\text{ with\ }E_{\text{p}}\propto (L_{\text{iso}%
})^{1/4}\text{\ }  \notag \\
&\propto &(L_{\text{iso}})^{1/8}\cdot (L_{\text{iso}}^{0.23}\cdot E_{\text{p}%
}^{-0.43})/(T_{90})^{1/8}  \notag \\
&\propto &L_{\text{iso}}^{0.36}\cdot E_{\text{p}}^{-0.43}/(T_{90})^{1/8},%
\text{ with }T_{90}\propto (L_{\text{iso}})^{-0.5}\text{ }  \notag \\
&\propto &L_{\text{iso}}^{0.36}\cdot E_{\text{p}}^{-0.43}/(L_{\text{iso}%
}^{-0.5})^{1/8}  \notag \\
&\propto &L_{\text{iso}}^{0.43}\cdot E_{\text{p}}^{-0.43}.
\end{eqnarray}%
Here, the adopted $T_{90}\propto (L_{\text{iso}})^{-0.5}$ correlation is
found from Figure 8(c) for the high-efficiency sub-sample ($\epsilon
_{\gamma }\gtrsim 80\%$).

In Figure 8(b), we compare the $\Gamma $ obtained from $\Gamma =10^{3.33}L_{%
\text{iso}}^{0.46}E_{\text{p}}^{-0.43}$ and the $\Gamma $ obtained from $%
\Gamma \propto E_{\text{iso}}^{1/8}E_{\text{p}}^{1/2}/(T_{90})^{1/4}$ for
the $\epsilon _{\gamma }\gtrsim 50\%$ sub-sample (17 bursts), the $\epsilon
_{\gamma }\lesssim 50\%$ sub-sample (18 bursts) and the high-efficiency
sub-sample ($\epsilon _{\gamma }\gtrsim 80\%$). Obviously, these two
estimations are well centered around the equal-value line and have linear
correlations. Furthermore, for the high-efficiency sub-sample, which has the
tightest $E_{\text{p}}\propto (E_{\text{iso}})^{1/4}$ correlation (with very
small dispersion; see Figure 3), these two estimations are almost identical.
For the $\epsilon _{\gamma }\gtrsim 50\%$ sub-sample and the $%
\epsilon _{\gamma }\lesssim 50\%$ sub-sample, which have larger
dispersions for the $E_{\text{p}}\propto (E_{\text{iso}})^{1/4}$ correlation,
the above two estimations show larger dispersion, also.

\subsection{The Distribution of the $E_{\text{iso}}/E_{\text{k}}$ for the
Whole Sample}

In Figure 10, we show the distribution of the $E_{\text{iso}}/E_{\text{k}}$
(indicating the efficiency $\epsilon _{\gamma }$) for the whole sample (117
bursts). The average value is around $\sim$ $10^{-0.2}$ to $10^{-0.3} $,
thus indicating an average efficiency of $\epsilon _{\gamma }$ $\sim$ $33\%$
to $40\%$. From Figure 10, we also find that the distribution seems to
consist of three Gaussian distributions. The high-efficiency peak ($\sim$ $%
10^{0.6}$) is almost consistent with the $\eta \propto 10^{0.35}\cdot E_{%
\text{p}}{}^{{}}$ and $\Gamma \propto 10^{-0.3}\cdot E_{\text{p}}$
correlations (namely, $E_{\text{iso}}/E_{\text{k}}=\eta /\Gamma $ $\sim$ $%
10^{0.65}$) in Figure 9(a). Theoretically, for the $\epsilon _{\gamma }>50\%
$ case, from Equation (\ref{de}) and $\eta \propto (L_{\text{iso}})^{1/4}$,
we should also have $\Gamma \propto $ $(L_{\text{iso}}/\eta )^{1/3}\propto $ 
$(L_{\text{iso}}/L_{\text{iso}}^{1/4}{}^{{}})^{1/3}\propto (L_{\text{iso}%
})^{1/4}$. So the high-efficiency peak is predicted to exist. For the $%
\epsilon _{\gamma }<50\% $ case, $E_{\text{iso}}/E_{\text{k}}=(R_{\text{ph}%
}/R_{s})^{-2/3}\propto (L_{\text{iso}}/\Gamma ^{4})^{-2/3}$. Thus, the
low-efficiency peak ($\sim$ $10^{-0.7}$) is the natural result of $\Gamma
\propto (L_{\text{iso}})^{1/4}$, where $\Gamma =\eta $.

\section{EVIDENCE FROM SHORT GRBS WITH $\protect\epsilon %
_{\protect\gamma }\gtrsim 50\%$ AND $\protect\epsilon _{\protect\gamma %
}\lesssim 50\%$.}

\label{sec:short}

For the short GRBs, similar to the long GRBs, we use the judgment of $E_{%
\text{iso,52}}=E_{\text{k,52}}=3.7\ast L_{\text{X,45}}$ ($\epsilon _{\gamma
}=50\%$) to obtain the $\epsilon _{\gamma }\gtrsim 50\%$ sample (4 bursts) and
the $\epsilon _{\gamma }\lesssim 50\%$ sample (8 bursts; see Table 8). 
In Figure 5(a), we show the $E_{\text{p}}$ and $E_{\text{iso}}$
distributions for these two distinguished samples. Obviously, the $\epsilon
_{\gamma }\gtrsim 50\%$ sample and the up-most distribution for the large
sample of short GRBs \citep{ZhangBB18b} (without $L_{\text{X,45}}$ detections
for most) do follow the $E_{\text{p}}\propto (E_{\text{iso}})^{1/4}$
correlation, well consistent with the prediction of the photosphere emission
model. Noteworthily, the distribution for GRB 170817A well fits the above
line (log ($E_{\text{p}}$) $=3.24+0.25\log $ ($E_{\text{iso}}$)), too. For
the $\epsilon _{\gamma }\lesssim 50\%$ sample, as predicted, the distribution
is below this line (since $E_{\text{p}}$ and $E_{\text{iso}}$ are smaller).
Similar to Figure 2(a) (to test whether the $E_{\text{p}}$ and $E_{\text{iso}%
}$ are both smaller by a factor of $E_{\text{iso}}/E_{\text{k}}=(R_{\text{ph}%
}/R_{s})^{-2/3}$), in Figure 5(b) we show the distribution of $E_{\text{ratio%
}} $ ($E_{\text{ratio}}=[(E_{\text{p}}/2.7k)^{4}\ast (4\pi
r_{1}^{2}ac)^{{}}/E_{\text{iso}}]^{1/3}$) and $E_{\text{iso}}/E_{\text{k}}$
for this $\epsilon _{\gamma }\lesssim 50\%$ sample. Note that we exclude 3
bursts with $E_{\text{iso}}\leq 10^{50}$ erg and that we have $%
r_{1}^{{}}=3.4\times 10^{7}$ cm here. Again as predicted, they are found to
be almost centered around the equal-value line, and have a linear
correlation.

Interestingly, we find that all the bursts of the $\epsilon _{\gamma
}\gtrsim 50\%$ sample (5 bursts, including GRB 170817A) have extended
emission, and the $E_{\text{p}}$ and $E_{\text{iso}}$ distribution in Figure
5(a) is for their main pulse. To further test this finding, in Figure 5(c)
we show the $E_{\text{p}}$ and $E_{\text{iso}}$ distribution of the main
pulse for 7 other bursts that have extended emission \citep{Minaev2019}. It
is found that, except for 3 bursts only detected by \textit{Swift} or HETE-2
(lacking the detections in the high-energy band), other 4 bursts do follow
the log ($E_{\text{p}}$) $=3.24+0.25\log $ ($E_{\text{iso}}$) correlation of
the $\epsilon _{\gamma }\gtrsim 50\%$ sample, supporting the above finding
again. The true $E_{\text{p}}$ values for these 3 outliers are likely to be
much larger. According to the above, the main pulse for the short GRBs with
extended emission is likely to be produced by the photosphere emission in
the unsaturated acceleration regime. Then, considering the smaller values of
both the $E_{\text{p}}$ and $E_{\text{iso }}$for their extended emission, we
think that the extended emission may be produced by the transition from the
unsaturated acceleration to the saturated acceleration ($E_{\text{p}}$
and $E_{\text{iso}}$ are both smaller by the same factor of $E_{\text{iso}%
}/E_{\text{k}}=(R_{\text{ph}}/R_{s})^{-2/3}$). To test this hypothesis, in
Figure 5(d), we show the comparison of the ratios of the $E_{\text{p}}$ and
the fluence of the main pulse and the extended emission for a large extended
emission sample (including the Swift/BAT bursts with redshift; \citealt{Gomp2020}%
; and the \textit{Fermi}/GBM bursts without redshift; \citealt{Lan2020}). As
predicted, these two ratios are found to be almost centered around the
equal-value line and they have linear correlations.

\section{Summary}

\label{sec:summa}

In this work, after obtaining the prompt emission efficiency of a large GRB
sample with redshift, we divide that GRB sample into three sub-samples ($%
\epsilon_{\gamma }\gtrsim 80\%$, $\epsilon _{\gamma }\gtrsim 50\%$, and $%
\epsilon _{\gamma }\lesssim 50\%$). Then, the well-known Amati relation \citep{Amati2002} is well explained by the photosphere emission model. Furthermore, for each
sub-sample, the X-ray and optical afterglow characteristics
are well consistent with the predictions of the photosphere emission model.
Ultimately, large amounts of convincing observational evidence for the
photosphere emission model are revealed for the first time.

\section*{Acknowledgements}

I thank the anonymous referee for the constructive
suggestions. I thank Bin-Bin Zhang and Liang Li for helpful discussions. Y.-Z.M. is
supported by the National Postdoctoral Program for Innovative Talents (grant
no. BX20200164). This work is supported by the National Key Research and
Development Programs of China (2018YFA0404204), the
National Natural Science Foundation of China (grant Nos.
11833003, U2038105, 12121003), the science research grants
from the China Manned Space Project with No.CMS-CSST-
2021-B11, and the Program for Innovative Talents, Entrepreneur
in Jiangsu. I also acknowledge the use of public data from the Fermi
Science Support Center, the Swift and the Konus-Wind.

\section*{Appendix A: the probability photosphere model and the dissipative photosphere model}

\section*{A.1. The probability photosphere model}

For the traditional photosphere model, the photosphere emission is all
emitted at the photospheric radius $R_{\text{ph}}$, where the optical depth
for a photon propagating towards the observer is equal to unity ($\tau =1$).
But, if only there is an electron at any position, the photon should have a
probability to be scattered there. For an expanding fireball, the photons
can be last scattered at any place in the fireball with a certain
probability. Thus, the traditional spherical shell photosphere is changed to
a probability photosphere, namely the probability photosphere model \citep{Pe2008}. Based
on careful theoretical derivation, the probability function $P(r,\Omega )$,
donating the probability for a photon to be last scattered at the radius $r$
and angular coordinate $\Omega $, can be given as (\citealt{Pe2008,Belo2011,Lund2013}) 
\begin{equation}
P(r,\Omega )=(1+\beta )D^{2}\times \frac{R_{\text{ph}}}{r^{2}}\exp \left( -%
\frac{R_{\text{ph}}}{r}\right) ,
\end{equation}%
where $\beta $ is the jet velocity and $D$ $=[\Gamma (1-\beta \cdot \cos
\theta )]^{-1}$ is the Doppler factor.

For the probability photosphere model, the observed photosphere spectrum is
the overlapping of a series of blackbodies with different temperatures, thus
its low-energy spectrum is broadened. After considering the jet with angular
structure \citep[e.g.,][]{Dai2001,Rossi2002,ZhMe2002}, the observed typical low-energy photon index $\alpha \sim -1.0$ \citep{Kan2006,ZhaBB2011},
spectral evolution and $E_{p}$ evolutions (hard-to-soft evolution or $E_{p}$%
-intensity tracking; \citealt{Lia1996,Lu2010,Lu2012}) can be reproduced \citep{Lund2013,Meng2019,Meng2021}.

\section*{A.2. The dissipative photosphere model}

The dissipative photosphere model (or the sub-photosphere model) considers
that there is an extra energy dissipation process in the area of moderate
optical depth ($1<\tau <10$; the sub-photosphere). Different dissipative
mechanisms have been proposed, such as shocks \citep{Ree2005}%
, magnetic reconnection \citep{Gian2007} and
proton--neutron nuclear collisions \citep{Vur2016,Belo2016}. Then,
relativistic electrons (with a higher temperature than that of the photons) are
generated that upscatter the thermal photons to obtain the non-thermal
(broadened) high-energy spectrum.

\section*{Appendix B: some theoretical descriptions for the afterglow}

\section*{B.1. The remaining kinetic energy in the afterglow phase ($E_{\text{k}%
}$) and the isotropic X-ray afterglow luminosity at 11 hours ($L_{\text{X,11h}}$).}

In the context of the standard afterglow model \citep{Pac1993,Mes1997}, since at a late
afterglow epoch (11 hours) the X-ray band is above the cooling frequency $\nu
_{c}$, the late-time X-ray afterglow luminosity ($L_{\text{X,11h}}$) only
sensitively depends on $E_{\text{k}}$ and $\epsilon _{e}$ (the electron
equipartition parameter). Furthermore, the fraction of energy in the
electrons ($\epsilon _{e}$) is quite centered around $0.1$, based on the
large-sample afterglow analysis. Thus, $E_{\text{k}}$ can be well estimated
by the $L_{\text{X,11h}}$ as following \citep{Lloy2004}:
\begin{eqnarray}
E_{\text{k}} &=&10^{52}\text{ ergs R}\left[ \frac{L_{\text{X,10h}}}{10^{46}%
\text{ ergs s}^{-1}}\right] ^{4/(2+p)}(\frac{1+z}{2})^{-1}  \notag \\
&&\times \epsilon _{e,-1}^{4(1-p)/(2+p)}\epsilon
_{B,-2}^{(2-p)/(2+p)}t_{10h}^{(3p-2)/(2+p)}\nu _{18}^{2(p-2)/(2+p)}  \notag
\\
&\simeq &10^{52}\text{ ergs R}\left[ \frac{1.1 L_{\text{X,11h}}}{10^{46}\text{
ergs s}^{-1}}\right] (\frac{1+z}{2})^{-1}   \notag \\
&&\times (\epsilon _{e,-1}^{{}})^{-1},\text{ (}p\simeq 2\text{)}  \notag \\
&&
\end{eqnarray}%
where $R$ = [$t$(10 h)/$t$(prompt)]$^{(17/16)\epsilon _{e}}$ $\sim$ $%
2.27$ is the radiative losses during the first 10 hours after the prompt
phase. Note that the derived $E_{\text{k}}$ is $9.2$ times larger in \citet{Fan2006},
since the $\nu _{m}$ (the characteristic frequency corresponding to the minimum
electron Lorentz factor) is about one and a half orders smaller. Previously,
it had been hard to judge which constant was better. Here, using the method ($%
\epsilon _{\gamma }=50\%$ and the maximum $\Gamma $; see Section 4.1) in this work, our result ($E_{\text{k,52}}\simeq 3.7\ast L_{\text{X,45}}$) is quite consistent with $E_{\text{k,52}}\simeq 2.5\ast L_{\text{X,45}}$ (\citealt{Fan2006}, without the inverse Compton effect).

The isotropic X-ray afterglow luminosity (in the 2-10 keV rest-frame
common energy band) at 11 hours (rest frame), $L_{\text{X,11h}}$, is computed
from the observed integral 0.3-10 keV unabsorbed fluxes at 11 hours ($F_{%
\text{X,11h}}$; estimated from the \textit{Swift}/XRT light curves) and the
measured spectral index $\Gamma _{1}$ (from the XRT spectra), along with
the luminosity distance $D_{L}$. The equation is as follows (\citealt{Avan2012}):
\begin{eqnarray}
L_{\text{X,11h}}(2-10\text{ keV}) &=&4\pi D_{L}^{2}\cdot F_{\text{X,11h}%
}(0.3-10\text{ keV})\cdot  \notag \\
&&\frac{(\frac{10}{1+z})^{2-\Gamma _{1}}\text{ }-(\frac{2}{1+z})^{2-\Gamma
_{1}}}{10^{2-\Gamma _{1}}\text{ }-\text{ }0.3^{2-\Gamma _{1}}}.
\end{eqnarray}%
Here, $F_{\text{X,11h}}$ is obtained by interpolating (or extrapolating)
the best-fit power law, for the XRT light curve within a selected time
range including (or close to) 11 hours, to the 11 hours.

\section*{B.2. The power-law shape of the X-ray afterglow predicted by the
classical hot fireball.}

The \textquotedblleft generic\textquotedblright\ afterglow model
(relativistic blastwave theory) for GRB predicts a power-law decaying
multi-wavelength afterglow \citep{Pac1993,Mes1997}, due to the self-similar nature of the
blastwave solution. The observed specific flux is
\begin{eqnarray}
f_{\nu } &\propto &E_{\text{k}}^{(p+2)/4}\epsilon _{e}^{p-1}\epsilon
_{B}^{(p-2)/4}t_{\text{obs}}^{-(3p-2)/4}\nu _{{}}^{-p/2}  \notag \\
&\simeq &\frac{(\epsilon _{e}^{{}}E_{\text{k}}^{{}})}{t_{\text{obs}}^{{}}}%
\nu _{{}}^{-1},(p\simeq 2).
\end{eqnarray}%
This power-law behavior ($f$ $\sim $ $t_{\text{obs}}^{-1}$) is well
consistent with the observations of the optical and radio afterglows. But
several surprising emission components (the steep decay $f$ $\sim $ $t_{%
\text{obs}}^{-3}$ phase, the plateau $f$ $\sim $ $t_{\text{obs}}^{-0.5}$
phase, and the flare) in the early X-ray afterglow are revealed by the \textit{Swift}
observations \citep{Zhang2006}, which are not predicted by the above standard (hot fireball)
model. These extra components imply that an extra energy injection (internal
or external) may exist, which can be magnetic-dominated.

\section*{B.3. The reverse shock in the optical afterglow predicted by the
classical hot fireball.}

For the classical hot fireball (the magnetic field in the ejecta is
dynamically unimportant, namely the magnetization parameter $\sigma \equiv
B^{\prime ^{2}}/(4\pi n_{p}^{\prime }m_{p}c^{2})\ll 1$), a strong reverse
shock (propagating back across the GRB ejecta to decelerate it) is predicted
in the early optical afterglow phase \citep{Mes1997,Sari1999}. This prediction is almost
confirmed by the discovery of a very bright optical flash in GRB 990123
while the GRB is still active. Later on, many more
reverse shock signals are found. The light curve of the reverse shock
declines more rapidly ($f$ $\sim $ $t_{{}}^{-2}$) than that of the forward shock ($f$
$\sim $ $t_{{}}^{-1}$), and rises more rapidly ($f$ $\gtrsim $ $t_{{}}^{3}$)
than that of the forward shock ($f$ $\sim $ $t_{{}}^{2}$) before the peak time.

\section*{B.4. The \textit{Fermi}/LAT (GeV emission) light curve from the radiative fireball.}

For the generic\ afterglow model, the total energy of the fireball remains
constant (the adiabatic case) after the forward shock starts to decelerate
(entering the self-similar phase). However, there could be another case that
the total energy of the fireball decreases (the radiative case), since a
large fraction of the dissipated energy is radiated away (by magnetic reconnection or electron-proton collisions) \citep{Ghise2010}. For this radiative fireball, the light curve (after the peak time)
declines more rapidly ($f$ $\sim $ $t_{{}}^{-10/7}$), $\Gamma $ $\propto $ $%
t_{{}}^{-3/7}$ ($\Gamma $ $\propto $ $t_{{}}^{-3/8}$ for the adiabatic case), and
the peak time is much earlier ($t_{\text{peak}}=$ 0.44$t_{\text{dec}}$, $t_{%
\text{peak}}=$ 0.63$t_{\text{dec}}$ for the adiabatic case; $t_{\text{dec}}$ is
the deceleration time).

%%%%%%%%%%%%%%%%%%%%%%%%%%%%%%%%%%Table 2%%%%%%%%%%%%%%%%%%%%%%%%%%%%%%%

\begin{table*}[tbp]
\caption{The observed quantities and inferred efficiency for the $\protect%
\epsilon _{\protect\gamma }\lesssim 50\%$ sub-sample after GRB 110213A.}
\label{Tab_2}
\begin{center}
%\begin{tiny}%
\par
\renewcommand
\arraystretch{0.48} 
\resizebox{\linewidth}{!}{
\begin{tabular}{cccccccccc}
\hline
\multicolumn{10}{c}{$\epsilon _{\gamma }\lesssim 50\%$ sub-sample (62 bursts)
} \\ \hline
GRB & z & $E_{p,z}$ & $E_{\text{iso}}$ & $L_{\text{X,11}}$ & $E_{\text{iso}}/[\frac{3.7\ast L_{\text{X,11}}}{(1+z)/2}]$ & $E_{\text{ratio}}$ & $(R_{\text{ph}}/R_{s})^{-2/3}$ & $\Gamma $ & $T_{90,i}$ \\ 
&  & (keV) & (10$^{52}$ erg) & (10$^{45}$ erg s$^{-1}$) &  &  &  &  & s \\ 
\hline
110213A & 1.46 & 183.83 $\pm $ 32.15 & 6.9 & 3.22 & 0.713 & 0.225 & 0.283 & 
157 & 11.9 \\ 
110213B & 1.083 & 256 $\pm $ 40 & 7.04 & 3.66 & 0.541 & 0.348 &  &  &  \\ 
110715A & 0.82 & 216.58 $\pm $ 12.74 & 4.97 & 9.99 & 0.122 & 0.313 &  &  & 
\\ 
111209A & 0.677 & 520 $\pm $ 89 & 5.2 & 2.95 & 0.399 & 0.991 &  &  &  \\ 
111228A & 0.7156 & 74 $\pm $ 53 & 1.65 & 3.24 & 0.118 & 0.056 &  &  &  \\ 
120326A & 1.798 & 115 $\pm $ 19 & 3.82 & 15.6 & 0.093 & 0.147 &  &  &  \\ 
120804A & 1.3 & 283 $\pm $ 62 & 0.657 & 1.09 & 0.187 & 0.877 &  &  &  \\ 
120811C & 2.671 & 203.98 $\pm $ 19.55 & 8.81 & 7.25 & 0.603 & 0.239 &  &  & 
\\ 
120907A & 0.97 & 241.16 $\pm $ 67.27 & 0.18 & 1.02 & 0.047 & 1.09 &  &  & 
\\ 
121211A & 1.023 & 202.76 $\pm $ 32.05 & 0.14 & 2.54 & 0.015 & 0.942 &  &  & 
\\ 
130420A & 1.297 & 131.59 $\pm $ 7.2 & 6.29 & 2.99 & 0.653 & 0.149 & 0.644 & 
149 & 45.7 \\ 
130612A & 2.006 & 186 $\pm $ 32 & 0.719 & 0.32 & 0.913 & 0.487 & 0.653 & 193
& 1.9 \\ 
130701A & 1.16 & 192.24 $\pm $ 8.64 & 2.62 & 1.32 & 0.580 & 0.33 &  &  &  \\ 
130702A & 0.145 & 17.2 $\pm $ 5.7 & 0.064 & 1.01 & 0.010 & 0.028 &  &  &  \\ 
130831A & 0.48 & 81.4 $\pm $ 13.32 & 0.805 & 1.02 & 0.157 & 0.156 & 0.154 & 
73 & 11.9 \\ 
130925A & 0.347 & 110.94 $\pm $ 3.1 & 15. & 14.6 & 0.187 & 0.089 &  &  &  \\ 
131117A & 4.042 & 222 $\pm $ 37 & 1.03 & 1.43 & 0.491 & 0.546 &  &  &  \\ 
131231A & 0.6439 & 292.52 $\pm $ 6.06 & 21.1 & 13. & 0.361 & 0.288 & 0.142 & 
145 & 17.7 \\ 
140213A & 1.2076 & 191.24 $\pm $ 7.85 & 8.88 & 13.4 & 0.197 & 0.218 &  &  & 
\\ 
140226A & 1.98 & 1234 $\pm $ 235 & 5.68 & 1.71 & 1.340 & 3.05 &  &  &  \\ 
140304A & 5.283 & 775.06 $\pm $ 173.37 & 10.3 & 5.19 & 1.690 & 1.34 &  &  & 
\\ 
140506A & 0.889 & 373.19 $\pm $ 61.49 & 1.4 & 5.88 & 0.060 & 0.986 &  &  & 
\\ 
140508A & 1.027 & 533.46 $\pm $ 28.44 & 22.5 & 14.4 & 0.428 & 0.629 &  &  & 
\\ 
140512A & 0.72 & 826 $\pm $ 201.24 & 7.25 & 6.98 & 0.241 & 2.64 &  &  &  \\ 
140515A & 6.32 & 376 $\pm $ 108 & 5.38 & 3.15 & 1.690 & 0.636 &  &  &  \\ 
140606B & 0.384 & 352 $\pm $ 46 & 0.25 & 1. & 0.047 & 1.62 &  &  &  \\ 
140620A & 2.04 & 230.19 $\pm $ 33.87 & 7.28 & 11.4 & 0.263 & 0.299 &  &  & 
\\ 
140801A & 1.32 & 250.6 $\pm $ 7. & 5.55 & 1.87 & 0.930 & 0.366 &  &  &  \\ 
140907A & 1.21 & 313 $\pm $ 21 & 2.71 & 3.58 & 0.226 & 0.626 &  &  &  \\ 
141109A & 2.993 & 763 $\pm $ 303 & 33.1 & 10. & 1.780 & 0.891 &  &  &  \\ 
141221A & 1.452 & 450.62 $\pm $ 87.15 & 2.46 & 1.11 & 0.734 & 1.05 & 0.965 & 
202 & 9.7 \\ 
141225A & 0.915 & 342.71 $\pm $ 52.13 & 0.859 & 0.494 & 0.450 & 1.04 &  &  & 
\\ 
150206A & 2.09 & 704.52 $\pm $ 71.07 & 61.9 & 21.1 & 1.220 & 0.651 &  &  & 
\\ 
150301B & 1.5169 & 460.51 $\pm $ 90.95 & 1.99 & 0.567 & 1.190 & 1.16 &  &  & 
\\ 
150323A & 0.59 & 151.05 $\pm $ 14.31 & 1.26 & 0.427 & 0.632 & 0.306 &  &  & 
\\ 
150403A & 2.06 & 1311.74 $\pm $ 53.09 & 116. & 43.9 & 1.100 & 1.21 &  &  & 
\\ 
150514A & 0.807 & 116.72 $\pm $ 10.19 & 0.878 & 1.26 & 0.171 & 0.245 &  &  & 
\\ 
150727A & 0.313 & 195.05 $\pm $ 25.18 & 0.2 & 0.062 & 0.575 & 0.794 &  &  & 
\\ 
150818A & 0.282 & 128 $\pm $ 37 & 0.1 & 0.09 & 0.193 & 0.571 &  &  &  \\ 
151027A & 0.81 & 366.76 $\pm $ 61.78 & 3.3 & 6.17 & 0.131 & 0.724 &  &  & 
\\ 
151029A & 1.423 & 82 $\pm $ 17 & 0.288 & 0.232 & 0.407 & 0.221 &  &  &  \\ 
160227A & 2.38 & 222 $\pm $ 55 & 5.56 & 14.6 & 0.174 & 0.311 &  &  &  \\ 
160509A & 1.17 & 770.74 $\pm $ 20.82 & 113. & 37.8 & 0.877 & 0.6 &  &  &  \\ 
160623A & 0.37 & 756.24 $\pm $ 19.18 & 25.3 & 14.8 & 0.317 & 0.963 &  &  & 
\\ 
160804A & 0.736 & 123.93 $\pm $ 7.25 & 2.7 & 1.16 & 0.543 & 0.182 &  &  & 
\\ 
161017A & 2.013 & 718.83 $\pm $ 122.83 & 8.3 & 4.35 & 0.777 & 1.31 &  &  & 
\\ 
161117A & 1.549 & 205.62 $\pm $ 7.76 & 13. & 7.04 & 0.637 & 0.212 &  &  & 
\\ 
161219B & 0.1475 & 71. $\pm $ 19.3 & 0.012 & 0.778 & 0.002 & 0.533 &  &  & 
\\ 
170113A & 1.968 & 333.92 $\pm $ 174.49 & 0.924 & 6. & 0.061 & 0.976 &  &  & 
\\ 
170604A & 1.329 & 512 $\pm $ 168 & 4.7 & 4.61 & 0.321 & 1. &  &  &  \\ 
170607A & 0.557 & 174.06 $\pm $ 14.06 & 0.915 & 3.1 & 0.062 & 0.411 &  &  & 
\\ 
170705A & 2.01 & 294.61 $\pm $ 23.01 & 18. & 25.3 & 0.289 & 0.307 &  &  & 
\\ 
170903A & 0.886 & 180 $\pm $ 25 & 0.865 & 3.02 & 0.073 & 0.438 &  &  &  \\ 
171205A & 0.0368 & 125$^{+141}_{-37}$ & 0.002 & 0.003 & 0.116 & 1.98 &  &  &  \\ 
171222A & 2.409 & 694 $\pm $ 12 & 8.94 & 5.83 & 0.706 & 1.22 &  &  &  \\ 
180205A & 1.409 & 205 $\pm $ 34 & 0.972 & 1.63 & 0.194 & 0.501 &  &  &  \\ 
180620B & 1.1175 & 372 $\pm $ 105 & 3.04 & 12.4 & 0.070 & 0.758 &  &  &  \\ 
180720B & 0.654 & 1052 $\pm $ 26 & 34. & 32. & 0.237 & 1.36 &  &  &  \\ 
180728A & 0.117 & 108 $\pm $ 8 & 0.233 & 1.11 & 0.032 & 0.343 &  &  &  \\ 
181201A & 0.45 & 220 $\pm $ 9 & 10. & 11.7 & 0.168 & 0.253 &  &  &  \\ 
190106A & 1.859 & 489 $\pm $ 257 & 9.96 & 14.1 & 0.272 & 0.735 &  &  &  \\ 
190114C & 0.4245 & 929.3 $\pm $ 9.4 & 27. & 9.3 & 0.560 & 1.24 &  &  &  \\ 
\hline
\end{tabular}
}
\end{center}
\end{table*}

%
%%%%%%%%%%%%%%%%%%%%%%%%%%%%%%%%%%%%%%%%%%%%%%%%%%%%%%%%%%%%%%%%%%%%%%%%%%%%%%

%%%%%%%%%%%%%%%%%%%%%%%%%%%%%%%%%%Table 3%%%%%%%%%%%%%%%%%%%%%%%%%%%%%%%

\begin{table*}[tbp]
\caption{The observed quantities and inferred efficiency for the $\protect%
\epsilon _{\protect\gamma }\gtrsim 50\%$ sub-sample after GRB 110213A.}
\label{Tab_3}
\begin{center}
%\begin{tiny}%
\par
\renewcommand\arraystretch{0.8} 
\resizebox{\linewidth}{!}{
\begin{tabular}{ccccccccc}
\hline
\multicolumn{9}{c}{$\epsilon _{\gamma }\gtrsim 50\%$ sub-sample (40 bursts)}
\\ \hline
GRB & z & $E_{p,z}$ & $E_{\text{iso}}$ & $L_{\text{X,11}}$ & $E_{\text{iso}}/[\frac{3.7\ast L_{\text{X,11}}}{(1+z)/2}]$ & $\eta /\Gamma $ & $\Gamma $ & 
$L_{\text{iso}}$ \\ 
&  & (keV) & (10$^{52}$ erg) & (10$^{45}$ erg s$^{-1}$) &  &  &  & (10$^{52}$
erg s$^{-1}$) \\ \hline
110422A & 1.77 & 429.35 $\pm $ 8.31 & 74.700 & 5.450 & 5.140 &  &  &  \\ 
110503A & 1.61 & 574.2 $\pm $ 31.32 & 21.300 & 4.390 & 1.710 &  &  &  \\ 
110731A & 2.83 & 1223 $\pm $ 75.4 & 31.500 & 4.650 & 3.510 & 1.722 & 500 & 
20.52 \\ 
110918A & 0.98 & 665.28 $\pm $ 79.2 & 271.000 & 39.100 & 1.850 &  &  &  \\ 
111008A & 5. & 624 $\pm $ 186 & 41.400 & 9.390 & 3.580 &  &  &  \\ 
120119A & 1.728 & 499.91 $\pm $ 21.71 & 40.200 & 4.100 & 3.610 &  &  &  \\ 
120711A & 1.405 & 2552 $\pm $ 91 & 204.000 & 39.800 & 1.660 & 7.587 & 258 & 
14.37 \\ 
120712A & 4.1745 & 642 $\pm $ 134.5 & 15.200 & 2.310 & 4.600 &  &  &  \\ 
120909A & 3.93 & 961.41 $\pm $ 125.42 & 69.000 & 7.570 & 6.070 & 5.060 & 288
& 14.88 \\ 
121128A & 2.2 & 244.19 $\pm $ 9.61 & 10.100 & 1.000 & 4.340 & 3.891 & 332 & 
6.64 \\ 
130408A & 3.76 & 1289.96 $\pm $ 204.68 & 32.400 & 5.570 & 3.740 &  &  &  \\ 
130427A & 0.34 & 1105.4 $\pm $ 7.3 & 89.000 & 17.000 & 0.947 & 1.050 & 471 & 
11.86 \\ 
130505A & 2.27 & 1939.11 $\pm $ 85.02 & 438.000 & 29.400 & 6.590 &  &  &  \\ 
130514A & 3.6 & 506 $\pm $ 193 & 49.500 & 5.730 & 5.370 &  &  &  \\ 
130518A & 2.488 & 1388.34 $\pm $ 55.23 & 216.000 & 6.000 & 17.000 &  &  & 
\\ 
130610A & 2.092 & 840.81 $\pm $ 344.98 & 5.780 & 0.645 & 3.750 & 6.465 & 204
& 1.3 \\ 
130907A & 1.24 & 866.88 $\pm $ 35.84 & 385.000 & 23.900 & 4.870 &  &  &  \\ 
131030A & 1.29 & 448.84 $\pm $ 13.74 & 32.700 & 7.060 & 1.430 &  &  &  \\ 
131105A & 1.686 & 713.18 $\pm $ 46.18 & 15.300 & 3.340 & 1.660 &  &  &  \\ 
140206A & 2.73 & 1780.44 $\pm $ 119.77 & 278.000 & 26.800 & 5.220 &  &  & 
\\ 
140419A & 3.96 & 1398.72 $\pm $ 188.48 & 228.000 & 10.800 & 14.200 &  &  & 
\\ 
140423A & 3.26 & 516.74 $\pm $ 64.73 & 43.800 & 4.720 & 5.350 & 6.019 & 303
& 6.05 \\ 
140629A & 2.275 & 282 $\pm $ 56 & 4.400 & 0.575 & 3.390 &  &  &  \\ 
140703A & 3.14 & 861.25 $\pm $ 148.3 & 18.400 & 3.280 & 3.130 &  &  &  \\ 
141004A & 0.573 & 231 $\pm $ 44 & 0.210 & 0.051 & 0.877 &  &  &  \\ 
141028A & 2.33 & 979.59 $\pm $ 53.39 & 51.000 & 7.460 & 3.070 &  &  &  \\ 
141220A & 1.3195 & 418.8 $\pm $ 24.17 & 2.290 & 0.322 & 2.230 &  &  &  \\ 
150314A & 1.758 & 957.28 $\pm $ 19.06 & 76.800 & 5.090 & 5.620 &  &  &  \\ 
151021A & 2.33 & 566.1 $\pm $ 43.29 & 113.000 & 4.470 & 11.400 &  &  &  \\ 
160131A & 0.97 & 1282.47 $\pm $ 453.1 & 87.000 & 6.990 & 3.310 &  &  &  \\ 
160625B & 1.406 & 1134.34 $\pm $ 15.51 & 510.000 & 39.200 & 4.230 &  &  & 
\\ 
161023A & 2.708 & 604.4 $\pm $ 137.2 & 68.000 & 5.900 & 5.780 &  &  &  \\ 
161129A & 0.645 & 240.84 $\pm $ 70.09 & 0.783 & 0.059 & 2.970 &  &  &  \\ 
170202A & 3.645 & 1147 $\pm $ 771 & 17.000 & 4.420 & 2.410 &  &  &  \\ 
171010A & 0.3285 & 227.17 $\pm $ 9.3 & 18.000 & 2.880 & 1.123 &  &  &  \\ 
180325A & 2.248 & 993.89 $\pm $ 162.4 & 23.000 & 1.300 & 7.770 &  &  &  \\ 
180329B & 1.998 & 146 $\pm $ 28 & 4.690 & 0.995 & 1.910 &  &  &  \\ 
180914B & 1.096 & 977 $\pm $ 61 & 370.000 & 7.600 & 13.800 &  &  &  \\ 
181020A & 2.938 & 1461 $\pm $ 225 & 82.800 & 4.560 & 9.660 &  &  &  \\ 
181110A & 1.505 & 120 $\pm $ 68 & 11.000 & 0.937 & 3.970 &  &  &  \\ \hline
\end{tabular}
}
\end{center}
\end{table*}

%
%%%%%%%%%%%%%%%%%%%%%%%%%%%%%%%%%%%%%%%%%%%%%%%%%%%%%%%%%%%%%%%%%%%%%%%%%%%%%%

%%%%%%%%%%%%%%%%%%%%%%%%%%%%%%%%%%Table 4%%%%%%%%%%%%%%%%%%%%%%%%%%%%%%%

\begin{table*}[tbp]
\caption{Fitting results of the time-integrated spectrum for the bursts with
extremely high efficiency (detected by \textit{Fermi})}
\label{Tab_4}
\begin{center}
%\begin{tiny}%
\par
\renewcommand\arraystretch{1} 
\resizebox{\linewidth}{!}{
\begin{tabular}{ccccccccccc}
\hline
GRB & $\alpha$ & $E_{\mathrm{c}}$ & $F$ & $\alpha$ & $\beta$ & $E_{\mathrm{p}}$ & $F$ & $\Delta$DIC & $p_{\mathrm{DIC}}$ & $p_{\mathrm{DIC}}$ \\ 
& (CPL) & (CPL) & (CPL) & (Band) & (Band) & (Band) & (Band) & (Band-CPL) & 
(CPL) & (Band) \\ \hline
110818A & -1.18$^{+0.15}_{-0.15}$ & 352$^{+169}_{-161}$ & 0.18$^{+0.12}_{-0.06}$$\times$10$^{-6}$ & 0.48$^{+0.50}_{-0.56}$ & -1.70$^{+0.07}_{-0.07}$ & 42$^{+7}_{-7}$ & 0.46$^{+1.54}_{-0.35}$$\times$10$^{-6}$
& -141.4 & -2.9 & -151.3 \\ 
120729A & -0.55$^{+0.04}_{-0.04}$ & 163$^{+23}_{-24}$ & 0.17$^{+0.05}_{-0.04} $$\times$10$^{-6}$ & -0.55$^{+0.04}_{-0.04}$ & -4.93$^{+2.44}_{-2.66}$ & 229$^{+30}_{-29}$ & 0.19$^{+0.07}_{-0.04}$$\times$10$^{-6}$ & -1.3 & 1.9 & 1.4 \\ 
131108A & -0.93$^{+0.02}_{-0.02}$ & 380$^{+25}_{-26}$ & 1.36$^{+0.12}_{-0.10} $$\times$10$^{-6}$ & 0.66$^{+0.08}_{-0.08}$ & -1.60$^{+0.01}_{-0.01}$ & 61$^{+0}_{-0}$ & 3.54$^{+0.69}_{-0.61}$$\times$10$^{-6}$
& 599.9 & 2.9 & 1.7 \\ 
150821A & -1.19$^{+0.02}_{-0.02}$ & 419$^{+35}_{-35}$ & 0.64$^{+0.05}_{-0.05} $$\times$10$^{-6}$ & -1.18$^{+0.03}_{-0.03}$ & -3.87$^{+1.71}_{-3.16}$ & 313$^{+31}_{-32}$ & 0.76$^{+0.19}_{-0.15}$$\times$10$^{-6}$ & -8.0 & 2.9 & -1.5 \\ 
161014A & -0.74$^{+0.10}_{-0.10}$ & 144$^{+24}_{-24}$ & 0.22$^{+0.07}_{-0.05} $$\times$10$^{-6}$ & 1.52$^{+0.34}_{-0.34}$ & -1.74$^{+0.04}_{-0.04}$ & 48$^{+2}_{-2}$ & 0.63$^{+0.82}_{-0.37}$$\times$10$^{-6}$
& 3.0 & 1.6 & -40.0 \\ 
170214A & -1.06$^{+0.01}_{-0.01}$ & 570$^{+26}_{-26}$ & 1.53$^{+0.08}_{-0.06} $$\times$10$^{-6}$ & -1.05$^{+0.01}_{-0.01}$ & -2.77$^{+0.36}_{-0.12}$ & 510$^{+22}_{-22}$ & 1.81$^{+0.15}_{-0.14}$$\times$10$^{-6}$ & -10.6 & 3.0 & 2.7 \\ \hline
\end{tabular}
}
\end{center}
\par
Note---For 120729A, 150821A and 170214A, the high-energy spectral indices
for BAND function are very small (-4.93, -3.87 and -2.77), while the
low-energy spectral indices and peak energy for BAND function and CPL model
are very similar. For 110818A, 131108A and 161014A, the low-energy spectral
indices for BAND function are extremely hard (0.48, 0.66, 1.52), and the
peak energy is extremely small (42, 61, 48). Thus, the CPL model is surely
the best-fit model.
\end{table*}

%
%%%%%%%%%%%%%%%%%%%%%%%%%%%%%%%%%%%%%%%%%%%%%%%%%%%%%%%%%%%%%%%%%%%%%%%%%%%%%%

%%%%%%%%%%%%%%%%%%%%%%%%%%%%%%%%%%Table 5%%%%%%%%%%%%%%%%%%%%%%%%%%%%%%%

\begin{table*}[tbp]
\caption{The bursts with almost same $E_{\text{ratio}}$ and $E_{\text{iso}%
}/E_{\text{k}}$ ($\protect\epsilon _{\protect\gamma }\lesssim 50\%$) and their power-law slope of the X-ray afterglow light
curve. }
\label{Tab_5}
\begin{center}
%\begin{tiny}%
\par
\renewcommand\arraystretch{1} 
\begin{tabular}{cccccccc}
\hline
\multicolumn{4}{c}{Bursts after GRB 110213A} & \multicolumn{4}{c}{Bursts
before GRB 110213A} \\ \hline
GRB & $E_{\text{iso}}/[\frac{3.7\ast L_{\text{X,11}}}{(1+z)/2}]$ & $E_{\text{%
ratio}}$ & X-ray slope & GRB & $E_{\text{iso}}/[\frac{3.7\ast L_{\text{X,11}}%
}{(1+z)/2}]$ & $E_{\text{ratio}}$ & X-ray slope \\ \hline
130420A & 0.653 & 0.644$^{a}$ & $-$0.900 & 050922C & 0.898 & 0.821 & $-$1.200
\\ 
130831A & 0.157 & 0.156 & $-$0.959 & 060210 & 0.751 & 0.597 & $-$0.970 \\ 
131117A & 0.491 & 0.546 & $-$0.998 & 060306 & 0.567 & 0.435 & $-$1.047 \\ 
140213A & 0.197 & 0.218 & $-$1.070 & 080603B & 0.629 & 0.528 & $-$0.850 \\ 
150301B & 1.190 & 1.16 & $-$1.179 & 091018 & 0.112 & 0.092 & $-$1.160 \\ 
150403A & 1.100 & 1.21 & $-$1.140 & 091020 & 0.912 & 0.831 & $-$1.090 \\ 
170705A & 0.289 & 0.307 & $-$0.969 & 100621A & 0.159 & 0.193 & $-$0.987 \\ 
\hline
\end{tabular}%
\end{center}
\par
$^{a}$ $E_{\text{iso}}/E_{\text{k}}=(R_{\text{ph}}/R_{s})^{-2/3}$%
\end{table*}

%%%%%%%%%%%%%%%%%%%%%%%%%%%%%%%%%%%%%%%%%%%%%%%%%%%%%%%%%%%%%%%%%%%%%%%%%%%%%%

%%%%%%%%%%%%%%%%%%%%%%%%%%%%%%%%%%Table 6%%%%%%%%%%%%%%%%%%%%%%%%%%%%%%%

\begin{table*}[tbp]
\caption{The observed quantities, inferred efficiency ($E_{\text{ratio}}$
and $(R_{\text{ph}}/R_{s})^{-2/3}$) and inferred $\Gamma $ (from $T_{p,op}$
and prompt emission) for the bursts with $T_{p,op}$ detection.}
\label{Tab_6}
\begin{center}
%\begin{tiny}%
\par
\renewcommand\arraystretch{0.8} 
\resizebox{\linewidth}{!}{
\begin{tabular}{ccccccccccccc}
\hline
\multicolumn{13}{c}{$\epsilon _{\gamma }\lesssim 50\%$ sub-sample (24 bursts)
} \\ \hline
GRB & z & $dE_{p,z}$/$E_{p,z}$ & $E_{p,z}$ & $E_{\text{iso}}$ & $E_{\text{ratio}}$ & $(R_{\text{ph}}/R_{s})^{-2/3}$ & $T_{90,i}$ & $T_{p,op}$ & $\Gamma _{op}$ & $\Gamma _{E}$ & $\Gamma _{L}$ & $L_{\text{iso}}$ \\ 
&  &  & (keV) & (10$^{52}$ erg) &  &  & s & s &  &  &  & (10$^{52}$ erg s$^{-1}$) \\ \hline
\multicolumn{13}{c}{$dE_{p,z}$/$E_{p,z}\lesssim 0.2$} \\ \hline
060124 & 2.3 & 0.176 & 635.0 $\pm $ 112.0 & 43.000 & 0.640 & 0.362 & 72.2 & 
631.0 & 220.0 & 272.0 & 451.0 & 14.200 \\ 
090618 & 0.54 & 0.059 & 156.0 $\pm $ 9.2 & 25.300 & 0.267 & 0.710 & 67.7 & 
91.2 & 319.0 & 128.0 & 340.0 & 2.050 \\ 
090926A & 2.11 & 0.028 & 908.0 $\pm $ 25.0 & 200.000 & 0.617 & 0.732 & 4.3 & 
8.1 & 851.0 & 798.0 & 828.0 & 74.000 \\ 
091020 & 1.71 & 0.037 & 507.0 $\pm $ 19.0 & 7.910 & 0.833 & 0.689 & 10.8 & 
135.0 & 295.0 & 316.0 & 254.0 & 3.300 \\ 
100728B & 2.11 & 0.116 & 404.0 $\pm $ 47.0 & 7.240 & 0.716 & 1.230 & 3.9 & 
33.9 & 462.0 & 360.0 & 216.0 & 1.860 \\ 
100814A & 1.44 & 0.093 & 344.0 $\pm $ 32.0 & 8.200 & 0.491 & 0.444 & 60.8 & 
589.0 & 164.0 & 170.0 & 167.0 & 0.920 \\ 
110213A & 1.46 & 0.087 & 241.0 $\pm $ 21.0 & 6.400 & 0.332 & 0.299 & 11.9 & 
324.0 & 199.0 & 207.0 & 284.0 & 2.090 \\ 
120922A & 3.1 & 0.167 & 156.0 $\pm $ 26.0 & 20.000 & 0.234 & 0.261 & 44.5 & 
891.0 & 190.0 & 138.0 & 398.0 & 2.900 \\ 
130420A & 1.3 & 0.054 & 331.0 $\pm $ 18.0 & 7.190 & 0.653 & 0.599 & 45.7 & 
356.0 & 190.0 & 176.0 & 109.0 & 0.350 \\ 
130612A & 2.01 & 0.172 & 186.0 $\pm $ 32.0 & 0.716 & 0.488 & 0.443 & 1.9 & 
110.0 & 246.0 & 219.0 & 212.0 & 0.875 \\ 
130831A & 0.48 & 0.164 & 80.9 $\pm $ 13.3 & 0.757 & 0.158 & 0.163 & 11.9 & 
724.0 & 93.3 & 92.0 & 185.0 & 0.296 \\ 
131231A & 0.64 & 0.042 & 288.0 $\pm $ 12.0 & 20.000 & 0.288 & 0.299 & 17.7 & 
100.0 & 240.0 & 237.0 & 239.0 & 1.700 \\ 
140629A & 2.28 & 0.199 & 282.0 $\pm $ 56.0 & 6.000 & 0.418 & 0.444 & 7.9 & 
151.0 & 293.0 & 246.0 & 298.0 & 2.700 \\ 
\multicolumn{13}{c}{$dE_{p,z}$/$E_{p,z}\gtrsim 0.2$} \\ 
050922C & 2.2 & 0.266 & 417.0 $\pm $ 111.0 & 4.530 & 0.773 & 0.739 & 2. & 
132.0 & 401.0 & 408.0 & 619.0 & 19.000 \\ 
060210 & 3.91 & 0.323 & 575.0 $\pm $ 186.0 & 41.500 & 0.568 & 0.408 & 51.9 & 
676.0 & 248.0 & 280.0 & 316.0 & 5.960 \\ 
060418 & 1.49 & 0.250 & 571.0 $\pm $ 143.0 & 12.800 & 0.832 & 0.738 & 41.4 & 
151.0 & 244.0 & 255.0 & 187.0 & 1.890 \\ 
060607A & 3.08 & 0.348 & 575.0 $\pm $ 200.0 & 10.900 & 0.886 & 0.774 & 25. & 
178.0 & 271.0 & 284.0 & 191.0 & 2.000 \\ 
070110 & 2.35 & 0.460 & 370.0 $\pm $ 170.0 & 5.500 & 0.618 & 0.200 & 26.4 & 
1170.0 & 135.0 & 207.0 & 117.0 & 0.451 \\ 
081007 & 0.53 & 0.246 & 61.0 $\pm $ 15.0 & 0.170 & 0.178 & 1.090 & 6.5 & 
123.0 & 152.0 & 77.1 & 85.8 & 0.043 \\ 
091029 & 2.75 & 0.287 & 230.0 $\pm $ 66.0 & 7.400 & 0.297 & 0.317 & 10.5 & 
407.0 & 218.0 & 213.0 & 234.0 & 1.320 \\ 
100906A & 1.73 & 0.348 & 158.0 $\pm $ 55.0 & 33.400 & 0.220 & 1.220 & 33.1 & 
100.0 & 396.0 & 160.0 & 366.0 & 2.450 \\ 
130215A & 0.6 & 0.408 & 248.0 $\pm $ 101.0 & 2.500 & 0.471 & 0.447 & 90. & 
741.0 & 111.0 & 113.0 & 63.9 & 0.084 \\ 
141109A & 2.93 & 0.404 & 750.0 $\pm $ 303.0 & 31.000 & 0.891 & 0.150 & 23.5
& 955.0 & 193.0 & 376.0 & 240.0 & 4.200 \\ 
141221A & 1.45 & 0.228 & 372.0 $\pm $ 85.0 & 1.900 & 0.888 & 0.723 & 9.7 & 
110.0 & 216.0 & 233.0 & 142.0 & 0.700 \\ \hline
\multicolumn{13}{c}{$\epsilon _{\gamma }\gtrsim 50\%$ sub-sample (23 bursts)}
\\ \hline
GRB & z & $dE_{p,z}$/$E_{p,z}$ & $E_{p,z}$ & $E_{\text{iso}}$ & $E_{\text{ratio}}$ & $\eta /\Gamma $ & $T_{90,i}$ & $T_{p,op}$ & $\Gamma _{op}$ & $\Gamma _{E}$ & $\Gamma _{L}$ & $L_{\text{iso}}$ \\ 
&  &  & (keV) & (10$^{52}$ erg) &  &  & s & s &  &  &  & (10$^{52}$ erg s$^{-1}$) \\ \hline
990123 & 1.6 & 0.043 & 2030.0 $\pm $ 88.0 & 239.000 & 1.700 & 1.150 & 23.9 & 
47.9 & 656.0 & 677.0 & 417.0 & 35.300 \\ 
060605 & 3.78 & 0.512 & 490.0 $\pm $ 251.0 & 2.830 & 1.120 & 3.610 & 16.6 & 
479.0 & 200.0 & 209.0 & 146.0 & 0.951 \\ 
061121 & 1.31 & 0.093 & 1290.0 $\pm $ 120.0 & 26.100 & 1.940 & 10.300 & 7.7
& 162.0 & 301.0 & 543.0 & 332.0 & 14.100 \\ 
071112C & 0.82 & 0.546 & 597.0 $\pm $ 326.0 & 1.600 & 1.760 & 1.950 & 2.9 & 
178.0 & 187.0 & 333.0 & 89.8 & 0.400 \\ 
080319B & 0.94 & 0.013 & 1260.0 $\pm $ 17.0 & 150.000 & 1.050 & 0.134 & 23.
& 17.4 & 810.0 & 508.0 & 281.0 & 9.590 \\ 
080804 & 2.2 & 0.056 & 809.0 $\pm $ 45.0 & 11.500 & 1.370 & 0.443 & 10.6 & 
63.1 & 437.0 & 358.0 & 189.0 & 2.690 \\ 
080810 & 3.35 & 0.121 & 1490.0 $\pm $ 180.0 & 39.100 & 2.060 & 1.330 & 24.4
& 117.0 & 453.0 & 460.0 & 257.0 & 9.270 \\ 
080916C & 4.35 & 0.089 & 2760.0 $\pm $ 246.0 & 560.000 & 1.930 & 0.035 & 11.5
& 6.2 & 2060.0 & 1050.0 & 600.0 & 104.000 \\ 
081203A & 2.1 & 1.200 & 1540.0 $\pm $ 1854.0 & 35.000 & 2.240 & 3.020 & 96.4
& 309.0 & 274.0 & 327.0 & 146.0 & 2.810 \\ 
090323 & 3.57 & 0.108 & 1900.0 $\pm $ 206.0 & 390.000 & 1.320 & 3.580 & 29.1
& 200.0 & 504.0 & 663.0 & 446.0 & 38.500 \\ 
090812 & 2.45 & 0.449 & 2020.0 $\pm $ 908.0 & 40.300 & 3.060 & 0.496 & 9.5 & 
47.9 & 583.0 & 681.0 & 229.0 & 9.550 \\ 
090902B & 1.82 & 0.015 & 2020.0 $\pm $ 31.0 & 440.000 & 1.380 & 0.094 & 6.9
& 8.5 & 1390.0 & 994.0 & 529.0 & 58.900 \\ 
100414A & 1.39 & 0.020 & 1490.0 $\pm $ 29.0 & 76.900 & 1.640 & 0.281 & 9.2 & 
34.7 & 622.0 & 638.0 & 226.0 & 7.000 \\ 
110731 & 2.83 & 0.057 & 1210.0 $\pm $ 69.0 & 40.000 & 1.550 & 0.041 & 1.75 & 
5.0 & 1410.0 & 805.0 & 459.0 & 27.000 \\ 
120711A & 1.4 & 0.039 & 2340.0 $\pm $ 91.0 & 150.000 & 2.400 & 7.910 & 17.2
& 240.0 & 328.0 & 744.0 & 266.0 & 15.200 \\ 
120909A & 3.93 & 0.067 & 1820.0 $\pm $ 123.0 & 72.900 & 2.190 & 2.400 & 23.3
& 288.0 & 366.0 & 557.0 & 210.0 & 7.190 \\ 
130427A & 0.34 & 0.009 & 1380.0 $\pm $ 13.0 & 80.900 & 1.450 & 1.260 & 46.1
& 21.9 & 599.0 & 413.0 & 435.0 & 27.000 \\ 
130610A & 2.09 & 0.419 & 912.0 $\pm $ 382.0 & 6.810 & 1.920 & 3.150 & 7. & 
204.0 & 260.0 & 395.0 & 171.0 & 2.400 \\ 
160629A & 3.33 & 0.074 & 1280.0 $\pm $ 95.0 & 47.000 & 1.580 & 0.688 & 14.9
& 81.3 & 531.0 & 494.0 & 272.0 & 9.100 \\ 
061007 & 1.26 & 0.030 & 902.0 $\pm $ 27.0 & 88.100 & 0.804$^{a}$ & 2.220 & 25.5
& 74.1 & 466.0 & 392.0 & 427.0 & 17.400 \\ 
110205A & 2.22 & 0.335 & 714.0 $\pm $ 239.0 & 56.000 & 0.685$^{a}$ & 8.540 & 79.8
& 813.0 & 205.0 & 248.0 & 193.0 & 2.500 \\ 
121128A & 2.2 & 0.050 & 198.0 $\pm $ 10.0 & 14.000 & 0.197$^{a}$ & 1.220 & 3.1 & 
74.1 & 422.0 & 247.0 & 516.0 & 6.400 \\ 
140423A & 3.26 & 0.122 & 533.0 $\pm $ 65.0 & 56.000 & 0.464$^{a}$ & 1.550 & 22.3
& 200.0 & 385.0 & 294.0 & 319.0 & 5.660 \\ \hline
\end{tabular}
}
\end{center}
\par
$^{a}$ $\epsilon _{\gamma }\gtrsim 50\%$ judged from $E_{\text{iso}}/E_{%
\text{k}}$.
\end{table*}

%
%%%%%%%%%%%%%%%%%%%%%%%%%%%%%%%%%%%%%%%%%%%%%%%%%%%%%%%%%%%%%%%%%%%%%%%%%%%%%%

%%%%%%%%%%%%%%%%%%%%%%%%%%%%%%%%%%Table 7%%%%%%%%%%%%%%%%%%%%%%%%%%%%%%%

\begin{table*}[tbp]
\caption{The observed quantities and the properties of $E_{\text{iso}}/L_{%
\text{X,11}}$ for the bursts with maximum $\Gamma $.}
\label{Tab_7}
\begin{center}
%\begin{tiny}%
\par
\renewcommand\arraystretch{0.8} 
\resizebox{\linewidth}{!}{ 
\begin{tabular}{ccccccccccccc}
\hline
GRB & z & $T_{90,i}$ & $dE_{p,z}$/$E_{p,z}$ & $E_{\text{ratio}}$ & $E_{p,z}$
& $E_{\text{iso}}$ & $L_{\text{X,11}}$ & $E_{\text{iso}}/L_{\text{X,11}}$ & $T_{p,op}$ & $\Gamma _{op}$ & $\Gamma _{op}/\Gamma _{\max }$ & $L_{\text{iso}} $ \\ 
&  & s &  &  & (keV) & (10$^{52}$ erg) & (10$^{45}$ erg s$^{-1}$) &  & s & 
&  & (10$^{52}$ erg s$^{-1}$) \\ \hline
100414A & 1.39 & 9.2 & 0.020 & 1.640 & 1490.0 $\pm $ 29.0 & 76.900 &  &  & 
34.7 & 622.0 & 1.040 & 7.000 \\ 
090812 & 2.45 & 9.5 & 0.449 & 3.060 & 2020.0 $\pm $ 908.0 & 27.18 & 6.525
& 4.166 & 47.9 & 583.0 & 0.902 & 9.550 \\ 
160629A & 3.33 & 14.9 & 0.074 & 1.580 & 1280.0 $\pm $ 95.0 & 47.000 &  &  & 
81.3 & 531.0 & 0.832 & 9.100 \\ 
100728B & 2.11 & 3.9 & 0.116 & 1.000 & 404.0 $\pm $ 47.0 & 2.66 & 0.841 & 
3.163 & 33.9 & 462.0 & 1.076 & 1.860 \\ 
050502A & 3.79 &  &  & 1.025 & 498.9 & 3.981 &  &  & 57.5 & 461.4 & 0.989 & 
2.600 \\ 
080804 & 2.2 & 10.6 & 0.056 & 1.370 & 809.0 $\pm $ 45.0 & 11.500 & 3.181 & 
3.615 & 63.1 & 437.0 & 0.928 & 2.690 \\ 
081008 & 1.97 &  &  & 0.438 & 267.3 & 4.19 & 0.915 & 4.579 & 162.2 & 261.5 & 
0.961 & 0.300 \\ 
080310 & 2.42 &  &  & 0.087 & 75.0 & 3.25 & 1.78 &  & 182.0 & 255.8 & 0.881
& 0.390 \\ 
&  &  &  &  &  &  &  &  & (2000) & (104) & (0.352) &  \\ 
081007 & 0.53 & 6.5 & 0.246 & 0.178 & 61.0 $\pm $ 15.0 & 0.170 &  &  & 
123.0 & 152.0 & 0.908 & 0.043 \\ \hline
\end{tabular}
}
\end{center}
\end{table*}

%
%%%%%%%%%%%%%%%%%%%%%%%%%%%%%%%%%%%%%%%%%%%%%%%%%%%%%%%%%%%%%%%%%%%%%%%%%%%%%%

%%%%%%%%%%%%%%%%%%%%%%%%%%%%%%%%%%Table 8%%%%%%%%%%%%%%%%%%%%%%%%%%%%%%%

\begin{table*}[tbp]
\caption{The observed quantities, inferred efficiency, and the properties of
extended emission for the short GRBs.}
\label{Tab_8}
\begin{center}
\renewcommand\arraystretch{0.8} 
\resizebox{\linewidth}{!}{
\begin{tabular}{ccccccccccc}
\hline
\multicolumn{8}{c}{$\epsilon _{\gamma }\lesssim 50\%$ sub-sample (8 bursts)}
&  &  &  \\ \hline
GRB & z & $E_{p,z}$ & $E_{\text{iso}}$ & $L_{\text{X,11}}$ & $E_{\text{iso}}/[\frac{3.7\ast L_{\text{X,11}}}{(1+z)/2}]$ & $E_{\text{ratio}}$ & $T_{90,i} $ &  &  &  \\ 
&  & (keV) & (10$^{52}$ erg) & (10$^{45}$ erg s$^{-1}$) &  &  & s &  &  & 
\\ \hline
051221A & 0.5465 & 677.0$_{-141.0}^{+200.0}$ & 0.91 & 0.277 & 0.687 & 0.296
& 0.14 &  &  &  \\ 
070724A & 0.457 & 119.5 $\pm $ 7.3 & 0.0016 & 0.0271 & 0.0116 & 0.243 & 
0.27 &  &  &  \\ 
070809 & 2.187 & 464.0 $\pm $ 223.0 & 0.104 & 0.338 & 0.133 & 0.368 & 0.44 & 
&  &  \\ 
130603B & 0.356 & 823.0$_{-71.0}^{+83.0}$ & 0.196 & 0.0773 & 0.465 & 0.64 & 
0.16 &  &  &  \\ 
131004A & 0.71 & 202.0 $\pm $ 51.0 & 0.068 & 0.145 & 0.109 & 0.14 & 0.9 &  & 
&  \\ 
140903A & 0.351 & 60.0 $\pm $ 22.0 & 0.0044 & 0.285 & 0.00281 & 0.0691 & 
0.22 &  &  &  \\ 
150423A & 0.22 & 146.0 $\pm $ 43.0 & 0.00075 & 0.0701 & 0.00176 & 0.408 & 
1.14 &  &  &  \\ 
160821B & 0.16 & 97.4 $\pm $ 22.0 & 0.012 & 0.0115 & 0.164 & 0.0944 & 0.41 & 
&  &  \\ 
\multicolumn{8}{c}{$\epsilon _{\gamma }\gtrsim 50\%$ sub-sample (4 bursts)}
&  &  &  \\ \hline
GRB & z & $T_{90,i}$ & $L_{\text{X,11}}$ & $E_{\text{iso}}/[\frac{3.7\ast L_{\text{X,11}}}{(1+z)/2}]$ & $E_{\text{iso}}$ & $E_{p,z}$ & $E_{\text{iso}}^{\text{Ext}}$ & $E_{p,z}^{\text{Ext}}$ & $E_{\text{iso}}/E_{\text{iso}}^{\text{Ext}}$ & $E_{p,z}/E_{p,z}^{\text{Ext}}$ \\ 
&  & s & (10$^{45}$ erg s$^{-1}$) &  & (10$^{52}$ erg) & (keV) & (10$^{52}$
erg) & (keV) &  &  \\ \hline
070714B & 0.923 & 0.65 & 0.040 & 4.127 & 0.640 & 1060$_{-215}^{+285}$ & 0.116
& 164.87 $\pm $ 73.13 & 5.517 & 6,429 \\ \hline
110402A & 0.805 & 2.8 & 0.255 & 1.454 & 1.520 & 1924$_{-451}^{+767}$ & 0.642
&  &  &  \\ \hline
150424A & 0.30 & 0.21 & 0.177 & 0.431 & 0.434 & 1191$_{-61}^{+64}$ & 0.0625
& 82.3 $\pm $ 82.1 & 6.944 & 14.349 \\ \hline
160410A & 1.717 & 0.58 & 0.441 & 7.749 & 9.300 & 3853$_{-973}^{+1429}$ & 1.55
& 495.3 $\pm $ 232.9 & 6.000 & 7.779 \\ \hline
170817A & 0.00968 & 0.50 &  &  & (4.7 $\pm $ 0.7) $\times $ 10$^{-6}$ & 65.6$_{-14.1}^{+35.3}$ & (1.6 $\pm $ 0.2) $\times $ 10$^{-6}$ & 38.4 $\pm $ 4.2 & 
3 & 1.708 \\ \hline
061006A & 0.4377 & 0.26 &  &  & 0.382 & 909$_{-191}^{+260}$ & 0.0674 & 150 & 
5.668 & 6.060 \\ \hline
071227A & 0.384 & 1.30 &  &  & 0.0591 & 875$_{-287}^{+790}$ & 0.0196 & 80 & 
3.015 & 10.938 \\ \hline
080123A & 0.495 & 0.27 &  &  & 0.32 & 2228$_{-1308}^{+1272}$ & 0.0398 & 53 & 
8.040 & 42.038 \\ \hline
061210A & 0.4095 & 0.07 &  &  & 0.0024 & 761$_{-264}^{+648}$ & 0.0422 &  & 
&  \\ \hline
060614A & 0.1254 & 4.4 &  &  & 0.24 & 340$_{-96}^{+241}$ $^{a}$ & 0.0765 &  &  & 
\\ \hline
050709A & 0.1606 & 0.06 &  &  & 0.0027 & 96.3$_{-13.9}^{+20.9}$ $^{a}$ &  &  &  & 
\\ \hline
050724A & 0.2576 & 2.4 &  &  & 0.009 & 138$_{-57}^{+503}$ $^{a}$ &  &  &  &  \\ 
\hline
\end{tabular}
}
\end{center}
\par
$^{a}$ Only detected by \textit{Swift} or HETE-2 (lacking detections in
the high-energy band).
\end{table*}


\begin{thebibliography}{Zhang \& M{\'e}sz{\'a}ros(2002)}
\bibitem[Abdo et al.(2009)]{Abdo2009} Abdo, A.~A., Ackermann, M., Ajello,
M., et al.\ 2009, \apjl, 706, L138

\bibitem[Abramowicz et al.(1991)]{Abra1991} Abramowicz, M.~A., Novikov,
I.~D., \& Paczynski, B.\ 1991, \apj, 369, 175

\bibitem[Acuner \& Ryde(2018)]{Acun2018} Acuner, Z., \& Ryde, F.\ 2018, %
\mnras, 475, 1708

\bibitem[Acuner et al.(2020)]{Acun2020} Acuner, Z., Ryde, F., Pe'er, A., et
al.\ 2020, \apj, 893, 128

\bibitem[Amati et al.(2002)]{Amati2002} Amati, L., Frontera, F., Tavani, M., et al.\ 2002, \aap, 390, 81

\bibitem[Axelsson et al.(2012)]{Axel2012} Axelsson, M., Baldini, L.,
Barbiellini, G., et al.\ 2012, \apjl, 757, L31

\bibitem[Axelsson \& Borgonovo(2015)]{AxBo2015} Axelsson, M., \& Borgonovo,
L.\ 2015, \mnras, 447, 3150

\bibitem[Band et al.(1993)]{Band1993} Band, D., Matteson, J., Ford, L., et
al.\ 1993, \apj, 413, 281

\bibitem[B{\'e}gu{\'e} \& Pe'er(2015)]{Be2015} B{\'e}gu{\'e}, D., \& Pe'er,
A.\ 2015, \apj, 802, 134

\bibitem[Beloborodov(2011)]{Belo2011} Beloborodov, A.~M.\ 2011, \apj, 737, 68

\bibitem[Beloborodov(2017)]{Belo2016} Beloborodov, A.~M.\ 2017, \apj, 838,
125

\bibitem[Bhattacharya et al.(2018)]{Bhat2018} Bhattacharya, M., Lu, W., Kumar, P., et al.\ 2018, \apj, 852, 24

\bibitem[Bhattacharya \& Kumar(2020)]{Bhat2020} Bhattacharya, M. \& Kumar, P.\ 2020, \mnras, 491, 4656

\bibitem[Burgess et al.(2017)]{Bur2017} Burgess, J.~M., Greiner, J., B{\'e}gu%
{\'e}, D., \& Berlato, F.\ 2017, arXiv:1710.08362

\bibitem[Burgess et al.(2020)]{Burgess2020} Burgess, J.~M., B{\'e}gu{\'e},
D., Greiner, J., et al.\ 2020, Nature Astronomy, 4, 174

\bibitem[Dai \& Gou(2001)]{Dai2001} Dai, Z.~G., \& Gou, L.~J.\ 2001, \apj,
552, 72

\bibitem[D'Avanzo et al.(2012)]{Avan2012} D'Avanzo, P., Salvaterra, R.,
Sbarufatti, B., et al.\ 2012, \mnras, 425, 506

\bibitem[Deng et al.(2022)]{Deng2022} Deng, L.-T., Lin, D.-B., Zhou, L., et al.\ 2022, \apjl, 934, L22

\bibitem[Deng \& Zhang(2014)]{Deng2014} Deng, W., \& Zhang, B.\ 2014, \apj,
785, 112

\bibitem[Dereli-B{\'e}gu{\'e} et al.(2020)]{Dere2020} Dereli-B{\'e}gu{\'e}, H., Pe'er, A., \& Ryde, F.\ 2020, \apj, 897, 145

\bibitem[Fan \& Piran(2006)]{Fan2006} Fan, Y., \& Piran, T.\ 2006, \mnras,
369, 197

\bibitem[Fan et al.(2012)]{Fan2012} Fan, Y.-Z., Wei, D.-M., Zhang, F.-W., \&
Zhang, B.-B.\ 2012, \apjl, 755, L6

\bibitem[Gao et al.(2015)]{Gao2015} Gao, H., Wang, X.-G., M{\'e}sz{\'a}ros,
P., et al.\ 2015, \apj, 810, 160

\bibitem[Geng et al.(2018)]{Geng18} Geng, J.-J., Huang, Y.-F., Wu, X.-F.,
Zhang, B., \& Zong, H.-S.\ 2018, \apjs, 234, 3

\bibitem[Geng et al.(2019)]{Geng2019} Geng, J.-J., Zhang, B., K{\"o}lligan,
A., Kuiper, R., \& Huang, Y.-F.\ 2019, \apjl, 877, L40

\bibitem[Ghirlanda et al.(2012)]{Ghirlan2012} Ghirlanda, G., Nava, L., Ghisellini, G., et al.\ 2012, \mnras, 420, 483

\bibitem[Ghirlanda et al.(2013)]{Ghir2013} Ghirlanda, G., Pescalli, A., \&
Ghisellini, G.\ 2013, \mnras, 432, 3237

\bibitem[Ghirlanda et al.(2018)]{Ghirlan2018} Ghirlanda, G., Nappo, F.,
Ghisellini, G., et al.\ 2018, \aap, 609, A112

\bibitem[Ghisellini et al.(2010)]{Ghise2010} Ghisellini, G., Ghirlanda, G.,
Nava, L., et al.\ 2010, \mnras, 403, 926

\bibitem[Giannios \& Spruit(2007)]{Gian2007} Giannios, D. \& Spruit, H.~C.\
2007, \aap, 469, 1

\bibitem[Goldstein et al.(2012)]{Gold2012} Goldstein, A., Burgess, J.~M.,
Preece, R.~D., et al.\ 2012, \apjs, 199, 19

\bibitem[Gompertz et al.(2020)]{Gomp2020} Gompertz, B.~P., Levan, A.~J., \&
Tanvir, N.~R.\ 2020, \apj, 895, 58

\bibitem[Goodman(1986)]{Good1986} Goodman, J.\ 1986, \apjl, 308, L47

\bibitem[Gruber et al.(2014)]{Grub2014} Gruber, D., Goldstein, A., Weller
von Ahlefeld, V., et al.\ 2014, \apjs, 211, 12

\bibitem[Guiriec et al.(2011)]{Gui2011} Guiriec, S., Connaughton, V.,
Briggs, M.~S., et al.\ 2011, \apjl, 727, L33

\bibitem[Guiriec et al.(2013)]{Gui2013} Guiriec, S., Daigne, F., Hasco{\"e}%
t, R., et al.\ 2013, \apj, 770, 32

\bibitem[Hou et al.(2018)]{Hou2018} Hou, S.-J., Zhang, B.-B., Meng, Y.-Z.,
et al.\ 2018, \apj, 866, 13

\bibitem[Iyyani \& Sharma(2021)]{Iyya2021} Iyyani, S. \& Sharma, V.\ 2021, \apjs, 255, 25

\bibitem[Kaneko et al.(2006)]{Kan2006} Kaneko, Y., Preece, R.~D., Briggs,
M.~S., et al.\ 2006, \apjs, 166, 298

\bibitem[Kobayashi et al.(1997)]{Koba1997} Kobayashi, S., Piran, T., \&
Sari, R.\ 1997, \apj, 490, 92

\bibitem[Lan et al.(2020)]{Lan2020} Lan, L., Lu, R.-J., L{\"u}, H.-J., et
al.\ 2020, \mnras, 492, 3622

\bibitem[Larsson et al.(2015)]{Lar2015} Larsson, J., Racusin, J.~L., \&
Burgess, J.~M.\ 2015, \apjl, 800, L34

\bibitem[Lazzati et al.(2013)]{Lazz2013} Lazzati, D., Morsony, B.~J.,
Margutti, R., \& Begelman, M.~C.\ 2013, \apj, 765, 103

\bibitem[Li et al.(2012)]{Li2012} Li, L., Liang, E.-W., Tang, Q.-W., et al.\ 2012, \apj, 758, 27

\bibitem[Li et al.(2018)]{Li2018} Li, L., Wang, Y., Shao, L., et al.\ 2018, \apjs, 234, 26

\bibitem[Li(2019a)]{Li2019a} Li, L.\ 2019a, \apjs, 242, 16

\bibitem[Li et al.(2019b)]{Li2019b} Li, L., Geng, J.-J., Meng, Y.-Z., et
al.\ 2019b, \apj, 884, 109

\bibitem[Li(2019c)]{Li2019c} Li, L.\ 2019c, \apjs, 245, 7

\bibitem[Li(2020)]{Li2019d} Li, L.\ 2020, \apj, 894, 100

\bibitem[Li et al.(2021)]{Li2021} Li, L., Ryde, F., Pe'er, A., et al.\ 2021, \apjs, 254, 35

\bibitem[Liang \& Kargatis(1996)]{Lia1996} Liang, E., \& Kargatis, V.\ 1996, %
\nat, 381, 49

\bibitem[Liang et al.(2013)]{Liang2013} Liang, E.-W., Li, L., Gao, H., et al.\ 2013, \apj, 774, 13

\bibitem[Liang et al.(2015)]{Liang2015} Liang, E.-W., Lin, T.-T., L{\"u},
J., et al.\ 2015, \apj, 813, 116

\bibitem[Lin et al.(2018)]{Lin2018} Lin, D.-B., Liu, T., Lin, J., et al.\
2018, \apj, 856, 90

\bibitem[Lloyd-Ronning \& Zhang(2004)]{Lloy2004} Lloyd-Ronning, N.~M., \&
Zhang, B.\ 2004, \apj, 613, 477

\bibitem[Lu et al.(2010)]{Lu2010} Lu, R.-J., Hou, S.-J., \& Liang, E.-W.\
2010, \apj, 720, 1146

\bibitem[Lu et al.(2012)]{Lu2012} Lu, R.-J., Wei, J.-J., Liang, E.-W., et
al.\ 2012, \apj, 756, 112

\bibitem[Lundman et al.(2013)]{Lund2013} Lundman, C., Pe'er, A., \& Ryde,
F.\ 2013, \mnras, 428, 2430

\bibitem[L{\"u} et al.(2012)]{Lv2012} L{\"u}, J., Zou, Y.-C., Lei, W.-H., et
al.\ 2012, \apj, 751, 49

\bibitem[Meng et al.(2018)]{Meng2018} Meng, Y.-Z., Geng, J.-J., Zhang,
B.-B., et al.\ 2018, \apj, 860, 72

\bibitem[Meng et al.(2019)]{Meng2019} Meng, Y.-Z., Liu, L.-D., Wei, J.-J.,
Wu, X.-F., \& Zhang, B.-B.\ 2019, \apj, 882, 26

\bibitem[Meng et al.(2022)]{Meng2021} Meng, Y.-Z., Geng, J.-J., \& Wu,
X.-F.\ 2022, \mnras, 509, 6047

\bibitem[M{\'e}sz{\'a}ros \& Rees(1997)]{Mes1997} M{\'e}sz{\'a}ros, P. \&
Rees, M.~J.\ 1997, \apj, 476, 232

\bibitem[M{\'e}sz{\'a}ros \& Rees(2000)]{Me2000} M{\'e}sz{\'a}ros, P., \&
Rees, M.~J.\ 2000, \apj, 530, 292

\bibitem[M{\'e}sz{\'a}ros(2002)]{Me2002} M{\'e}sz{\'a}ros, P.\ 2002, \araa,
40, 137

\bibitem[Minaev \& Pozanenko(2020)]{Minaev2019} Minaev, P.~Y. \& Pozanenko,
A.~S.\ 2020, \mnras, 492, 1919

\bibitem[Nava et al.(2012)]{Nava2012} Nava, L., Salvaterra, R., Ghirlanda,
G., et al.\ 2012, \mnras, 421, 1256

\bibitem[Paczynski(1986)]{Pac1986} Paczynski, B.\ 1986, \apjl, 308, L43

\bibitem[Paczynski \& Rhoads(1993)]{Pac1993} Paczynski, B. \& Rhoads, J.~E.\
1993, \apjl, 418, L5

\bibitem[Parsotan \& Lazzati(2022)]{Pars2022} Parsotan, T. \& Lazzati, D.\ 2022, \apj, 926, 104

\bibitem[Pe'er(2008)]{Pe2008} Pe'er, A.\ 2008, \apj, 682, 463

\bibitem[Pe'er \& Ryde(2011)]{Pe2011} Pe'er, A., \& Ryde, F.\ 2011, \apj,
732, 49

\bibitem[Pe'er et al.(2015)]{Pe2015} Pe'er, A., Barlow, H., O'Mahony, S., et
al.\ 2015, \apj, 813, 127

\bibitem[Piran(1999)]{Pi1999} Piran, T.\ 1999, \physrep, 314, 575

\bibitem[Rees \& Meszaros(1994)]{Rees1994} Rees, M.~J., \& Meszaros, P.\
1994, \apjl, 430, L93

\bibitem[Rees \& M{\'e}sz{\'a}ros(2005)]{Ree2005} Rees, M.~J., \& M{\'e}sz{%
\'a}ros, P.\ 2005, \apj, 628, 847

\bibitem[Rossi et al.(2002)]{Rossi2002} Rossi, E., Lazzati, D., \& Rees,
M.~J.\ 2002, \mnras, 332, 945

\bibitem[Ruffini et al.(2013)]{Ru2013} Ruffini, R., Siutsou, I.~A., \&
Vereshchagin, G.~V.\ 2013, \apj, 772, 11

\bibitem[Ryde(2004)]{Ry2004} Ryde, F.\ 2004, \apj, 614, 827

\bibitem[Ryde(2005)]{Ry2005} Ryde, F.\ 2005, \apjl, 625, L95

\bibitem[Ryde \& Pe'er(2009)]{Ry2009} Ryde, F., \& Pe'er, A.\ 2009, \apj,
702, 1211

\bibitem[Ryde et al.(2010)]{Ry2010} Ryde, F., Axelsson, M., Zhang, B.~B., et
al.\ 2010, \apjl, 709, L172

\bibitem[Ryde et al.(2017)]{Ry2017} Ryde, F., Lundman, C., \& Acuner, Z.\
2017, \mnras, 472, 1897

\bibitem[Sari \& Piran(1999)]{Sari1999} Sari, R. \& Piran, T.\ 1999, \apj,
520, 641

\bibitem[Song \& Meng(2022)]{Song2022} Song, X.-Y. \& Meng, Y.-Z.\ 2022, \mnras, 512, 5693

\bibitem[Tang et al.(2021)]{Tang2021} Tang, Q.-W., Wang, K., Li, L., et al.\ 2021, \apj, 922, 255

\bibitem[Thompson(1994)]{Thom1994} Thompson, C.\ 1994, \mnras, 270, 480

\bibitem[Tsvetkova et al.(2017)]{Tsve2017} Tsvetkova, A., Frederiks, D.,
Golenetskii, S., et al.\ 2017, \apj, 850, 161

\bibitem[Uhm \& Zhang(2014)]{Uhm2014} Uhm, Z.~L. \& Zhang, B.\ 2014, Nature
Physics, 10, 351

\bibitem[Vereshchagin \& Siutsou(2020)]{Vere2020} Vereshchagin, G.~V. \& Siutsou, I.~A.\ 2020, \mnras, 494, 1463

\bibitem[Vereshchagin et al.(2022)]{Vere2022} Vereshchagin, G., Li, L., \& B{\'e}gu{\'e}, D.\ 2022, \mnras, 512, 4846

\bibitem[von Kienlin et al.(2020)]{Kien2020} von Kienlin, A., Meegan, C.~A.,
Paciesas, W.~S., et al.\ 2020, \apj, 893, 46

\bibitem[Vurm \& Beloborodov(2016)]{Vur2016} Vurm, I., \& Beloborodov,
A.~M.\ 2016, \apj, 831, 175

\bibitem[Vyas et al.(2021)]{Vyas2021} Vyas, M.~K., Pe'er, A., \& Eichler, D.\ 2021, \apj, 908, 9

\bibitem[Wang et al.(2020)]{Wang2020} Wang, K., Lin, D.-B., Wang, Y., et al.\ 2020, \apj, 899, 111

\bibitem[Wang et al.(2022)]{Wang2022} Wang, Y., Zheng, T.-C., \& Jin, Z.-P.\ 2022, arXiv:2205.08427

\bibitem[Wygoda et al.(2016)]{Wygo2016} Wygoda, N., Guetta, D., Mandich,
M.~A., \& Waxman, E.\ 2016, \apj, 824, 127

\bibitem[Xue et al.(2019)]{Xue2019} Xue, L., Zhang, F.-W., \& Zhu, S.-Y.\
2019, \apj, 876, 77

\bibitem[Yamazaki et al.(2020)]{Yama2020} Yamazaki, R., Sato, Y., Sakamoto, T., et al.\ 2020, \mnras, 494, 5259

\bibitem[Yang et al.(2020)]{Yang2020} Yang, J., Chand, V., Zhang, B.-B., et
al.\ 2020, \apj, 899, 106

\bibitem[Yi et al.(2020)]{Yi2020} Yi, S.-X., Wu, X.-F., Zou, Y.-C., et al.\
2020, \apj, 895, 94

\bibitem[Yu et al.(2015)]{Yu2015} Yu, H.-F., van Eerten, H.~J., Greiner, J.,
et al.\ 2015, \aap, 583, A129

\bibitem[Yu et al.(2016)]{Yu2016} Yu, H.-F., Preece, R.~D., Greiner, J., et
al.\ 2016, \aap, 588, A135

\bibitem[Zhang \& M{\'e}sz{\'a}ros(2002)]{ZhMe2002} Zhang, B., \& M{\'e}sz{%
\'a}ros, P.\ 2002a, \apj, 571, 876

\bibitem[Zhang et al.(2006)]{Zhang2006} Zhang, B., Fan, Y.~Z., Dyks, J., et
al.\ 2006, \apj, 642, 354

\bibitem[Zhang et al.(2007)]{ZhaB2007} Zhang, B., Liang, E., Page, K.~L., et
al.\ 2007, \apj, 655, 989

\bibitem[Zhang(2011)]{ZhaB2011} Zhang, B.\ 2011, Comptes Rendus Physique,
12, 206

\bibitem[Zhang \& Yan(2011)]{ZhangYan2011} Zhang, B., \& Yan, H.\ 2011, \apj%
, 726, 90

\bibitem[Zhang(2020)]{ZhaB2020} Zhang, B.\ 2020, Nature Astronomy, 4, 210

\bibitem[Zhang et al.(2011)]{ZhaBB2011} Zhang, B.-B., Zhang, B., Liang,
E.-W., et al.\ 2011, \apj, 730, 141

\bibitem[Zhang et al.(2018a)]{ZhaBB2018} Zhang, B.-B., Zhang, B.,
Castro-Tirado, A.~J., et al.\ 2018a, Nature Astronomy, 2, 69

\bibitem[Zhang et al.(2018b)]{ZhangBB18b} Zhang, B.-B., Zhang, B., Sun, H.,
et al.\ 2018b, Nature Communications, 9, 447

\bibitem[Zhang et al.(2021)]{ZhangBB2021} Zhang, B.-B., Liu, Z.-K., Peng, Z.-K., et al.\ 2021, Nature Astronomy, 5, 91

\bibitem[Zhang et al.(2021b)]{ZhangZ2021} Zhang, Z.~J., Zhang, B.-B., \& Meng, Y.-Z.\ 2021, arXiv:2109.14252

\bibitem[Zhao et al.(2022)]{Zhao2022} Zhao, P.-W., Tang, Q.-W., Zou, Y.-C., et al.\ 2022, \apj, 929, 179

\end{thebibliography}
\end{document}